\documentclass[9pt,lineno]{article}

\usepackage[margin=1.2in]{geometry}
\usepackage{url}
\usepackage{hyperref}
\usepackage{graphicx}
\usepackage[version=4]{mhchem}
\usepackage{siunitx}
\usepackage{amsmath}
\usepackage{amssymb}
\usepackage{authblk}
\newcommand\numberthis{\addtocounter{equation}{1}\tag{\theequation}}

\title{How adaptive immunity constrains the composition and fate of large bacterial populations}

\author[1]{Madeleine Bonsma-Fisher}
\author[1]{Dominique Souti\`ere} 
\author[1,2]{Sidhartha Goyal}
\date{}

\affil[1]{Department of Physics, University of Toronto, 60 St George St, Toronto, ON M5S 1A7}
\affil[2]{Institute of Biomaterials \& Biomedical Engineering, University of Toronto, 164 College Street, Toronto, ON M5S 3G9} 
\begin{document}

\maketitle

\begin{abstract}
Features of the CRISPR-Cas system, in which bacteria integrate small segments of phage genome (spacers) into their DNA to neutralize future attacks, suggest that its effect is not limited to individual bacteria but may control the fate and structure of whole populations. Emphasizing the population-level impact of the CRISPR-Cas system, recent experiments show that some bacteria regulate CRISPR-associated genes via the quorum sensing (QS) pathway. Here we present a model that shows that from the highly stochastic dynamics of individual spacers under QS control emerges a rank-abundance distribution of spacers that is time-invariant, a surprising prediction that we test with dynamic spacer-tracking data from literature. This distribution depends on the state of the competing phage-bacteria population, which due to QS-based regulation may coexist in multiple stable states that vary significantly in their phage-to-bacterium ratio, a widely used ecological measure to characterize microbial systems.
\end{abstract}

\section{Introduction}

Complex communities of microorganisms are important ecological forces in almost every environment from hot springs \cite{Ward2006} to humans \cite{Schwabe2013, Collins2014, Korem2015, Muhlebach2018, OToole2017}. Phages, viruses which infect bacteria, are integral components of microbial populations: phage predation has been shown to strongly influence bacterial evolution, diversity, and numbers \cite{Heidelberg2009, Suttle2007}. To counter phages, bacteria have evolved many and complex immune mechanisms \cite{Doron2018}. CRISPR-Cas is one such defense mechanism which is both adaptive and heritable, \textit{i.e.} it not only learns from past infections but also passes this knowledge to future generations. Many models have addressed the effects of CRISPR-Cas on microbial populations, but a conceptual vacuum remains: What experimental features of natural populations should be measured to compare with model predictions? 

CRISPR-Cas machinery for adaptive immunity allows bacteria to acquire unique genetic elements (called spacers) from prior phage encounters to specifically target and evade recurrent attacks. 
The spacers are 10s of nucleotides long, and at each encounter may be acquired from any of the 100s of possible locations on the infecting phage genome (called protospacers). 
Since individual spacers are distinguishable and because they are integrated in the genome, the result is a lineage of cells that can be identified by its spacer(s). The fate of an individual lineage, however, is subject to large fluctuations due to the stochastic dynamics of individual bacteria in a large rapidly evolving population. Experiments show that the abundance of individual spacers in a bacterial population under phage attack is indeed highly dynamic and varies over several orders of magnitude from one spacer to the next \cite{Andersson2008a, Tyson2008, Heidelberg2009, Paez-Espino2013, Paez-Espino2015}. This leads to a natural question: What controls spacer diversity and abundance; in other words, how does recurrent phage attack alter the structure and composition of interacting spacer-marked lineages in a bacterial population? 

Several previous models have addressed the role and dynamics of observed diversity of spacer types \cite{He2010, Weinberger2012a, Childs2012, Haerter2012, Han2013a, Childs2014, Bradde2017, Han2017} in a qualitative way: (1) how system parameters such as phage adsorption rate \cite{Han2017}, spacer acquisition rate \cite{Childs2012, Han2017}, and phage mutation and recombination \cite{Han2013a} affect spacer diversity, (2) how increasing diversity promotes population stability \cite{Childs2012, Childs2014}, and (3) have reproduced the observed asymmetry in diversity along the locus in natural populations \cite{He2010,Weinberger2012a, Han2013a} by modelling biased acquisition at the leader end of the CRISPR locus. Most recently, Bradde {\it et al.} \cite{Bradde2017} showed a connection between spacer acquisition rates and spacer effectiveness to spacer diversity. To make a direct connection with data, we analyzed sequencing data from Paez-Espino {\it et al.} \cite{Paez-Espino2013}, a co-evolution experiment with phage and bacteria which tracked spacer dynamics. Our analysis shows that despite rapid turnover of individual spacer types the spacer rank-abundance distribution quickly stabilizes, which is a new and striking observation that previous models have not addressed. 

Recently, similar questions about diversity in the adaptive immune system have gained traction in the context of vertebrates which generate and maintain a large population of specialized immune cells that, as a group, contain an extremely diverse set of binding sites that individually recognize different viruses.  Like spacer abundance, the abundance of individual binding sites is highly variable \cite{Weinstein2009, Zarnitsyna2013, Desponds2016}. This observation has led to the suggestion that a broad abundance distribution of binding sites may strike a balance between generating a rapid response against likely invaders with capturing new invaders \cite{Desponds2016}. Although this is hard to test in vertebrates, laboratory experiments that alter bacterial population composition synthetically show that bacteria are more successful at fending off phages as their population-level spacer diversity increases \cite{Houte2016}. How the dynamics of individual bacterial lineages shape spacer diversity and how diversity in spacer sequences or \textit{types} relates to diversity in spacer \textit{abundances} remains unanswered.

Beyond the role of individual spacer lineages in shaping population structure, recent experiments have shown that bacterial populations exert top-down control on the CRISPR system: two species of bacteria have been observed to regulate their CRISPR-Cas systems in response to cell density \cite{Hoyland-Kroghsbo2016, Patterson2016}. Interestingly, this control acts via the quorum sensing pathway, a pathway which also controls population-level responses such as virulence. This suggests a different paradigm where the effects of CRISPR-Cas need to be considered at the collective population level, rather than at the level of individual cells. Previous population-level models have not addressed this effect \cite{Heilmann2010, Levin2010, Haerter2011, Weinberger2012c, Iranzo2013a, Levin2013, Santos2014, Berezovskaya2014, Westra2015, Bradde2017, Han2017, Ali2017, Weissman2017}, and modelling efforts addressing CRISPR-Cas regulation have focused on the relevant gene circuits and production of transcribed spacers called CRISPR RNAs (crRNAs), not on the population-level effects of regulation \cite{Djordjevic2012, Djordjevic2013, Guzina2017}.

We build a model that addresses the two aforementioned fundamental and unaddressed aspects of the CRISPR-Cas system: (1) our model shows how stable rank-abundance distributions may arise despite rapid turnover of individual spacer types that are identical in their ability to provide immunity, and (2) our model shows that density-dependent regulation of CRISPR-Cas admits a bistable state at the population level where the phage-bacterial population can be stable with two different configurations under the same external conditions. We further argue how having the knowledge of spacer diversity along with bistable states may shed light on the fate of natural microbial populations.

\section*{Model}
\begin{figure}
\centering
\includegraphics[width=0.65\linewidth]{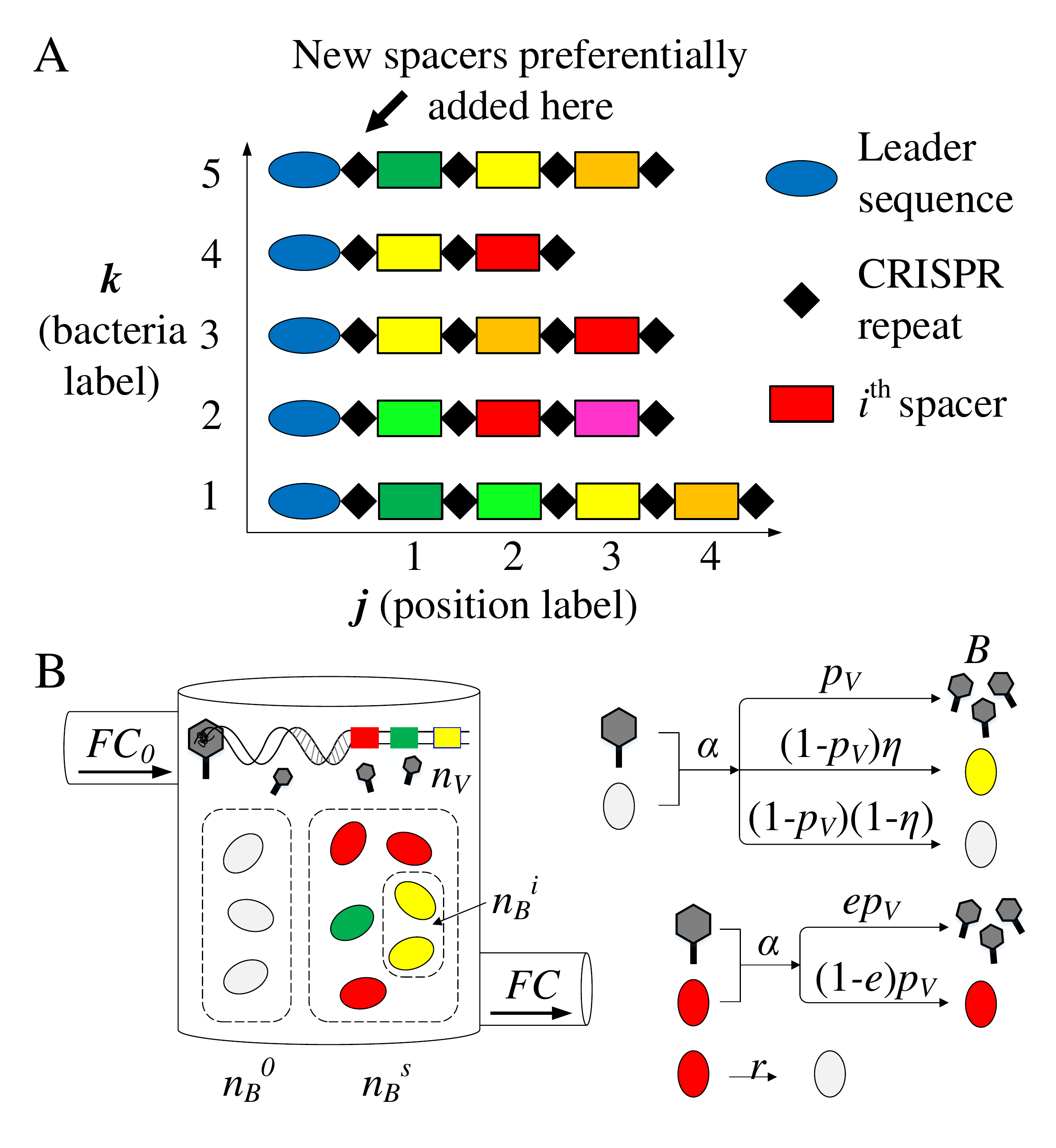} 
\caption{
(A) CRISPR locus:
Small ($\sim$ 30 nt) samples of invasive phage DNA called spacers (colored rectangles) are incorporated into the CRISPR genetic locus. 
Spacers are separated by short ($\sim$ 30 nt) sequences called repeats (black diamonds). 
Multiple spacers can be stored at a CRISPR locus, resulting in a genetic record of immunization \cite{Barrangou2014}. 
In our analysis of the experimental data shown in Figure \ref{simulation_exp}A-C, we identify spacers with a type $i$, a locus position $j$, and a bacterium $k$.  
(B) In our model, bacteria and phages interact in a chemostat (flow cell) with a constant inflow and outflow rate $F$. 
Nutrients flow into the chemostat at a fixed concentration $C_0$.
Phages are assumed to be identical with a large, fixed number of possible protospacers. 
Phages adsorb to bacteria with rate $\alpha$ and successfully infect and kill naive bacteria with probability $p_V$. 
Each bacterium can acquire a single spacer ($j = 1$). 
Spacers are tracked in the population as the number of bacteria containing a spacer of type $i$, $n_B^i$. 
If a naive bacterium survives an infection, it can acquire a spacer with probability $\eta$. 
All spacers are assumed to be equally effective: the probability of phage success in an infection is reduced by $e$ if a bacterium has a spacer. 
Bacteria with spacers revert to naive bacteria by losing a spacer with rate $r$.
} \label{mean_field_diagram}
\end{figure}

Adaptive immunity in bacteria is controlled by a set of Cas proteins, 
which in a nutshell accomplish two different tasks. (1) When an invading phage inserts its genome into a bacterial cell but is not successful in killing the bacterium, Cas proteins take a small piece of phage genome and insert it into the bacterial genome at a specific site called the CRISPR locus. (2) During a subsequent phage attack, the bacterium can use the information stored in the CRISPR locus to recognize the invading phage and neutralize it. Multiple spacers can be stored at a CRISPR locus, providing a genetic record of immunization that is inherited during DNA replication. The immunization record in principle can be read via next generation sequencing and provides a rich presence/absence observable: the binary variable $s_{ijk}$ indicating whether spacer type $i$ is in locus position $j$ in host bacterium $k$ (Figure \ref{mean_field_diagram}A, SI equation \ref{sijk_2}).

We model the abundance of the $i^{\textrm th}$ spacer, $n_B^i(t)$, which is obtained by summing over all bacteria and locus positions, {\it i.e.} $n_B^i(t) = \sum_{j,k}s_{ijk}(t)$. An important simplifying assumption of our model is that each locus has at most one spacer, {\it i.e.} $j = 1$; this assumption is borne out of analysis of a laboratory experiment that shows that spacer dynamics stabilize rapidly within tens of generations with each bacterium predominantly having one new spacer (see SI section \ref{app1-2} for details of data analysis) \cite{Paez-Espino2013}. Additionally, a model that allowed more than one spacer also found that only the most recently acquired spacers dominate the dynamics \cite{Childs2012}. 
With this assumption, the abundance of individual spacer types can be mapped to the number of bacteria with a particular spacer, $n_B^i$. In addition, we assume each spacer to have equal effectiveness; this is both a simplifying assumption and also acknowledges our lack of experimental knowledge about differences among spacers and their effectiveness. 

To capture the inherent stochastic nature of spacer dynamics, we model the probability distribution $P(n_B^0, \{n_B^i\}, n_V, C, t)$, which is the probability at time $t$ of observing $n_B^0$ bacteria without spacers, $\{n_B^i\}$ bacteria with spacer type $i$, $n_V$ phages, and a nutrient concentration of $C$. Interactions included in the model are illustrated in Figure \ref{mean_field_diagram}B and described in detail in SI section \ref{app2-1}. This construction highlights another important simplifying assumption which is also valid for short timescales: lack of phage diversity, { \it i.e.} all phages are assumed to be identical. In addition, we model the phage-bacteria population in a flow cell or \textit{chemostat}, a well-stirred vessel in which nutrients flow in at a constant rate and concentration and the mixture flows out with the same rate. A chemostat is not only comparable to periodic dilution experiments in the laboratory, it is also a reasonable approximation of real-world microbial populations from a gutter to a gut.  In many of these natural environments, nutrients and waste flow in and out --- the environment is not static like a petri dish. 
Additionally, the chemostat flow rate $F$ is an experimental ``knob'' that can be used to tune a population-level bifurcation we describe later.

Our stochastic model has a corresponding mean-field or population-level description for average values of the different random variables, each represented by the same symbol as their corresponding random variable. At the mean-field level, all the spacer-containing bacteria can be pooled into a single variable $n_B^s = \sum_i n_B^i$, and the number of bacteria without spacers is $n_B^0$. 
The mean-field equations are given below. 
Parameter descriptions can be found in Figure \ref{mean_field_diagram} and SI Table \ref{tab:params}. We assume that the bacterial growth rate is linear with the concentration of nutrients $C$; relaxing this assumption does not qualitatively change our results (see SI section \ref{app3-2}).

%\small
\footnotesize
\label{eqn:mean_field}
\begin{align}
\frac{dC}{dt} = &
       \underbrace{FC_0}_\text{flow in}
       - \underbrace{g C(n_B^s + n_B^0)}_\text{bacterial growth}
       -\underbrace{FC}_\text{flow out} \nonumber \\
\frac{dn_V}{dt} = & 
       \underbrace{-\alpha n_V(n_B^s + n_B^0)}_\text{phage adsorption}
       + \underbrace{\alpha Bp_V n_V (n_B^s(1-e) + n_B^0)}_\text{phage burst and bacterial lysis} - {Fn_V} \numberthis \\
\frac{dn_B^0}{dt} = & {gCn_B^0}
       -{\alpha p_Vn_Vn_B^0}
       -\underbrace{\alpha (1-p_V)\eta n_V n_B^0}_\text{spacer acquisition}   + \underbrace{r n_B^s}_\text{spacer loss}
	-{F n_B^0}  \nonumber \\
\frac{dn_B^s}{dt}=& {gCn_B^s}
	-{\alpha p_V (1-e) n_V n_B^s}
	+{\alpha (1-p_V)\eta n_V n_B^0}
	-{r n_B^s}
	-{F n_B^s} \nonumber
\end{align}

\normalsize

\section*{Results}

\begin{figure}
	\centering
	\includegraphics[width=0.95\linewidth]{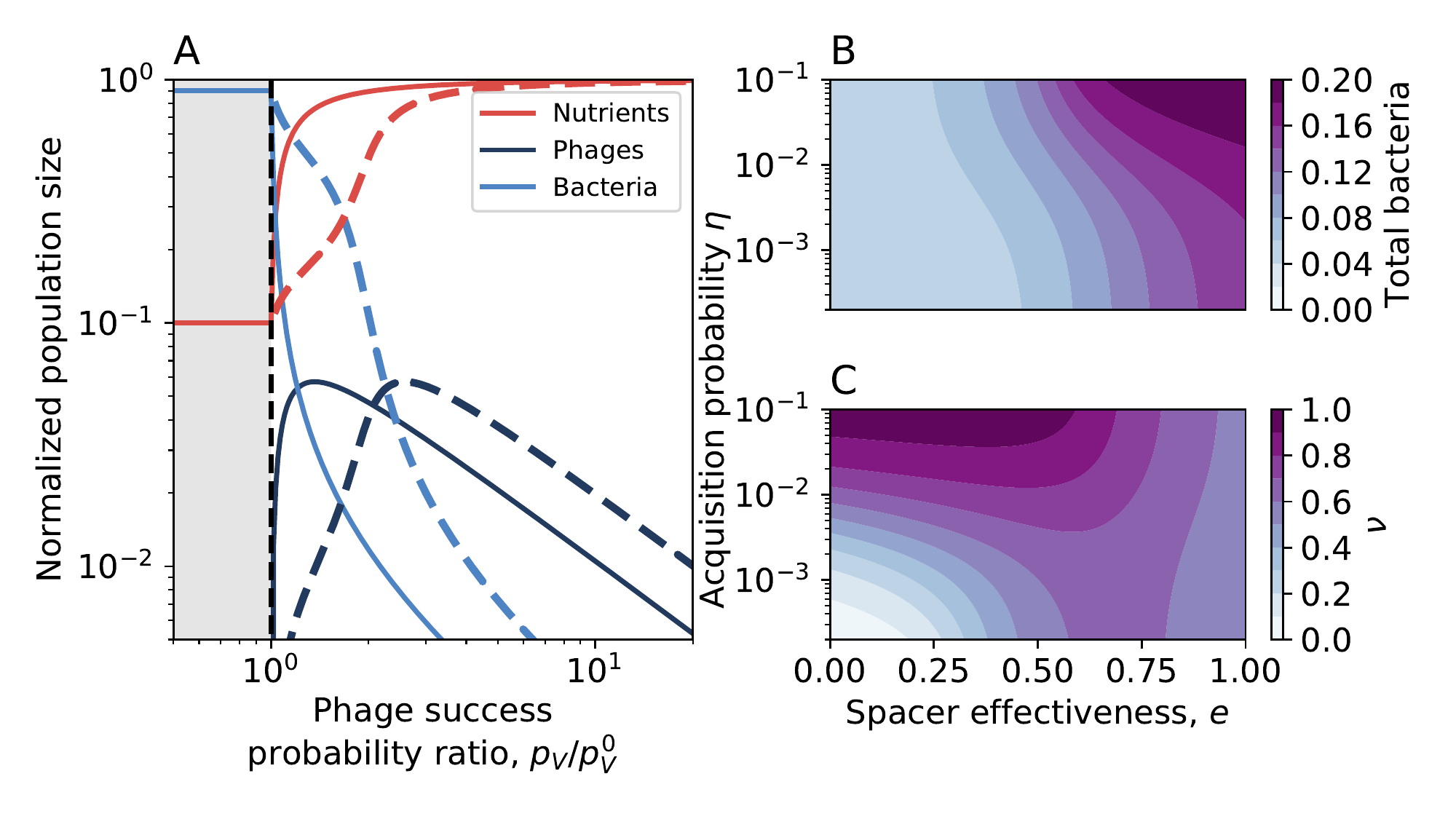} 
	\caption{(A) Bacteria, phage, and nutrients at steady state as a function of the  probability of phage success $p_V$ for a model without CRISPR (spacer effectiveness $e=0$, solid lines) and for a model where bacteria have CRISPR systems and are able to acquire spacers ($e=0.5$, dashed lines). 
Population sizes are normalized by the inflow nutrient concentration $C_0$, and phages are additionally scaled by the burst size $B$.
As the probability of phage success $p_V$ increases, bacteria decrease in number. 
Below $p_V = p_V^0$, phages cannot persist and the fraction of bacteria with spacers is $0$.  
Phages increase with increasing $p_V$ and then decrease at high $p_V$ because the bacterial population is too small to support more phages.
(B) Normalized total bacteria as a function of spacer acquisition probability $\eta$ and spacer effectiveness (equal for all spacers). 
(C) Fraction of bacteria with spacers ($\nu$) as a function of $\eta$ and $e$.}
	
	 \label{zyx_vs_pv0_CRISPR}
\end{figure}

\subsection{Mean-field steady-states} 
For phages to invade a bacterial population that is stable in a chemostat, their probability of successfully infecting bacteria without the benefits of adaptive immunity, $p_V$, needs to be above a certain minimum value given by $p_V^0 = \frac{1}{B}\left(\frac{gf}{(1-f)\alpha} +1\right)$, where $f=F/(gC_0)$.
For $\frac{gf}{(1-f)\alpha} << 1$ (satisfied at the parameters we use for low flow rates), $p_V^0$ is approximately $1/B$: phages must succeed approximately every $1/B$ interactions in order to persist in the population. $p_V^0$ is surprisingly small for realistic values of the burst size $B$; for example if $B=100$, then $p_V^0 \approx 0.01$.
As $p_V$ rises above the threshold value, 
the steady-state phage population, $n_V$, first rises while the bacterial population decreases as they get killed by phages. 
Interestingly, the steady-state bacterial population keeps decreasing with increasing $p_V$, but the phage population exhibits a non-monotonic behavior with a maximum population size at an intermediate value of $p_V^\star = p_V^0 + \sqrt{\frac{p_V^0}{f}(p_V^0 - \frac{1}{B})}$. This steady-state behavior is qualitatively the same for bacteria with adaptive immunity ($e>0$) as for bacteria without adaptive immunity ($e = 0$). Quantitatively, however, bacteria always fare better in the presence of adaptive immunity (Figure \ref{zyx_vs_pv0_CRISPR}A). One surprising observation is that the minimum success probability required for phages to invade a bacterial culture is independent of adaptive immunity. This is because there are no bacteria with spacers at steady state below $p_V = p_V^0$, and as a result, phage invasion occurs independently of the CRISPR system (SI Figure \ref{nu_phage_vs_pv}). 

Much like increasing $p_V$, an increasing spacer effectiveness $e$ causes the total number of bacteria at steady-state to increase monotonically (Figure \ref{zyx_vs_pv0_CRISPR}B), since a bacterium with a spacer is less likely to be killed by phages as $e$ increases. However, even for $e>0$, not all bacterial cells in a population have a spacer, and the steady state fraction of the bacterial population with spacers, $\nu$, is governed by a balance of spacer acquisition, $\eta$ spacer loss, $r$, and the effect of $e$ on the bacterial population. As a result, the steady-state level of bacteria can increase by either increased spacer acquisition or improved spacer effectiveness; contours in Figure \ref{zyx_vs_pv0_CRISPR}B show the tradeoff between $\eta$ and $e$ that maintains bacterial population size. 

In contrast to total bacterial population, $\nu$ first increases as $e$ increases but reaches a maximum at an intermediate value of $e$ (Figure \ref{zyx_vs_pv0_CRISPR}C). This can be understood as $\nu$ qualitatively tracking the phage population size, which shows a peak at intermediate spacer effectiveness (SI Figure \ref{large-alpha}). Qualitatively, this behavior is similar to the total phage population having a non-monotonic behavior with increasing $p_V$. 

\subsection{Spacer rank-abundance distributions}

\begin{figure}
\centering
\includegraphics[width=0.8\textwidth]{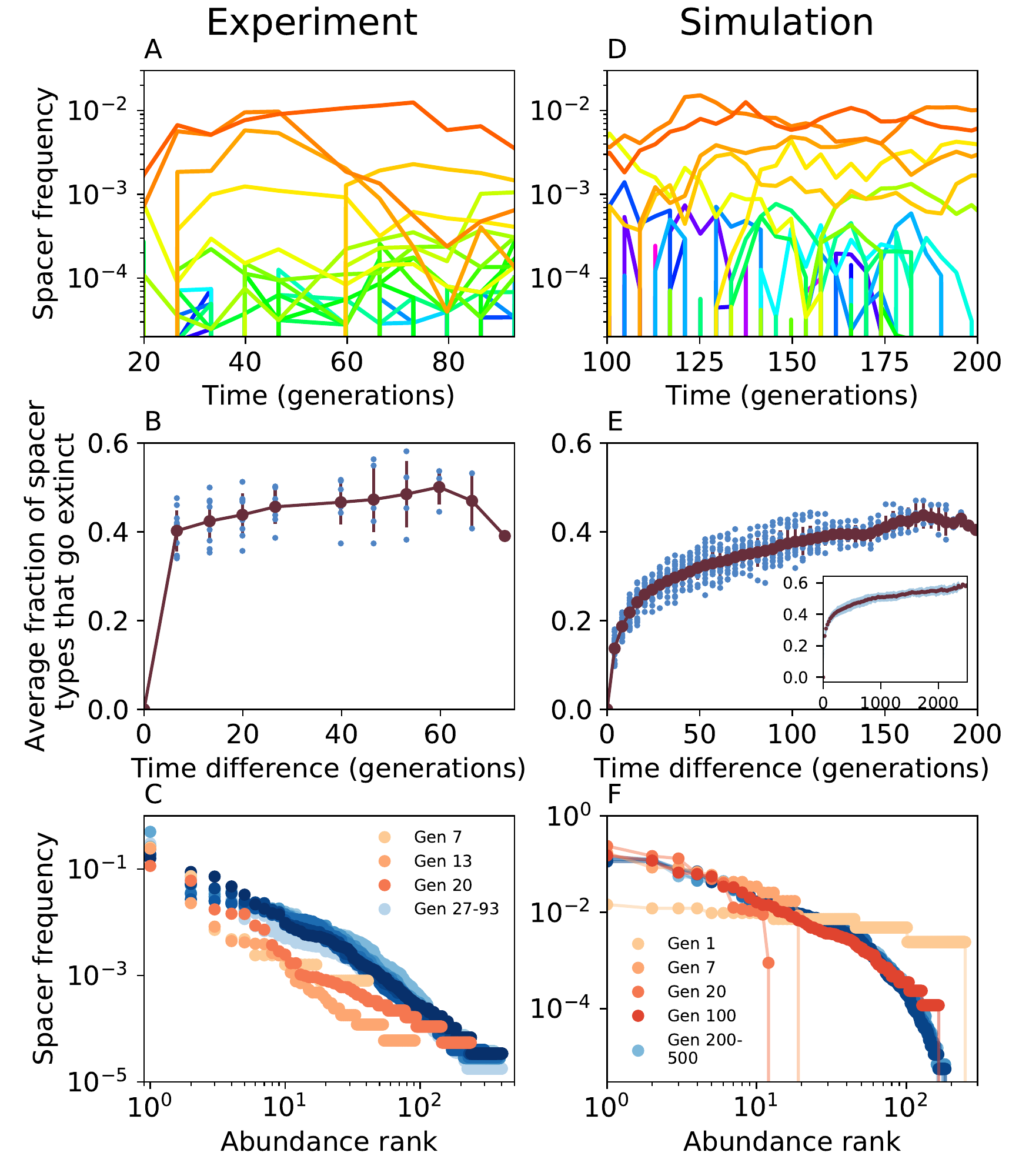} 
\caption{Comparison of spacer distributions between simulations (D-F) and experimental data from \cite{Paez-Espino2013} (A-C). 
(A) and (D) Subset of spacer type trajectories over time in generations for experimental data (A) and simulated data with $\eta = 10^{-5}$ and $e=0.387$ (D). Qualitative simulation results are insensitive to the choice of $e$ and $\eta$. Individual spacer abundances fluctuate throughout the experiment and simulation.  
(B) As a function of time difference at steady-state (Day 4 / generation 26 onwards), we calculated the fraction of spacer types that have gone extinct (blue points), averaged over all times (red line). Error bars are standard deviation.
(E) Same as (B) but for simulated data from generation 300 to 500. A large fraction of spacer types go extinct during the course of the experiment and simulation. Inset: fraction of spacer types that go extinct for a long simulation from generation 500 to 3000. The fraction that go extinct continues to increase with time.
(C) The rank-abundance distribution of spacer clone sizes reaches a steady state in the experiment after about 20 generations (Day 3 of the experiment). Darker blue indicates later times.
(F) The distribution of spacer clone sizes reaches a steady state in the simulation after about 100 generations. Plotted is the same quantity as in (C).
Even after the distribution of clone sizes has reached steady state, individual spacer types experience continual turnover.} \label{simulation_exp}
\end{figure} 

Even at steady-state with stable populations of phage and bacteria, the individual spacer abundances in the bacterial population are highly dynamic and vary significantly over time. This has been seen most directly in laboratory experiments \cite{Paez-Espino2013, Paez-Espino2015} but has also been observed in natural samples such as a hypersaline lake \cite{Emerson2013}, human saliva \cite{Pride2011}, and acid mine drainage \cite{Andersson2008a, Weinberger2012a}. 
This continual spacer turnover is influenced by bacterial reproduction and death, spacer acquisition, and spacer loss, all of which have been observed in natural and laboratory populations. 
 
In our stochastic model, we keep track of individual spacer acquisition and loss events. Not surprisingly, we find that spacer abundances fluctuate over time (see Figure \ref{simulation_exp}D, Figure \ref{simulation_exp}E, and SI section \ref{app4-2}). However, we also find that the spacer rank-abundance distribution reaches a stationary state from an initial state with no spacers, shown in Figure \ref{simulation_exp}F and SI section \ref{app2-3}. Not only does the spacer distribution in our simple model reach a stationary state while individual spacers turn over rapidly, it also shows 1000-fold variation in spacer abundances despite the fact that all spacers are functionally identical in our model and provide resistance to the same phage. The exact shape of the distribution depends on various parameters (see SI section \ref{app2-3}) and is well-approximated by a gamma distribution which has been used to describe species abundance distributions in ecology \cite{Dennis1984, Engen1996, Diserud2000, Plotkin2002} (SI section \ref{app2-4}).

To test predictions with data, we analyzed experimental data reported by Paez-Espino et al. \cite{Paez-Espino2013} from a bacterial population under constant phage attack. We summarized their raw sequencing data into the presence/absence tensor $s_{ijk}$ as shown in Figure \ref{mean_field_diagram}A, and we tracked dynamics of individual spacers $n_B^i(t) = \sum_{j,k} s_{ijk}(t)$.  Our analysis showed that the abundance of individual spacer types fluctuated throughout the 15 days ($\sim 80$ generations) of the experiment (Figure \ref{simulation_exp}A), with more than $40\%$ of spacers going extinct within a time difference of a few generations from any starting time (Figure \ref{simulation_exp}B). In contrast, we find that the spacer rank-abundance distribution reaches a stationary state, as shown in Figure \ref{simulation_exp}C. Notably, the rank-abundance distribution is broad with some spacers having a roughly 1000-fold higher abundance than others. However, in contrast to the intuition that highly abundant spacers may be more effective, these high-abundance spacers also experience continual turnover, shown in SI Figure \ref{spacer_turnover} (SI section \ref{app4-1}). Both the simulated and experimental data show similar mean times to extinction as a function of spacer abundance (SI section \ref{app4-2}), another indication of continual spacer turnover at steady-state.

In general our analysis highlights that individual spacer identity and abundance may not themselves be important but collectively may provide a time-invariant observable in the form of steady-state rank-abundance distributions. And somewhat counterintuitively, spacers need not be functionally different in their effectiveness or acquisition probability to get large variability in spacer abundances.

\subsection{Regulation of \textit{cas} expression}

\begin{figure}
\centering
\includegraphics[width=1\linewidth]{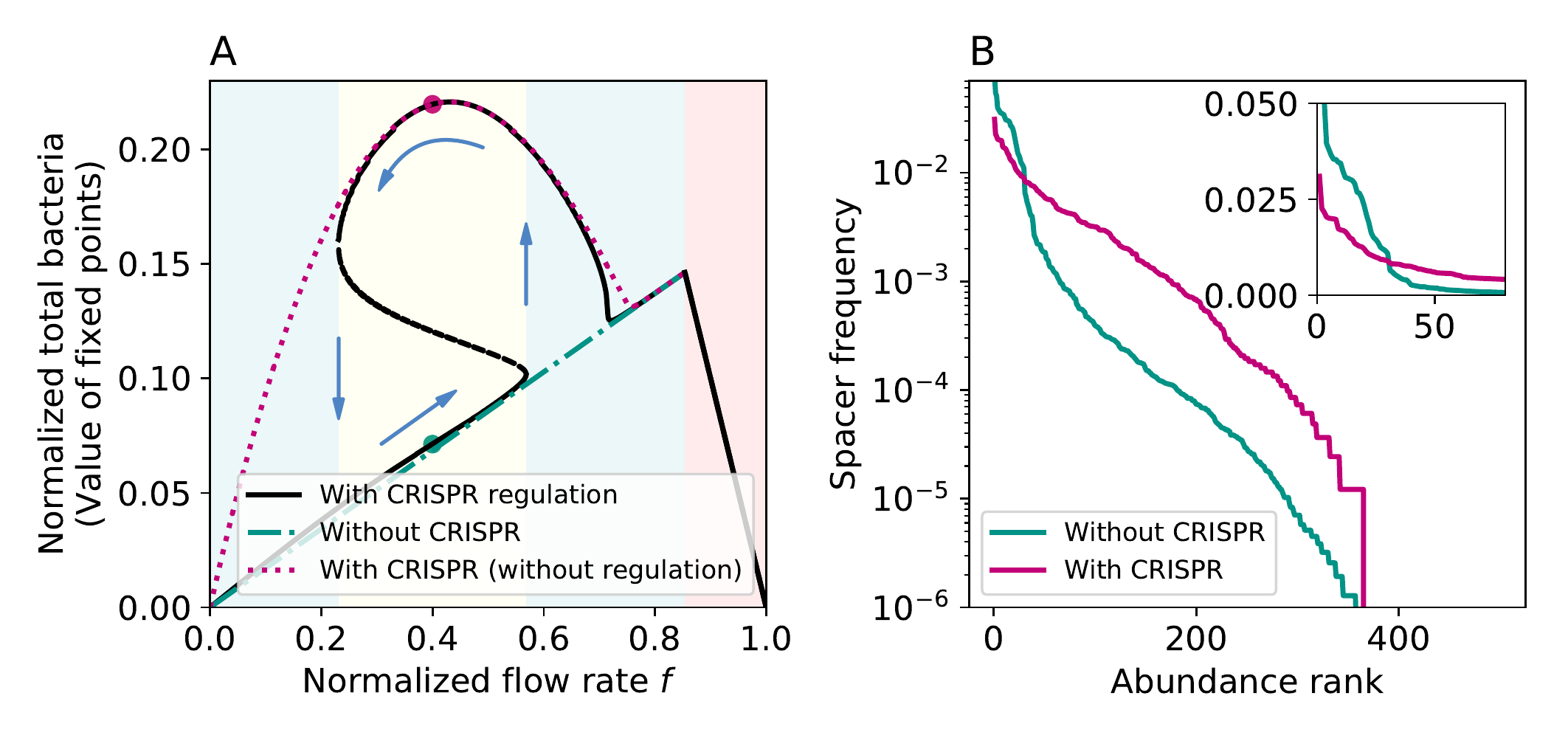} 
\caption{Bacterial upregulation of \textit{cas} gene expression at high density can induce bistability (yellow shaded area) as a function of the normalized chemostat flow rate $f = F/(gC_0)$, a parameter that is easy to tune experimentally. The blue shaded region is monostable, and in the pink shaded region phages cannot persist.
(A) The bacterial population size (solid black lines) exhibits hysteresis (blue arrows) between a low-expression, low-density state and a high-expression, high-density state. 
(B) The spacer rank-abundance distribution shape depends on the ecological state of the population. Plotted are two rank-abundance distributions from simulations of the high and low expression states respectively; population sizes for each distribution indicated by dots in (A). Inset: linear frequency scale.} \label{regulation}
\end{figure} 

Merely having an effective spacer, however, is not enough: to effectively neutralize phage, bacteria need to express \textit{cas} genes when under attack. Experimental work has shown that bacteria can regulate their CRISPR-Cas systems in response to cell density, controlled under the quorum sensing pathway \cite{Hoyland-Kroghsbo2016, Patterson2016}. A cell increases its expression of Cas proteins at high cell density in response to a high concentration of quorum sensing molecules and down-regulates its expression of Cas proteins at low cell density. To understand the role of cell density-dependent regulation of the CRISPR-Cas system, we made spacer effectiveness $e$ to be a function of cell-density:  $e(x) = e_{min} + (e_{max} - e_{min})\left(\frac{x^n}{x^n + x_0^n}\right)$, where $x$ is the normalized bacterial population size. This function is characterized by three numbers: minimum effectiveness, $e_{min}$, maximum effectiveness, $e_{max}$, and typical population size where the behavior changes from low to high effectiveness (see SI section \ref{app5-1}). Regulation of {\it cas} genes may also alter other parameters of the model such as acquisition, spacer loss, and growth rates, but we show in SI section \ref{app5-3} that adding regulation to other possible parameters independent of effectiveness has little effect, and that in conjunction with density-dependent effectiveness they do not change the qualitative features we describe below.

Notably, the dependence of spacer effectiveness on population size changes both the number and value of the steady-state fixed points. We find that the whole bacteria-phage-nutrient system undergoes a saddle-node bifurcation and is bistable for a range of parameters. The bistability results from a positive feedback that is established under QS control of the CRISPR-Cas system but is absent otherwise: the total bacterial population size increases with increasing spacer effectiveness, and in turn effectiveness increases as CRISPR-Cas is upregulated by higher bacterial density (SI Figure \ref{e_vs_x}). There are various parameters that can be used as the bifurcation parameter, but one that can be easily controlled in experimental systems and perhaps plays a role in natural systems is the normalized chemostat flow rate $f = F/(gC_0)$, which can also be thought of as the inverse of nutrient availability. Figure \ref{regulation} shows how bacterial and phage abundance varies as flow rate is changed in the presence of density-dependent regulation of the CRISPR-Cas system. At the two extremes, for low flow rate the system behaves with no adaptive immunity and bacterial (and phage) population size is low, while at high flow rate adaptive immunity kicks in and bacteria can maintain a higher population size. The phage population remains low at high flow rate both because bacteria are more resistant and because phages are removed from the system at a higher rate. At very high flow rate, phages go extinct and the bacterial population starts decreasing linearly with flow rate. For intermediate flow rate, the low and high states are both stable, allowing the system to be in either state. In principle these two population-level states could coexist and interact.

This bistable system may also exhibit hysteresis, which may have important ecological consequences, possibly functioning as a memory of past phage pressure or providing a switch-like behavior between ``on'' and ``off'' states of the CRISPR system.
Not only can the phage-to-bacterium ratio (called VPR) be significantly different between the two states but spacer composition and diversity can also be quite different (see Figure \ref{regulation}B and SI section \ref{app2-3}). 

Our model exhibits bistability quite generically for large parameter ranges but requires choosing an appropriately steep function for effectiveness (see discussion in SI section \ref{app5-1}). 

\section*{Discussion}   \label{discussion}

CRISPR-Cas is a unique system in that adaptive immunity is both hereditary and acquired. Its impacts on population dynamics are thus unlike any other immune system, and experimental observations must be interpreted with theory specific to the CRISPR-Cas system. Our analysis of experimental data yielded a striking result: rank-abundance spacer distributions are stable over time paralleling population-level stability, despite what looks like ongoing turnover in the abundances of individual spacer types. This overall stability suggests a need for a population-level approach in which questions about spacer diversity are addressed alongside questions about CRISPR-Cas regulation. In this framework, communities of bacteria function collectively more like a {\it single organism} capable of complex signalling and behavior than like a collection of individual bacteria undergoing selective dynamics. 

In this work, we propose and analyze a simplified model of interacting bacteria and phage in which bacteria regulate the CRISPR-Cas system in a density-dependent way, which in turn controls the spacer-marked {\it clonal} composition of the bacterial population under phage attack.
We find that the bacteria-phage population exhibits bistability with the possibility of co-existence between two ecologically different states. These two stable states may differ by orders of magnitude in the phage-to-bacterium ratio as well as differing in the spacer diversity and composition of the population. Our model also provides a framework where large variability in spacer abundance may arise due to population dynamics rather than due to individual parameters of spacers, since our model is neutral with no selective advantage for particular spacers. And finally, our model shows how a stable spacer rank-abundance distribution may emerge while individual spacer types turn over rapidly.

Sequencing provides an easy way to track spacers, which in turn provide a direct record of past interactions between a bacterium and its phages. Although there has not been much effort towards spacer tracking in individual bacteria, population-level spacer dynamics is becoming readily accessible both from laboratory experiments \cite{Paez-Espino2013, Paez-Espino2015, Weissman2017} and natural populations \cite{Heidelberg2009, Andersson2008a}. In the laboratory, both large variability and rapid turnover of individual spacer types have been observed. Understanding these dynamics is certainly interesting but requires much higher sampling and resolution than what is currently available \cite{Levy2015}. Also, acquiring such data, especially time-resolved, for natural systems such as microbial mats and acid mine drainage may not be practical. Here we show that the spacer rank-abundance distribution may provide a more useful time-invariant observable for understanding the underlying dynamics in both natural and laboratory systems; our work predicts that measuring spacer abundances in natural populations may reveal abundance distributions that are stable in time and potentially indicative of the environmental conditions despite differences on the level of spacer sequences between populations and over time. 

Even without phage diversity and phage mutations in our model, we reproduce important features of the spacer dynamics observed in recent laboratory experiments \cite{Paez-Espino2013}. 
In the presence of mutant phages, the net effectiveness of different spacers in providing immunity against phages may vary from one spacer to the next. We expect that a spacer's effectiveness will depend on the fraction of the phage population with a matching protospacer. This fitness difference between spacers will have consequences for the population dynamics, and some aspects have been addressed in experiments \cite{Deveau2008, Sun2013, Paez-Espino2015} and models \cite{Childs2012, Weinberger2012c, Han2013, Levin2013, Iranzo2013a, Childs2014, Weissman2017} and reviewed in \cite{England2013}. 

Multistability at the level of cellular states, where a fraction of the population switches to an alternate state, has been explored at length with implications from bet-hedging to lytic-lysogenic switching to antibiotic resistance and persistence \cite{Gore2009, Eldar2010, Norman2015, Tarnita2015, Symmons2016}. Similarly-structured populations are now being explored in contexts from healthy regenerating tissues to pathologies such as cancer \cite{Shaffer2017}. 
While recent models for large interacting microbial populations using a statistical mechanical approach \cite{Bunin2016, Tikhonov2016, Tikhonov2017, Biroli2017} show that ecological multistability akin to what is seen in a spin glass may be present in such populations, they remain experimentally inaccessible. A notable exception is Gore \textit{et al.}, who observed population-level bistability and coexistence between two cooperating yeast strains in a mixed culture \cite{Gore2009}. 
Here we provide one of the first examples of multistable, multispecies ecological states that may be readily accessible in experiments. 
We show that for a population of bacteria and phages the flow rate of a chemostat or dilution rate of a serially diluted population can serve as a bifurcation parameter. 
Since both nutrient concentration (which controls population density) and dilution rate are easy to control experimentally, ecological states in our phage-bacteria population should be readily accessible (see SI section \ref{app5-4}). 

\begin{figure}
\includegraphics[width=0.95\linewidth]{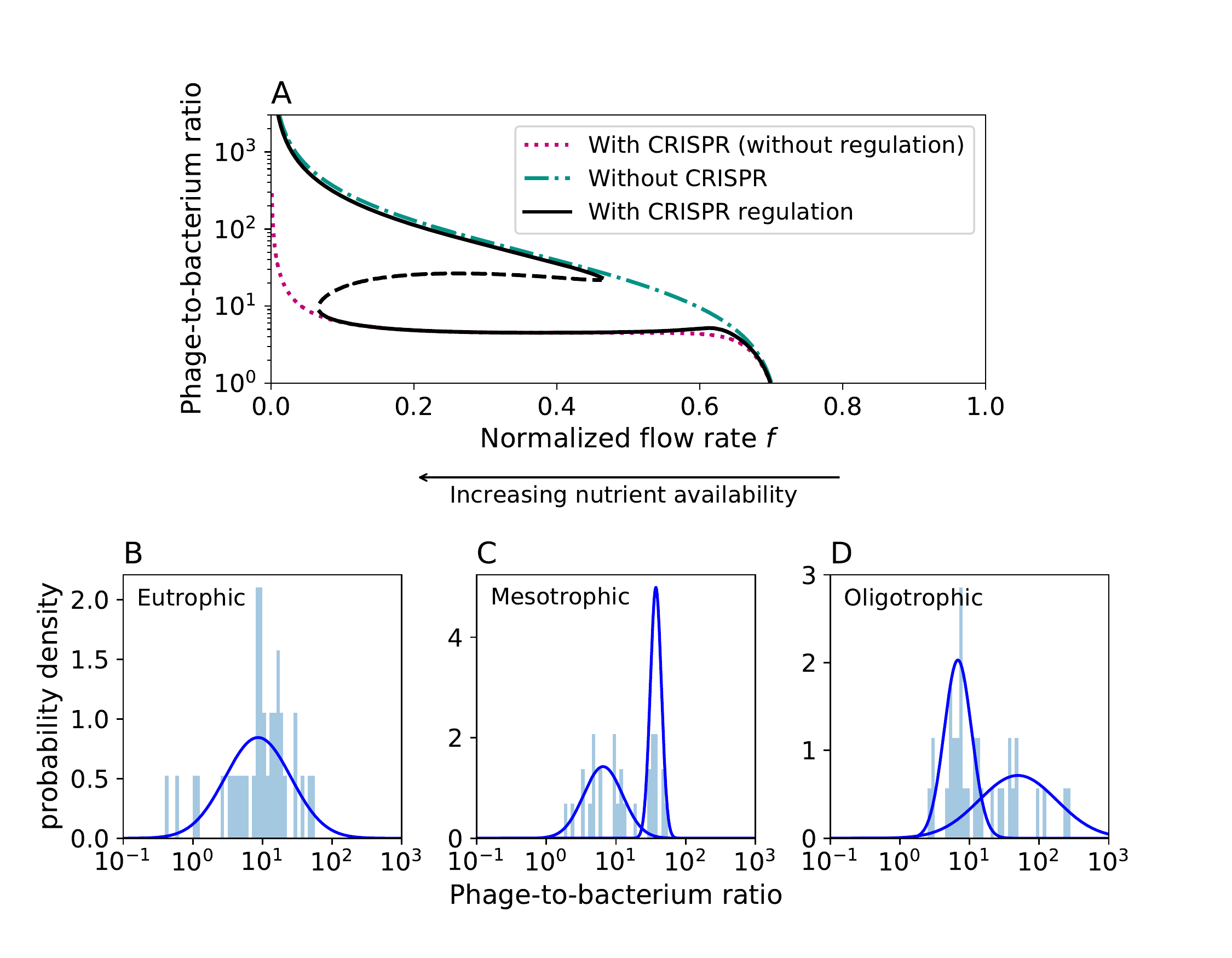} 
\caption{
(A) The phage-to-bacterium ratio (virus-to-prokaryote ratio, VPR) can differ by more than ten-fold between the two bistable states in the model (solid black lines). These values reflect the two underlying ecological states: VPR is low when bacteria are at high density and upregulate CRISPR-Cas expression, and VPR is high at low bacterial density and low CRISPR-Cas expression. 
(B)-(D) VPR histogram and fitted log-normal distributions for organisms from eutrophic (high nutrient), mesotrophic (moderate nutrient), and oligotrophic (low nutrient) environments (data from Parikka \textit{et al.} \cite{Parikka2017}). We fit one-dimensional Gaussian mixture models with one and two Gaussian distributions respectively to the data and chose the best-fitting model (blue line) using the Akaike Information Criterion (AIC). 
The data was fit better by a single Gaussian distribution for the eutrophic data ($\Delta$ AIC $=45.7$) and two Gaussian distributions for the mesotrophic ($\Delta$ AIC $=10.2$) and oligotrophic data ($\Delta$ AIC $=8.1$). For each fit, we calculated the likelihood that the not-chosen model was a better fit:  $e^{-\Delta \text{AIC}/2}$ \cite{Burnham2002}. This likelihood is $1.2 \times 10^{-10}$, $0.006$, and $0.017$ for the eutrophic, mesotrophic, and oligotrophic data respectively.}
\label{VPR}
\end{figure} 

In natural populations where phages and bacteria coexist, the phage-to-bacterium ratio, also called virus-to-prokaryote ratio (VPR), has been measured and reported for a wide range of conditions. While viruses are generally assumed to outnumber bacteria by a factor of ten \cite{Brussow2002, Suttle2007, Held2013, Santos2014}, the measured ratio can vary between samples by as much as a factor of $10^6$ \cite{Parikka2017}. 
The underlying factors and ecological significance of observed VPR values are not well understood. 
Our model predicts a variable phage-to-bacterium ratio for different parameters. Notably, the VPR for the low-expression branch of the bistable system is approximately ten times higher than for the high-expression branch (Figure \ref{VPR}A). 
These values reflect the two underlying ecological states: VPR is low when bacteria are at high density and upregulate CRISPR-Cas expression, and VPR is high when bacteria are at low density and have turned down CRISPR-Cas expression. 
This suggests that low observed VPR values may be indicative of an active bacterial defense system, while high VPR may correspond to a bacterial population strongly controlled by phages. With deep metagenomic sequencing it will be possible to measure VPR in natural environments for phage-bacteria species pairs that are known to interact, shedding more light on the significance of phage pressure in natural microbial communities. 

In our model, the normalized chemostat flow rate $f$ is inversely proportional to the inflow nutrient concentration $C_0$, which suggests that the model's VPR predictions and the ecological conditions under which CRISPR-Cas is advantageous may be impacted by nutrient availability. A study by Payet and Suttle \cite{Payet2013} found that phage production and phage-induced mortality of bacteria were both highest in marine samples when the water was most productive and nutrient-rich, while lysogens were more common when the water was oligotrophic. This is also consistent with the finding that phage infection risk is higher at high bacterial density \cite{Kasman2002, Knowles2016, Hoyland-Kroghsbo2016}.  
 
To connect this qualitative feature of our model to natural populations, we analyzed VPR data from Parikka \textit{et al.} \cite{Parikka2017} and found that the distribution of measured VPR values appears bimodal in low and moderate nutrient environments. It may be the case that at high nutrient levels where bacteria live in dense communities and are at high risk of lytic phage predation, most or all bacteria employ a highly-expressed CRISPR-Cas system and VPR is peaked at a single low value in that environment (Figure \ref{VPR}B). Conversely, at low to moderate nutrient levels, different bacteria may use different immune strategies and so VPR values may span a wider range (Figure \ref{VPR}C-D). Note that at very low $f$ and high nutrient availability, our model predicts monostability in the low-density, low-expression stable state corresponding to high VPR, yet we observe a unimodal low VPR in high nutrient environments (Figure \ref{VPR}B). In these conditions when phages are a large threat, bacteria may use another signal besides density to upregulate the CRISPR-Cas system. 
In this work we provide an intuitive connection between an observed quantity such as VPR and a non-trivial insight into the ecological state of interacting bacteria and phages. 

\section{Methods and Materials}

\subsection*{Data Analysis}

We analyzed data from an experiment in which S. \textit{thermophilus} bacteria were mixed with phages and sequenced to track the expanding portion of the CRISPR locus over fifteen days \cite{Paez-Espino2013} by labelling spacers with a type $i$ corresponding to a unique spacer sequence, a locus position $j$, and a bacteria label $k$. 
All spacers within an edit distance of 2 from each other were grouped into the same type. See SI section \ref{app1-1} for details.

We compared data reported by Parikka \textit{et al.} \cite{Parikka2017} with our model. When plotting VPR values, we combined average VPR measurements and individual VPR measurements (`VPR av' and `VPR' columns) to create a combined dataset of VPR values. 

Our processed data can be found on GitHub at \url{https://github.com/mbonsma/CRISPR-immunity}.

\subsection*{Model Analysis}

The mean-field model was solved exactly at steady state in Mathematica. Steady-state values with regulation added were calculated numerically. See SI section \ref{app3-1} for stability analysis.

\subsection*{Simulations}
Simulations were written in $\text{C}^{++}$ and performed using the tau leaping method \cite{Cao2006}. See SI section \ref{app2-2} for details. Simulation code can be found on GitHub at \url{https://github.com/mbonsma/CRISPR-immunity}.

\section{Acknowledgments}

We thank David Paez-Espino for discussions surrounding data from \cite{Paez-Espino2013}. We thank Devaki Bhaya and Anton Zilman for helpful discussions. 
We acknowledge funding from the Natural Sciences and Engineering Research Council of Canada and Vanier Canada Graduate Scholarships.

%\nocite{*} % This command displays all refs in the bib file
%\bibliography{Bonsma_Soutiere_Goyal_arxiv}
%\bibliographystyle{unsrt}
%\bibliography{Biophys-Paper2017.bib}

\appendix

\label{first:app}

\section{Data analysis}
\label{app1-1}

We used data from \cite{Paez-Espino2013} which is publicly available in the NCBI Sequence Read Archive under the accession \hyperref[https://www.ncbi.nlm.nih.gov/sra/?term=SRA062737]{SRA062737}. It includes four data files (SRR630110, SRR630111, SRR630412, and SRR630413) which we used for our analysis. We extracted the data corresponding to the MOI2 deep sequencing experiment and separated it into time points by checking each read for matches to the primers identified in the supplementary information of \cite{Paez-Espino2013}. Any reads with a mismatch between the annotation of the forward and reverse primers were discarded. Any remaining unsorted reads were excluded from the following analysis. 

\subsection{Identifying and sorting spacers}

\begin{figure}
\centering
	\includegraphics[width=13cm]{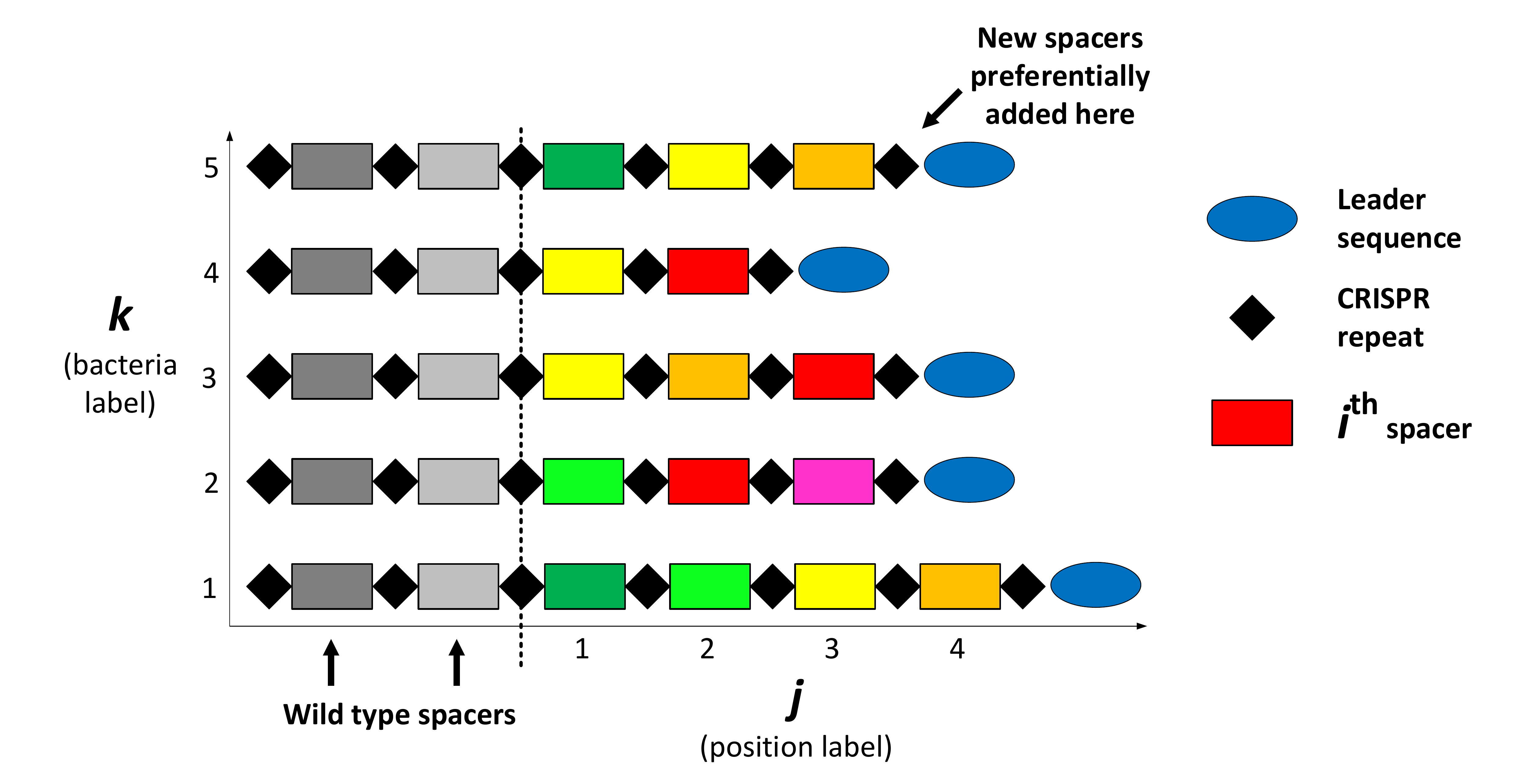}
	\caption{Schematic of the portion of the S. \textit{thermophilus} CRISPR locus sequenced in \cite{Paez-Espino2013}. We identified spacers with a type $i$, a locus position $j$, and bacteria number $k$. Coloured rectangles to the right of the dashed line represent spacers sequenced as the locus expands. Wild type spacers are shown in greyscale.}
	\label{CRISPR_array}
\end{figure}

\begin{figure}
\centering
	\includegraphics[width=0.7\textwidth]{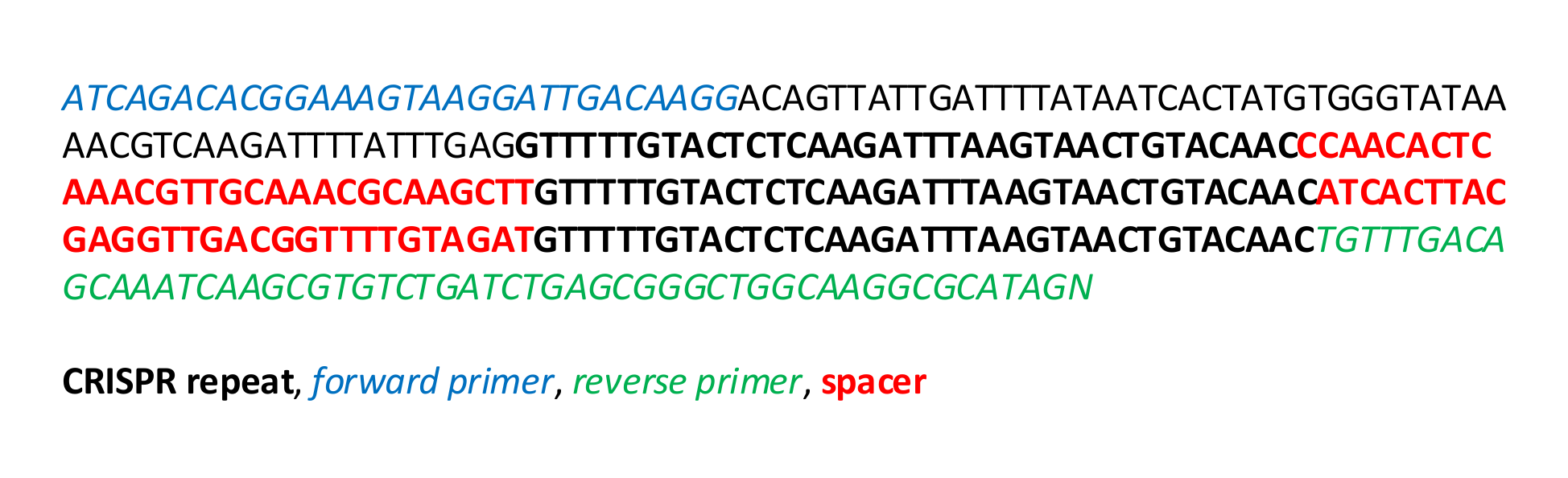}
	\caption{Example read covering the expanding CRISPR locus. The forward primer which overlaps with the leader sequence is shown in blue italics. The reverse primer which overlaps the first wild type spacer is shown in green italics. CRISPR repeats are shown in bold red and spacers in bold black.}
	\label{example_read}
\end{figure}

We extracted and catalogued spacers from the published raw read data of \cite{Paez-Espino2013}. Since only the expanding CRISPR end was sequenced, each read represents the longest possible sequence from wild type to leader end and so further assembly was not required (SI Figure \ref{CRISPR_array} and SI Figure \ref{example_read}). 

Because of this very specialized data structure, detecting CRISPR spacers and inferring their order was conceptually straightforward. A spacer was defined as any sequence flanked by two repeats. Since each read was bordered by wild type sequence and leader end sequence, all repeat sequences were complete and not truncated. SI Figure \ref{example_read} shows a typical read in more detail. Note that in this orientation, the spacer numbered ``1'' is found at the end of the read. To collect spacers, we (1) detected repeat sequences, reversing the read if the repeats were reversed, (2) inferred spacers as sequences between repeats, and (3) categorized spacers by comparing to previously detected spacers. 

Repeat sequence variation was present due to sequencing errors or naturally occurring SNPs. We used a regular expression to match variations on the number of Ts in a 5-T region of the repeat - the forward repeat was matched with "GTTT*GTACTCTCAAGATTTAAGTAACTGTACAAC" and the reverse repeat was matched with "GTTGTACAGTTACTTAAATCTTGAGAGTACAAA*C". These expressions match an identical string with three or more Ts or As in the region of the asterisk. This is a reasonable allowance to make since the 454 sequencing platform used to sequence this data is known to have high insertion and deletion rates in homopolymer regions \cite{Gilles2011}. 

To detect the most possible spacers, we developed methods to deal with repeat sequence variation beyond simple insertions and deletions in the homopolymer region. We inferred the presence of an undetected repeat by measuring the length of sequence before the first detected repeat, after the last detected repeat, and between two repeats. If any of these lengths exceeded its threshold (determined based on the known primer lengths and average spacer length, respectively), a more careful search for repeats was performed using the \textit{pairwise2} module in Biopython which performs a local pairwise alignment between the ideal repeat sequence and the read in question. 

\subsubsection{Pairwise alignment settings}
If the alignment with the true repeat (36 nucleotides long) was less than 31 nucleotides long, the alignment was discarded. The scoring system was as follows: match score of 1, mismatch score of -1, gap open score of -0.8 for the target sequence, gap open score of -0.7 for the repeat, and gap extend penalty of -1 for each sequence. The gap open scores were chosen to be different for the repeat and read so that the algorithm could identify how many gaps were opened and in which each sequence, in order to properly identify the start and end of each spacer.  

If no good match was found in a region between two repeats, the remaining ``long" spacer was discarded and a placeholder was inserted to preserve position information. Using this method, the number of detected repeats increased from $550931$ to $622067$, a $12.9$\% increase.

Repeats detected in this second search sometimes contained gaps with respect to the read or vice versa. In these cases, conventional labelling of nucleotide position prevented accurate detection of the start and end of adjacent spacers. We detected how many gaps were present and whether they occurred in the repeat or the read and then adjusted the indices of adjacent spacers accordingly. The scoring scheme was carefully chosen so that the number and placement of gaps could be inferred from the score. 

\subsubsection{Spacer type assignment}
We compared newly detected spacers to a growing list of previously detected spacers to assign it a type. If it matched an existing spacer exactly, it was assigned that type. Otherwise, a global pairwise alignment was performed between the new spacer and all existing spacers. If a match was found for which the score subtracted from the spacer length was within a chosen cutoff, the new spacer was assigned that type. This definition of cutoff is equivalent to the number of allowed SNPs between spacers under the scoring scheme used. If no match was found in either case, the new spacer was assigned a new type.  

\begin{figure}
\centering
	\includegraphics[width=0.7\textwidth]{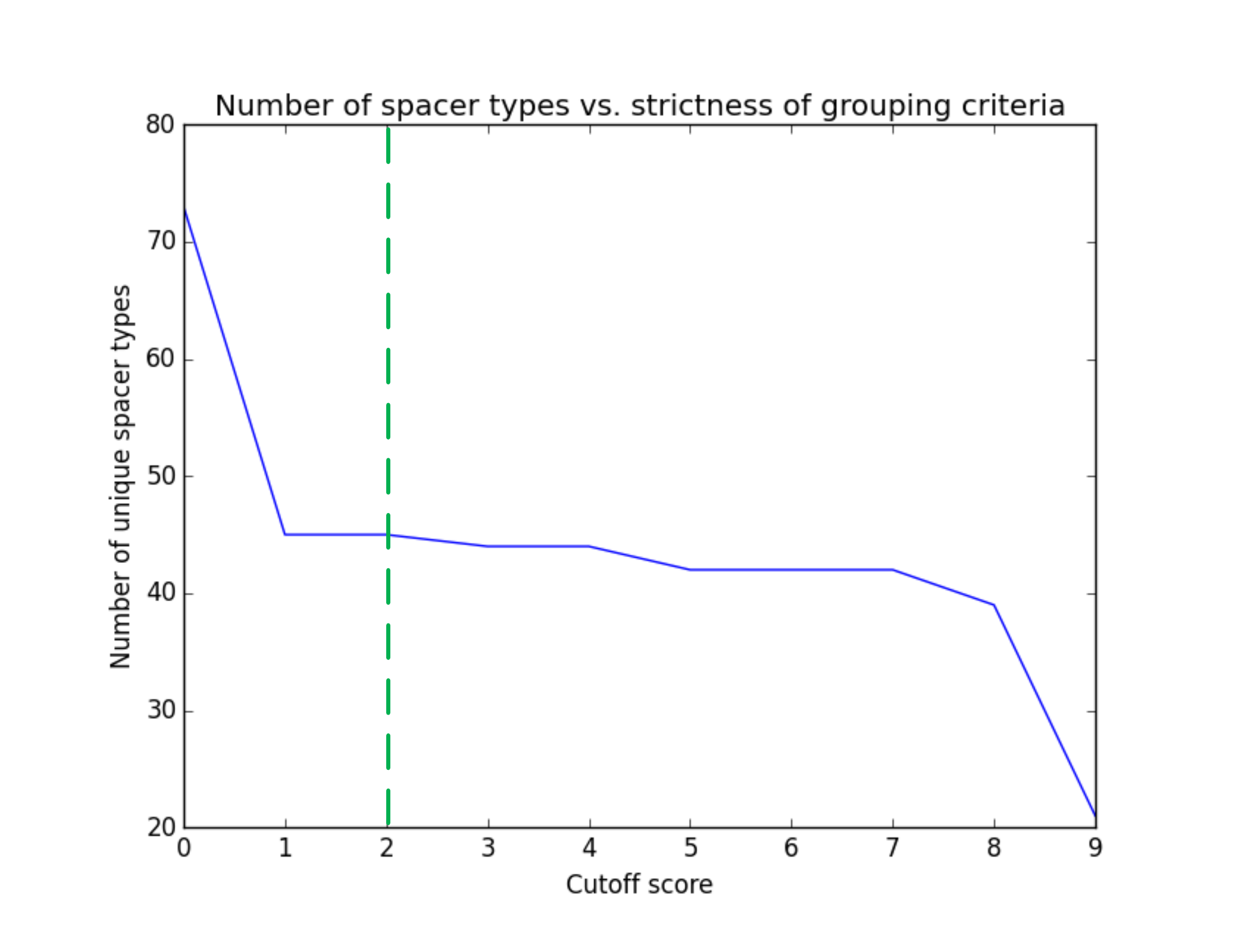}
	\caption{Number of unique spacer types vs. cutoff for 190 reads from time point 11. The green vertical dashed line indicates the selected cutoff.}
	\label{spacer_cutoff}
\end{figure}

To choose an appropriate tolerance for spacer alikeness, we tested this spacer sorting algorithm on a small sample of data (190 reads) as the cutoff was increased from 0 to 9. SI Figure \ref{spacer_cutoff} shows the number of unique spacer types detected as the cutoff is changed. It can be seen that there is a clear plateau between cutoff values of 1 and 8, which indicates that the system is insensitive to the cutoff if it falls in this range. We chose a cutoff of 2 for the analysis. 

In this way, we created a master dataset for each time point that contained each detected spacer, a number indicating the source read, the spacer position in the read, and the assigned spacer type. The definition of spacer type was consistent across time points, or in other words the same comparison list was carried through all time points.  

\subsection{Analysis}
\label{app1-2}

We extracted CRISPR spacers from the raw reads at each time point by finding sequences flanked by an S. \textit{thermophilus} CRISPR repeat (SI Figure \ref{CRISPR_array}). Newly detected spacers were added to an existing group if they were within an edit distance of $2$ of another spacer in that group. Data was organized into an array $s_{ijk}$ (equation \ref{sijk_2}).

\begin{equation}
  s_{ijk}(t) =
  \begin{cases}
  1 & \text{if spacer type } i \text{ is at position } j \text{ in bacterium } k \\
  0 & \text{otherwise}
  \end{cases} \label{sijk_2}
\end{equation}

We tracked individual spacer types, or ``clones'', $n_B^i(t)$, by summing over all bacteria and all locus positions: $n_B^i(t) = \sum_{j,k} s_{ijk}(t)$.

\begin{figure}
\centering
	\includegraphics[width=0.7\textwidth]{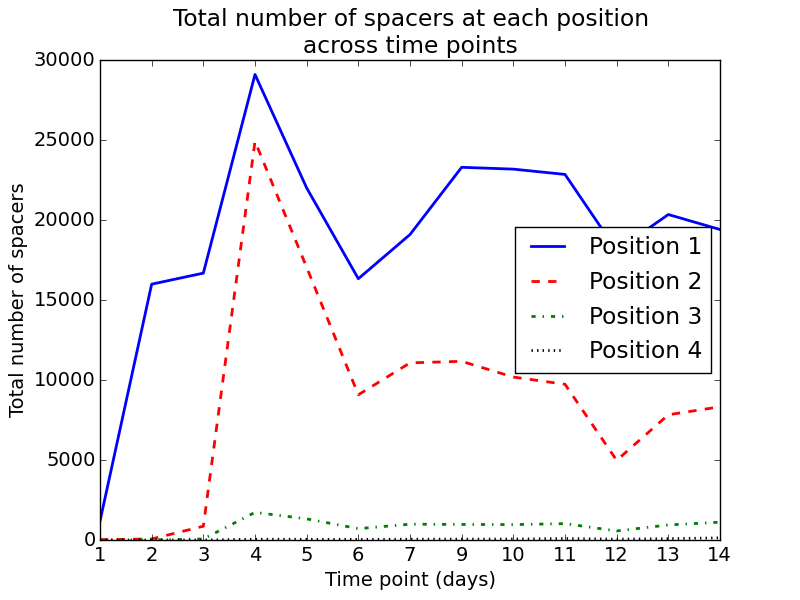}
	\caption{Total number of spacers at each time point in the experiment. Position 1 represents the oldest spacer (closest to the wild type spacers). Over half of all bacteria that acquired spacers, even at the end of the experiment, only acquired a single spacer.}
	\label{total_spacers}
\end{figure}

Most bacteria acquired only a single spacer; over half of bacteria from days 4-14 which had acquired 1 or more spacers only acquired a single spacer (SI Figure \ref{total_spacers}). 

\label{second:app}

\section{Model description}
\label{app2-1}

\begin{table}[ht]
\centering
\caption{\label{tab:params} Model parameters}
\begin{tabular}{llr}
%\toprule
Parameter & Description & Value \\
%\midrule
$\frac{1}{gC_0}$ & Bacterial doubling time & $41.7 \text{ min}$ \\
$C_0$ & Inflow nutrient concentration in \\
& units of bacterial cell density \\
$\alpha$ & Phage adsorption rate & $2 \times 10^{-10} \text{ min}^{-1}$  \\
$B$ & Phage burst size & $170$ \\
$F$ & Chemostat flow rate & \\
$p_V$ & Probability of phage success \\
& for bacteria without spacers &  \\  
$e$ & Spacer effectiveness &  \\
$r$ & Rate of spacer loss &  \\
$\eta$ & Probability of spacer acquisition &  \\
%\bottomrule
\end{tabular}

\medskip 
Parameter values are as above unless otherwise indicated. Representative values estimated for Streptococcus thermophilus bacteria in lab conditions.

\end{table}

\begin{table}[ht]
\caption{Model reactions}
\begin{tabular}{r l}
$ b^{0,i}+C\xrightarrow{g} 2 b^{0,i} $ & bacterium divides \\ 
$ b^{0,i}\xrightarrow{F} \emptyset  $ & bacterium flows out \\ 
$ V\xrightarrow{F} \emptyset  $ & phage flows out \\ 
$ \emptyset \xrightarrow{F C_0} C  $ & nutrients flow in \\ 
$ C \xrightarrow{F} \emptyset  $ & nutrients flow out \\ 
$ b^{0}+V\xrightarrow{\alpha p_V} B V  $ & interaction, phage wins \\ 
$ b^{0}+V\xrightarrow{\alpha (1-p_V)(1-\eta)} b^{0}   $ & interaction, bacterium survives \\ 
$ b^{0}+V\xrightarrow{\alpha (1-p_V)\eta/m } b^{i}   $ & interaction, bacterium survives and acquires a spacer \\ 
$ b^{i}+V\xrightarrow{\alpha p_v^s} B V  $ & interaction, phage wins \\ 
$ b^{i}+V\xrightarrow{\alpha (1-p_v^s)} b^i  $ & interaction, bacterium survives \\ 
$ b^{i}\xrightarrow{r} b^{0}  $ & bacterium loses spacer \\ 
\end{tabular} 
\label{tab:reactions}
\end{table}

We model bacteria and phages interacting in a chemostat. The populations we track are nutrient concentration $C$, phages $n_V$, and bacteria $n_b$ which can either have no spacer ($n_b^0$) or a spacer of type $i$ ($n_b^i$). Nutrients flow in at concentration $C_0$ with rate $F$, and all species flow out with rate $F$. The total number of bacteria with a spacer is $ n_b^s $ and the total number of bacteria is $ n_B $. The phage in the solution are all clonal and have $ m $ distinct protospacers. Bacteria grow at rate $ gC $. With rate $ \alpha $, a phage interacts with a bacterium. With probability $ p_V $, the phage will kill bacteria without spacers and produce a burst of new phages with size $ B $, while for bacteria with spacers that probability is reduced to $ p_v^s=(1-e) p_V$ ($ 0\leq e \leq 1 $). Bacteria without spacers that survive an attack have a chance to acquire a spacer with probability $ \eta $. Bacteria with a spacer lose their spacer at rate $ r $. Parameter descriptions and default values are shown in SI Table \ref{tab:params}.

\subsubsection{Reactions}

Table \ref{tab:reactions} lists all the interactions present in our model between individual bacteria ($b$), phages ($V$) and nutrients ($C$). 

\pagebreak

\subsection{Master equation}

The reactions in Table \ref{tab:reactions} can be formulated as a master equation describing the probability of observing $n_b^0$ bacteria without spacers, the set ${n_b^i}$ bacteria with spacers of type $i$, $n_V$ phages, and a nutrient concentration of $C$ at time $t$ (equation \ref{eqn:master}). 

\begin{equation}
\begin{aligned}
  \frac{d P(n_b^0,\{n_b^i\},n_V,C,t)}{dt}&= g (C+1)(n_b^0-1)P(n_b^0-1,\{n_b^i\},n_V,C+1,t)\\
  &+\sum_{j=1}^{m}g (C+1) (n_b^j-1)P(n_b^0,\{n_b^{i\neq j}\},n_b^j-1,n_V,C+1,t)\\
  &+F (n_b^0+1)P(n_b^0+1,\{n_b^i\},n_V,C,t)\\
  &+\sum_{j=1}^{m}F  (n_b^j+1)P(n_b^0,\{n_b^{i\neq j}\},n_b^j+1,n_V,C,t)\\
  &+F (n_V+1)P(n_b^0,\{n_b^i\},n_V+1,C,t)\\
  &+F (C+1) P(n_b^0,\{n_b^i\},n_V,C+1,t)\\
  &+F C_0 P(n_b^0,\{n_b^i\},n_V,C-1,t)\\
  &+\alpha p_V (n_b^0+1) (n_V-B+1) P(n_b^0+1,\{n_b^i\},n_V-B+1,C,t)\\
  &+\alpha (1-p_V)(1-\eta)n_b^0 (n_V+1) P(n_b^0,\{n_b^i\},n_V+1,C,t)\\
  &+\sum_{j=1}^{m} \frac{\alpha(1-p_V)\eta}{m} (n_b^0+1) (n_V+1) P(n_b^0+1,\{n_b^{i\neq j}\},n_b^j-1,n_V+1,C,t)\\
  &+\sum_{j=1}^{m}\alpha p_v^s (n_b^j+1) (n_V-B+1) P(n_b^0,\{n_b^{i\neq j}\},n_b^j+1,n_V-B+1,C,t)\\
  &+\sum_{j=1}^{m}\alpha (1-p_v^s) n_b^j (n_V+1) P(n_b^0,\{n_b^{i\neq j}\},n_b^j,n_V+1,C,t)\\
  &+\sum_{j=1}^{m} r (n_b^j+1)  P(n_b^0-1,\{n_b^{i\neq j}\},n_b^j+1,n_V,C,t)\\
  &-\left(F(n_b^0+\sum_{j=1}^{m}n_b^j+n_V+C+C_0)+gC(n_b^0+\sum_{j=1}^{m}n_b^j)\right. \\
  & \left. +\alpha n_V(n_b^0+\sum_{j=1}^{m}n_b^j)+r\sum_{j=1}^{m}n_b^j\right)P(n_b^0,\{n_b^i\},n_V,C,t)\\
\end{aligned}
\label{eqn:master}
\end{equation}

The 1st term is included only for $ n_b^0>1 $, the 2nd term if $ n_b^j>1 $, the 7th term for $ C\geq1 $, 8th term if $ n_V>B-1 $, the 10th term for $ n_b^j\geq1 $, the 11th term for $ n_V>B-1 $ and the 13th term for $ n_b^0\geq1 $.

\subsection{Mean-field dynamics} 

We can also write equations for the averages of the microscopic quantities (equations \ref{eqn:nbo1} to \ref{eqn:c1}). 

\subsubsection{Microscopic equations}
\begin{equation}
\frac{d\left< n_b^0\right>}{dt}=-F\left< n_b^0\right>+g\left< C n_b^0\right>-\alpha p_V \left< n_b^0 n_V\right>-\alpha(1-p_V)\eta \left< n_b^0 n_V\right> + \sum_{j=1}^{m} r \left< n_b^j\right>
\label{eqn:nbo1}
\end{equation}

\begin{equation}
\frac{d\left< n_b^j\right>}{dt}=-F\left< n_b^j\right>+g\left< C n_b^j\right>-\alpha p_v^s \left< n_b^j n_V\right>-  r \left< n_b^j\right>+\frac{\alpha(1-p_V)\eta}{m} \left< n_b^0 n_V\right> 
\label{eqn:nbi}
\end{equation}

\begin{equation}
	\begin{aligned}
	\frac{d\left< n_V\right>}{dt}=&-F\left< n_V\right>+\alpha p_V (B-1)\left< n_b^0 n_V\right>-\alpha (1-p_V) \left< n_b^0 n_V\right>+\\
&\sum_{j=1}^{m}\alpha p_v^s (B-1)\left< n_b^j n_V\right>-\sum_{j=1}^{m}\alpha (1-p_v^s) \left< n_b^j n_V\right>\\
	\end{aligned}
\label{eqn:nv1}
\end{equation}

\begin{equation}
\frac{d\left< C\right>}{dt}=F(\left< C\right>-C_0)-g\left< C\left( n_b^0+\sum_{j=1}^{m}n_b^j\right)\right>
\label{eqn:c1}
\end{equation}

We approximate the correlations $ \left< X Y\right>\approx \left< X\right> \left<Y\right> $. 

\begin{equation}
\frac{d\left< n_b^0\right>}{dt}=-F\left< n_b^0\right>+g\left< C \right>\left< n_b^0\right>-\alpha p_V \left< n_b^0 \right>\left< n_V\right>-\alpha(1-p_V)\eta \left< n_b^0 \right>\left< n_V\right> + \sum_{j=1}^{m} r \left< n_b^j\right>
\end{equation}

\begin{equation}
\frac{d\left< n_b^j\right>}{dt}=-F\left< n_b^j\right>+g\left< C \right>\left< n_b^j\right>-\alpha p_v^s \left< n_b^j \right>\left< n_V\right>-  r \left< n_b^j\right>+\frac{\alpha(1-p_V)\eta}{m} \left< n_b^0 \right>\left< n_V\right> 
\end{equation}

\begin{equation}
\begin{aligned}
\frac{d\left< n_V\right>}{dt}=&-F\left< n_V\right>+\alpha p_V (B-1)\left< n_b^0 \right>\left< n_V\right>-\alpha (1-p_V) \left< n_b^0 \right>\left< n_V\right>+\\
&\sum_{j=1}^{m}\alpha p_v^s (B-1)\left< n_b^j \right>\left< n_V\right>-\sum_{j=1}^{m}\alpha (1-p_v^s) \left< n_b^j \right>\left< n_V\right>\\
\end{aligned}
\end{equation}

\begin{equation}
\frac{d\left< C\right>}{dt}=F(\left< C\right>-C_0)-g\left< C\right>\left( \left<n_b^0\right>+\sum_{j=1}^{m}\left<n_b^j\right>\right)
\end{equation}

Then, we replace means by deterministic variables $n_b^0$, $n_b^j$, $n_V$, and $C$. 

\begin{equation}
\frac{d n_b^0}{dt}=-F n_b^0+g C  n_b^0-\alpha p_V  n_b^0  n_V-\alpha(1-p_V)\eta  n_b^0  n_V + \sum_{j=1}^{m} r  n_b^j
\end{equation}

\begin{equation}
\frac{d n_b^j}{dt}=-F n_b^j+g C n_b^j-\alpha p_v^s  n_b^j  n_V-  r  n_b^j+\frac{\alpha(1-p_V)\eta}{m}  n_b^0  n_V
\end{equation}

\begin{equation}
\begin{aligned}
\frac{d n_V}{dt}&=-F n_V+\alpha p_V (B-1)n_b^0 n_V-\alpha (1-p_V)  n_b^0  n_V\\
&+\sum_{j=1}^{m}\alpha p_v^s (B-1) n_b^j n_V-\sum_{j=1}^{m}\alpha (1-p_v^s)  n_b^j  n_V\\
\end{aligned}
\end{equation}

\begin{equation}
\frac{d C}{dt}=F( C-C_0)-g C\left( n_b^0+\sum_{j=1}^{m}n_b^j\right)
\end{equation}

\subsubsection{Macroscopic equations}

We can define new variables, $ n_b^s=\sum_{j=1}^{m} n_b^j $, $ n_B=n_b^0+n_b^s $, $\nu = n_B^s/n_B$ ($1-\nu=n_B^0/n_B$), and $p_V^s = (1-e)p_V$. 

\begin{equation}
\frac{d n_b^0}{dt}=-F n_b^0+g C  n_b^0-\alpha p_V  n_b^0  n_V-\alpha(1-p_V)\eta  n_b^0  n_V + r  n_b^s
\label{eqn:nb0dot_1}
\end{equation}

\begin{equation}
\frac{d n_b^s}{dt}=-F n_b^s+g C n_b^s-\alpha (1-e)p_V  n_b^s  n_V-  r  n_b^s+\alpha(1-p_V)\eta  n_b^0  n_V
\label{eqn:nbsdot_1}
\end{equation}

\begin{equation}
\frac{d n_V}{dt}=-F n_V-\alpha n_B n_V +\alpha p_V(1-e\nu) B n_B n_V
\label{eqn:nvdot_1}
\end{equation}

\begin{equation}
\frac{d C}{dt}=F( C-C_0)-g C n_B
\label{eqn:cdot_1}
\end{equation}

\begin{equation}
\frac{d n_B}{dt}=-F n_B+g C n_B-\alpha p_V(1-e\nu)  n_B  n_V
\end{equation}

\subsection{Description of simulations}
\label{app2-2}

\begin{figure}
	\centering
	\includegraphics[width=0.7\textwidth]{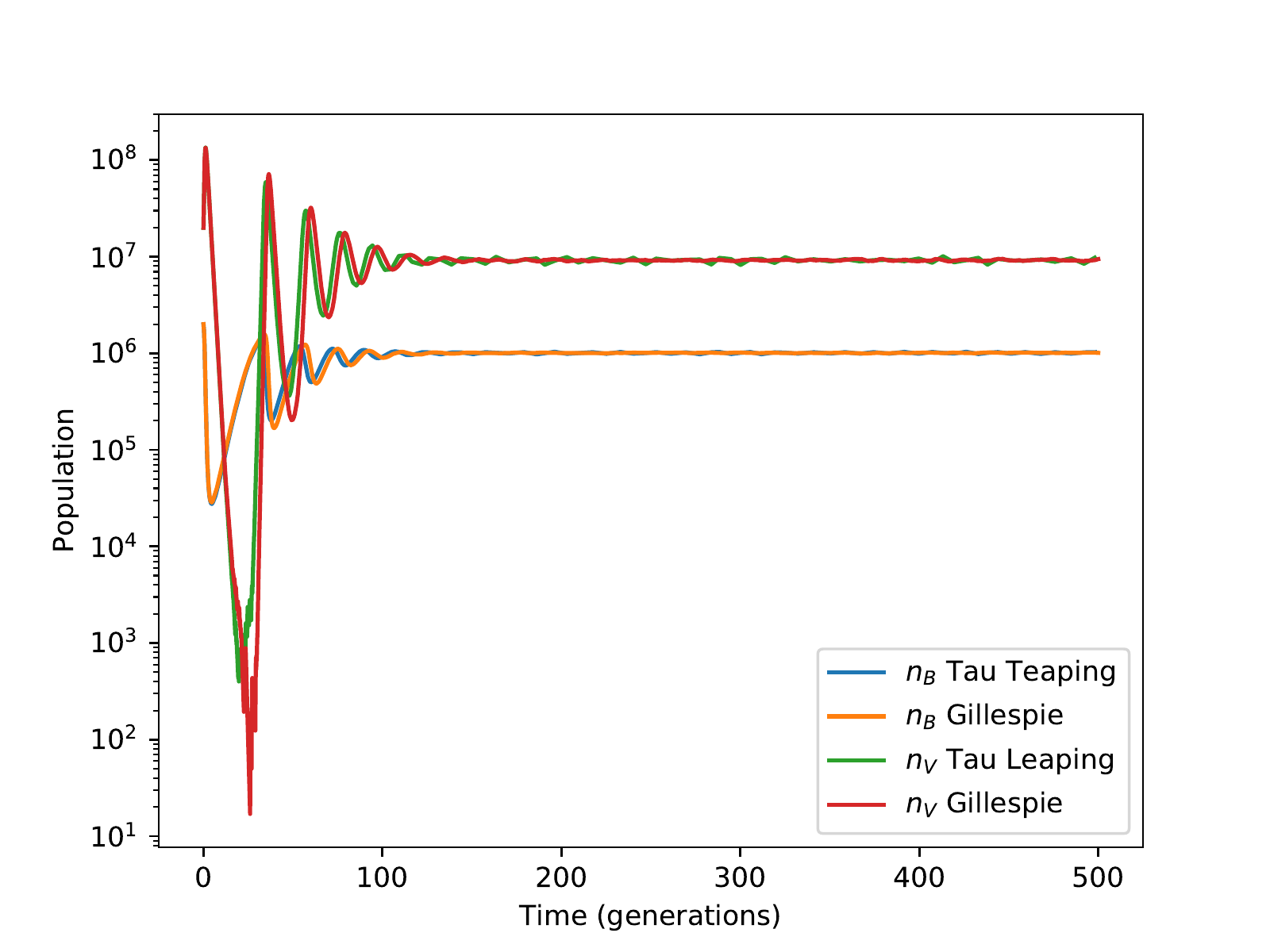}
	\caption{Total bacteria ($n_B$) and total phage ($n_V$) as a function of time for a Gillespie simulation and a tau leaping simulation. The two simulation techniques produce very similar results.}
	\label{tau_vs_gillespie}
\end{figure}

\begin{figure}
\centering
\includegraphics[width=1.0\textwidth]{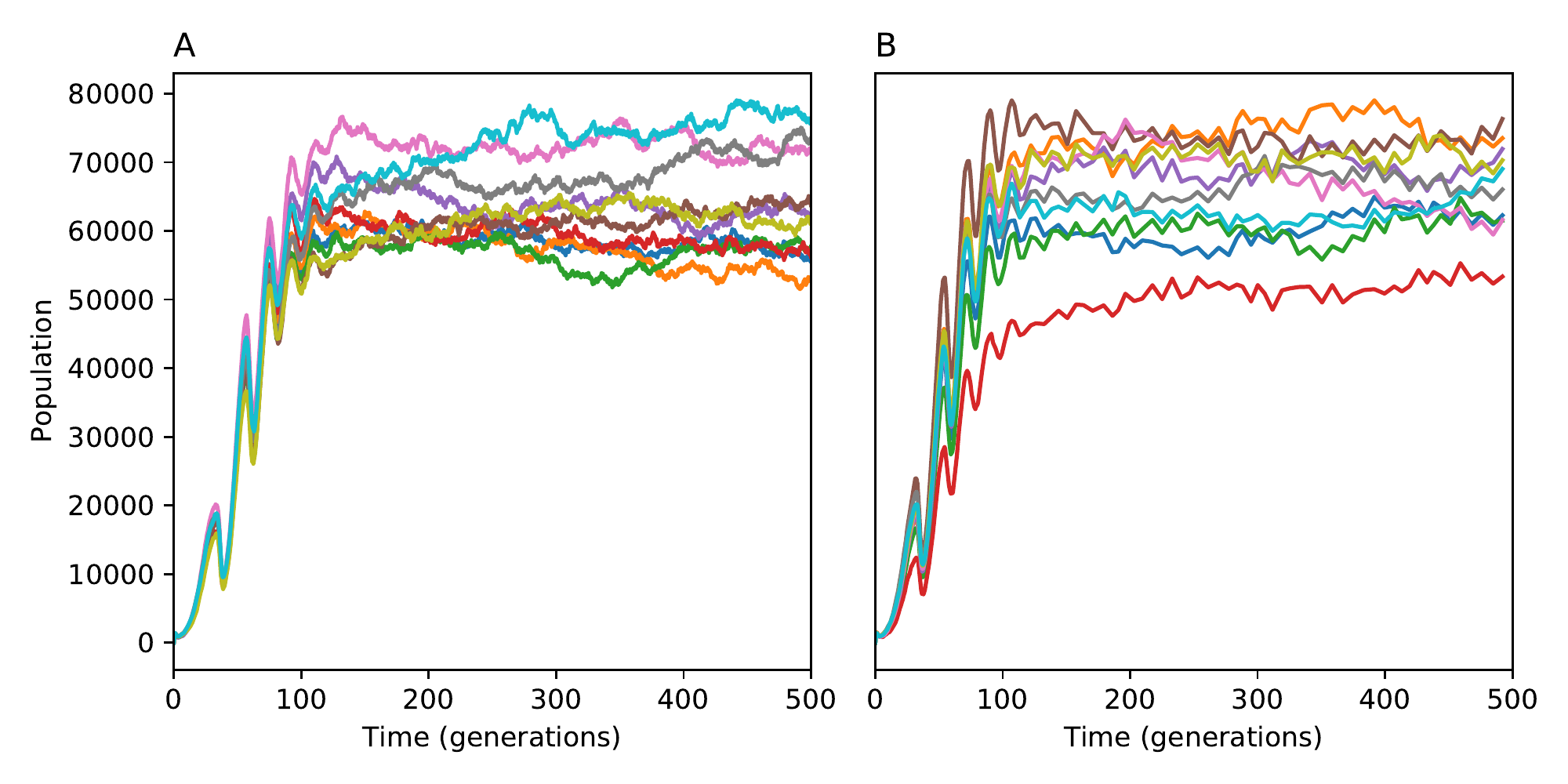}
\caption{Comparison of individual spacer type trajectories using tau leaping and Gillespie simulation techniques. (A) 10 spacer type trajectories vs time using Gillespie simulation methods. (B) 10 spacer type trajectories vs time using tau leaping simulation methods.}
\label{fig:tau_vs_gillespie_traj}
\end{figure}

Simulations were written in C++ and performed on a Lenovo ideapad Y700 and on \hyperref[https://www.scinethpc.ca/]{SciNet}. We primarily used the tau leaping method \cite{Cao2006} and compared with Gillespie simulations for some cases. Both methods showed good agreement for the mean-field behaviour of bacteria and phages (SI Figure \ref{tau_vs_gillespie}) and produced the same qualitative behaviour for individual spacer types (SI Figure \ref{fig:tau_vs_gillespie_traj}).

\subsection{Parameter choices}

Burst size for phage that target S. thermophilus is between 140-200 \cite{Lucchini1999}. The rate of adsorption for phage is of the order of $10^{-8} \text{ min}^{-1}$ ml \cite{Delbruck1940}. Using a volume of $V = 50 ml$, our total adsorption rate is $\alpha = 2 \times 10^{-10} \text{ min}^{-1}$ per bacteria and phage.

\cite{Vaningelgem2004} measured the maximum growth rate of S. thermophilus in milk at $42^{\circ}$C to be $2.4 \times 10^{-2} \text{ min}^{-1}$. This corresponds to $gC_0$ in our model.

The other parameters were picked in order to get a stable fixed point where phage and bacteria coexist, with population sizes relevant to experiments such as \cite{Paez-Espino2013}.

\subsection{Simulation results}
\label{app2-3}

\begin{figure}
\centering
\includegraphics[width=0.7\textwidth]{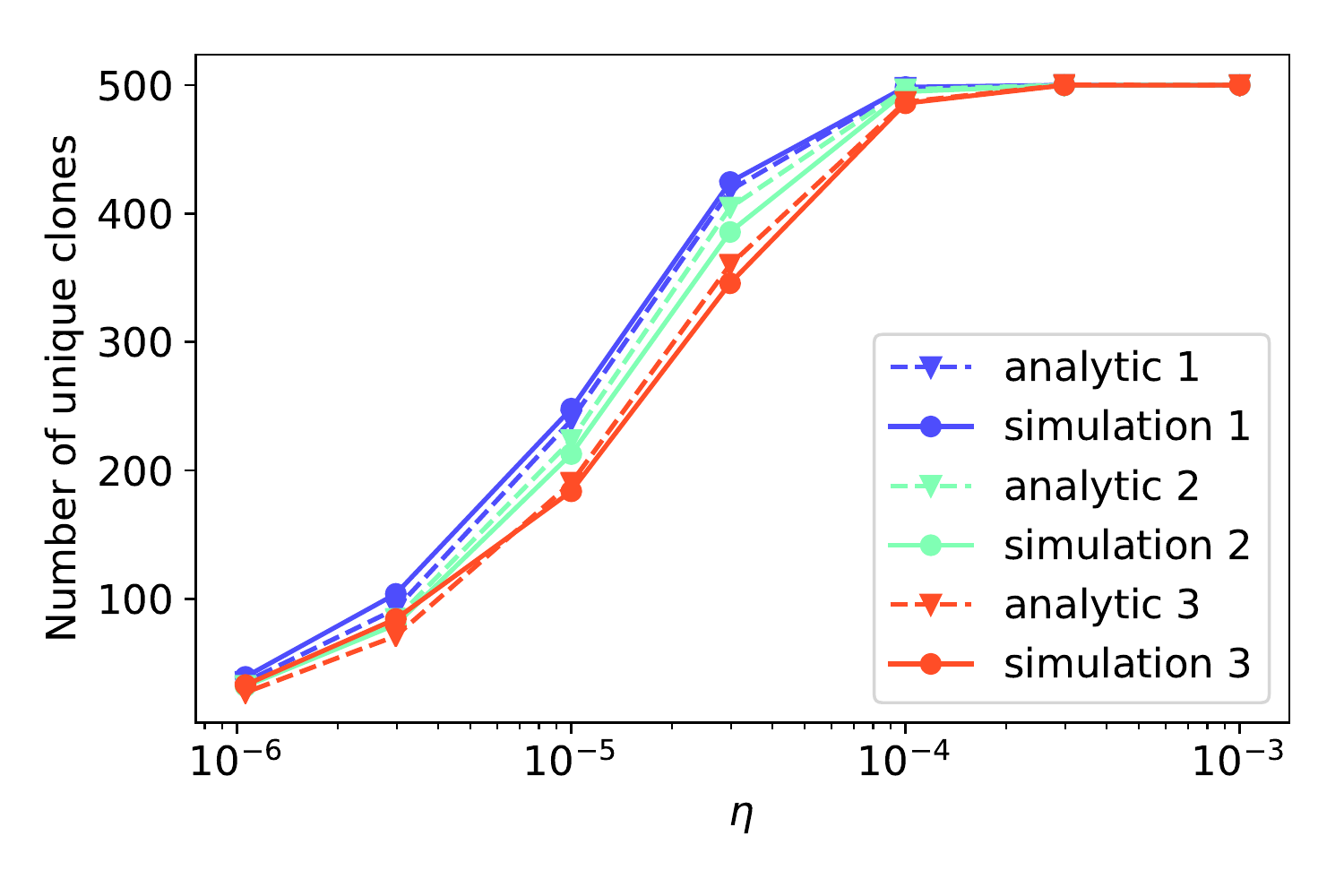} 
\caption{The average number of unique spacer types present at steady-state in simulations (circles and solid lines) increases with increasing $\eta$. The simulation results are well-matched by the analytic prediction from equation \ref{bk_any_clone_size} (triangles and dashed lines). The parameters $\eta$ and $e$ were chosen for each simulation so that all points on each colored curve correspond to a constant total bacterial population size of $0.15C_0$ (blue points), $0.1C_0$ (green points) and $0.05C_0$ (red points).}
\label{unique_clones}
\end{figure}

Our simulations were performed with a maximum of $m = 500$ spacer types that can be acquired by bacteria. This upper limit on the number of spacer types limits the total diversity of spacer types that can be observed and only impacts the spacer abundance distribution at large $\eta$. 
The qualitative simulation results, namely a continuous turnover of individual spacers and the presence of a non-trivial steady-state spacer abundance distribution, are insensitive to the choice of $\eta$ provided not all $m$ spacer types are acquired. This puts an upper bound on $\eta$ of $\approx 10^{-4}$ in our simulation, but simulations with higher $\eta$ can be performed with large values of $m$. SI Figure \ref{unique_clones} shows the average total number of unique spacer types at steady state as a function of $\eta$ for simulation data, compared to the analytic prediction given by equation \ref{bk_any_clone_size}: the number of unique spacer types predicted at steady state is $\sum_k b_k$.

\begin{figure}
\centering
\includegraphics[width=1\textwidth]{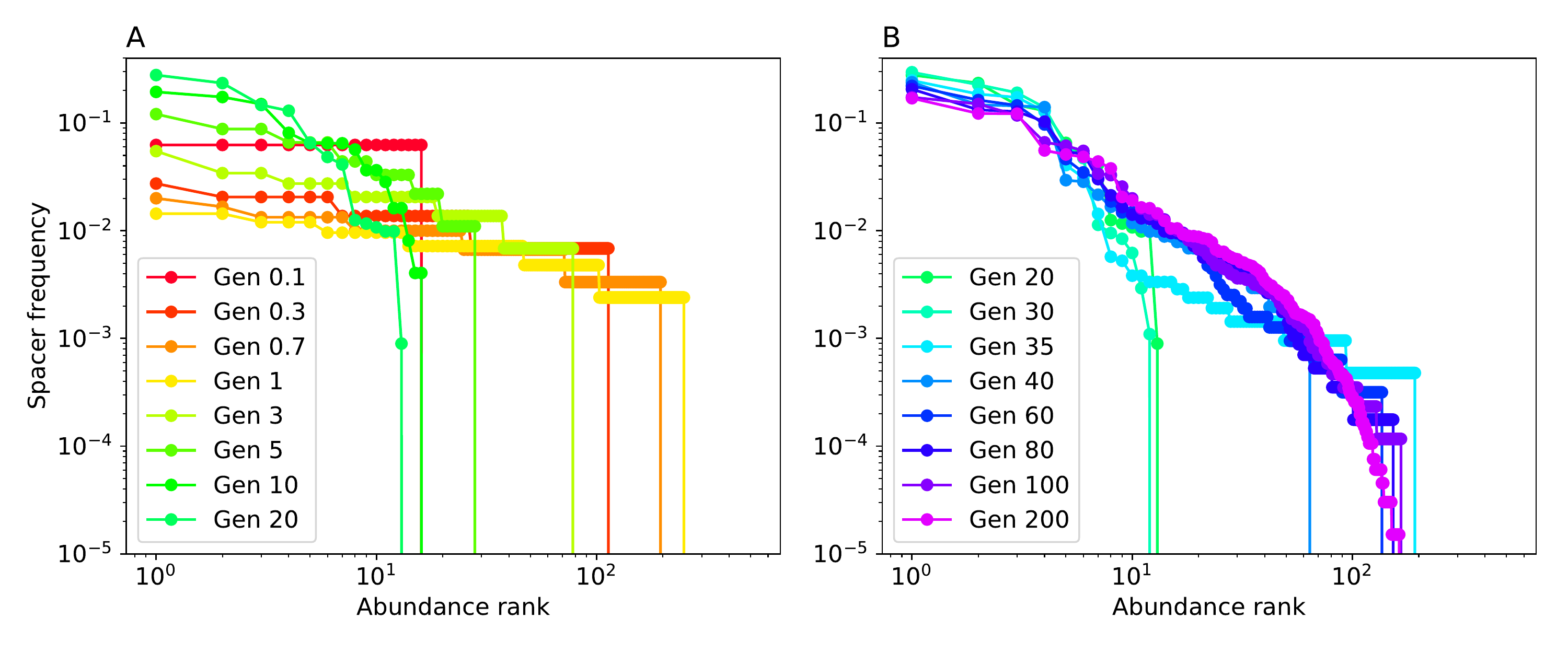} 
\caption{Spacer rank-abundance distributions for a simulation with $\eta = 10^{-5}$ and $e=0.387$. Time is rescaled into units of bacterial generations. (A) The distribution at early times begins as a flat distribution with a few bacteria having a single spacer (0.1 generations). As time progresses, more bacteria acquire spacers and some spacer types grow to larger sizes, making the distribution steeper and broader. (B) At longer times, the distribution reaches its steady-state shape at about generation 200.}
\label{rank_abundance_times}
\end{figure}

We initialized each simulation with no bacteria with spacers, or in other words the rank-abundance distribution is uniform at 0 abundance at the start of all simulations. The steady-state distribution evolves from a very different shape at early times. SI Figure \ref{rank_abundance_times} shows the spacer rank-abundance distribution for various time points of a simulation run. 

\subsection{Origin of rank-abundance curve}
\label{app2-4}
The spacer rank-abundance distribution resulting from our simulations can be analytically derived from the following master equation, which describes $b_k$, the number of spacer types, or clones, of size $k$. The size of a clone can increase through bacterial division with rate $gC$ (first term) or decrease through flow ($F$), spacer loss ($r$), and phage predation ($\alpha n_V p_v(1-e)$). The third term in equation \ref{any_clone_size} describes spacer acquisition: since in our simulations the total number of protospacers is fixed at $m = 500$, a newly acquired spacer will be added to an existing clone of size $k$ with probability $\eta / m$, where $\eta$ is the probability of acquiring any spacer in an interaction in which the phage does not succeed (which happens with probability $1-p_V$). 

\begin{equation} \label{any_clone_size}
\begin{gathered}
\partial_t b_k = gC[(k-1)b_{k-1} - kb_k] + (F+r + \alpha n_V p_V (1-e)) [(k+1)b_{k+1} - kb_k] \nonumber \\
+ \alpha n_b^0 n_V (1-p_V) \frac{\eta}{m} [b_{k+1} - b_k]
\end{gathered} \numberthis
\end{equation}

The variables $n_V$, $n_B^0$, and $C$ evolve according to their mean-field equations (\ref{eqn:nb0dot_1}, \ref{eqn:nvdot_1}, \ref{eqn:cdot_1}).
The total number of bacteria with spacers $n_B^s = \sum_k k b_k$; $\partial_t \sum_k k b_k$ is equivalent to equation \ref{eqn:nbsdot_1}.

At steady-state, all the population variables are constant, and equation \ref{any_clone_size} can be solved using a generating function and the method of characteristics. 

\subsubsection{Generating function solution} \label{gf}

The generating function for the probability distribution $b_k(t)$ is $G(z,t) = \sum_k z^k b_k(t)$.  

Let $\beta = gC$, $\mu = F+r + \alpha n_V p_V (1-e)$, and $D = \alpha \eta n_B^0 n_V (1-p_V)$. Multiplying equation \ref{any_clone_size} by $\sum_k z^k$ and noting that $\partial_z G(z,t) = \sum_k k z^{k-1}b_k(t)$, we get the following differential equation:

\begin{equation} \label{G_PDE_any_size}
\partial_t G(z,t) = \partial_z G(z,t) \left( z^2 \beta -z(\beta + \mu) +\mu \right) + G(z,t) \frac{D}{m}(z-1)
\end{equation}

Equation \ref{G_PDE_any_size} can be solved with the method of characteristics \cite{VanKampen1981}. We parametrize the function $G(z,t)$ with a new variable $s$. Applying the chain rule:

\begin{equation} 
\partial_s G(z(s),t(s)) = \frac{\partial G}{\partial z}\frac{\partial z}{\partial s} + \frac{\partial G}{\partial t}\frac{\partial t}{\partial s} 
\end{equation}

And by comparison with equation \ref{G_PDE_any_size}, the characteristic equations are 

\begin{equation}  \label{teq}
\frac{\partial t}{\partial s} = 1
\end{equation}

\begin{equation} 
\frac{\partial z}{\partial s} = (1-z)(\beta z - \mu)
\end{equation}

\begin{equation} 
\frac{\partial G}{\partial s} = G \frac{D}{m}(z-1)
\end{equation}

From equation \ref{teq} we see $t = s+c_1$, so we can choose $t_0 = c_1 = 0$ and replace $s$ with $t$ going forward. 

Solving the characteristic equation for $z$ by integrating both sides gives equation \ref{z_sol}.

\begin{equation} \label{z_sol}
\frac{1-z}{\mu - \beta z}\text{e}^{(\beta - \mu)t} = c_2 
\end{equation}

At $t = 0$, $z$ will pass through some point $z_0$, so we have the initial condition $z(0) = z_0$. With $z_0$ in equation \ref{z_sol} at $t=0$, we get equation \ref{z0_sol}, where $c_2$ is given by equation \ref{z_sol}.

\begin{equation} \label{z0_sol}
z_0 = \frac{c_2 \mu -1}{c_2 \beta - 1}
\end{equation}

The variation of $G$ along the $z-t$ curve is 

$$
\frac{\partial G}{\partial z} = -\frac{\frac{GD}{m}(z-1)}{z^2\beta -z(\beta + \mu) + \mu} = -\frac{GD}{m(\beta z - \mu)}
$$

Integrating both sides, we get

$$G(z) = \Omega(c_2)(\beta z - \mu)^{-\frac{D}{\beta m}}$$

The constant $\Omega$ is a function of the characteristic $z$-$t$ curve (equation \ref{z_sol}). To find the particular form of $\Omega(c_2)$, we apply the initial condition $G(z,0) = zN_0$, meaning that we start with $N_0$ clones of size 1 at time $t=0$. 

$$G(z,0) = zN_0 = \Omega\left( \frac{1-z}{\mu - \beta z} \right) (\beta z - \mu)^{-\frac{D}{\beta m}}$$

Let $\xi = \frac{1-z}{\mu - \beta z}$, therefore $z = \frac{\xi\mu - 1}{\xi\beta -1}$.

$$\Omega (\xi)(\beta \left(\frac{\xi\mu - 1}{\xi\beta -1}\right) - \mu)^{-\frac{D}{\beta m}} = \left(\frac{\xi\mu - 1}{\xi\beta -1}\right)N_0$$

Solving for $\Omega (\xi)$:

$$\Omega (\xi) = \left(\frac{\xi\mu - 1}{\xi\beta -1}\right)N_0 (\beta \left(\frac{\xi\mu - 1}{\xi\beta -1}\right) - \mu)^{\frac{D}{\beta m}}$$

The full solution for $G(z,t)$ can be written by replacing the constant $\Omega (c_2)$ with the expression for $\Omega (\xi)$ and replacing $\xi$ with $\xi \epsilon$, where $\epsilon = \text{e}^{(\beta-\mu)t}$ is the time-dependent part of the $z-t$ curve.

$$
G(z,t) = N_0(\beta z - \mu)^{-\frac{D}{\beta m}} \left(\frac{\xi\epsilon\mu - 1}{\xi\epsilon\beta -1}\right) (\beta \left(\frac{\xi\epsilon\mu - 1}{\xi\epsilon\beta -1}\right) - \mu)^{\frac{D}{\beta m}}
$$

Finally, replacing $\xi$ with $ \frac{1-z}{\mu - \beta z}$, we get

$$
G(z,t) = N_0(\beta z - \mu)^{-\frac{D}{\beta m}} \left( \frac{(1-z)\epsilon \mu + \beta z - \mu}{(1-z)\epsilon \beta + \beta z - \mu} \right) (\beta \left( \frac{(1-z)\epsilon \mu + \beta z - \mu}{(1-z)\epsilon \beta + \beta z - \mu} \right) - \mu)^{\frac{D}{\beta m}}
$$

$G(1,t) = N_0$, meaning that the total population remains conserved, consistent with our assumption that all the population variables are at steady-state. 

The limit as $t \to \infty$ of $G(z,t)$ is 

$$G(z) = N_0 \left(\frac{\beta z - \mu}{\beta-\mu}\right)^{-\frac{D}{\beta m}}$$

We can construct $b_k$ by taking successive derivatives of $G(z)$: $b_k = \frac{1}{k!}\frac{\partial G}{\partial z}\vert_{z=0}$

\begin{equation} 
b_k = \frac{N_0 \prod_{i=1}^k [D/m + (i-1)\beta] (\frac{\mu - \beta}{\mu})^{D/(\beta m)}}{k! \mu^k}
\end{equation}

$$b_0 = N_0\left(\frac{\mu - \beta}{\mu}\right)^{D/(\beta m)}$$

We can re-write this expression using Stirling's approximation for $k!$ to facilitate evaluation at large $k$. 

\begin{equation} \label{bk_any_clone_size}
b_k = \frac{N_0}{\sqrt{2 \pi k}} 
\text{exp} \left[ \frac{D}{\beta m} \text{ln}\left(\frac{\mu-\beta}{\mu}\right) 
+ \sum_{i=1}^k \text{ln}\left( \frac{e}{k\mu}(\frac{D}{m} + (i-1)\beta) \right) \right] 
\end{equation}

\begin{figure}
\centering
\includegraphics[width=0.8\textwidth]{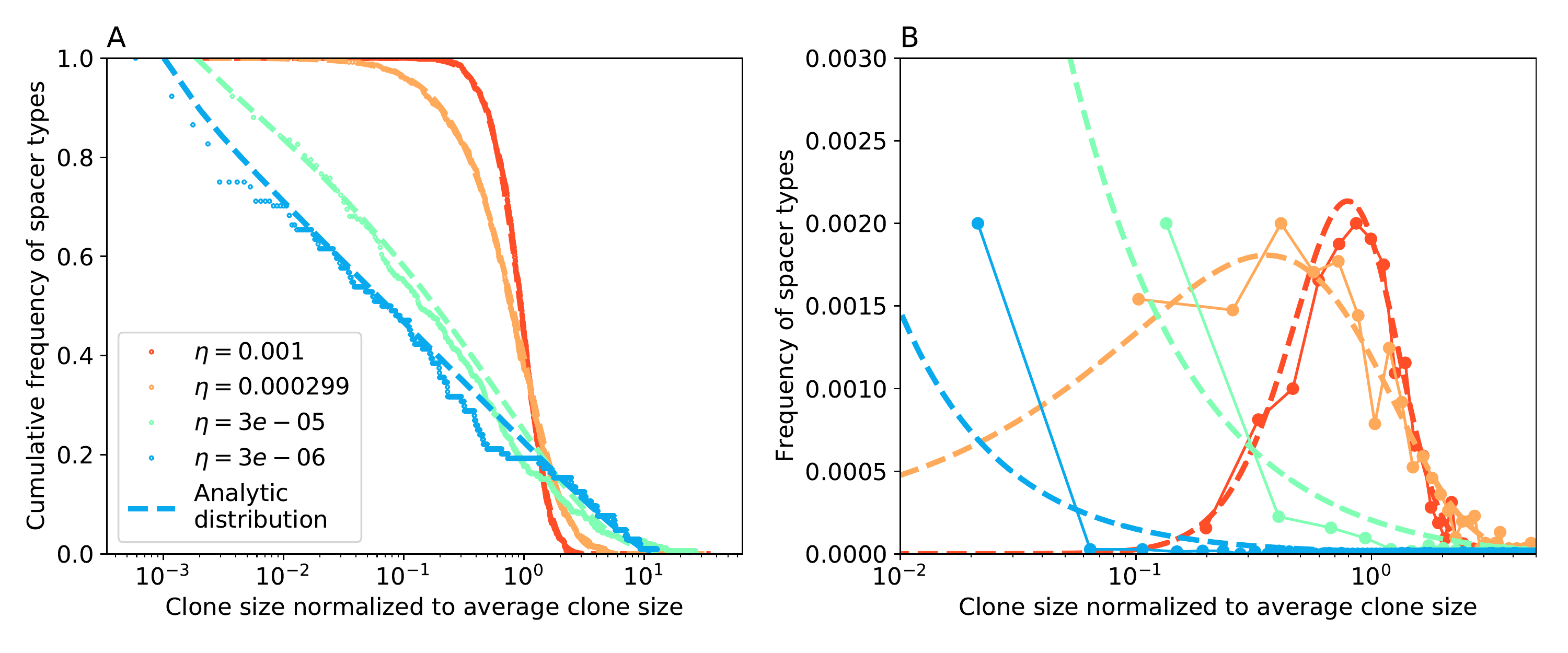}
\caption{Equation \ref{bk_any_clone_size} (dashed lines) compared with spacer clone size distributions from simulations at steady-state (dots). (A) Distributions in cumulative form. (B) Distributions shown as a histograms (dots). The analytic steady-state distribution matches well with the simulation results, except in the cases where the acquisition rate $\eta$ is very low and the choice of bin size has a large effect. Here $N_0 = m$, the total number of unique clones.}
\label{fig:bk_any_size}
\end{figure}

Equation \ref{bk_any_clone_size} is an analytic expression describing the steady-state spacer abundance distribution that results from our simulations. SI Figure \ref{fig:bk_any_size} compares the analytic distribution to the steady-state spacer clone size distribution from our simulations at several values of the spacer acquisition probability $\eta$, with $e$ chosen such that the total number of bacteria is the same for all cases. The corresponding rank-abundance distribution can be obtained from the cumulative distribution (SI Figure \ref{fig:bk_any_size}A) by flipping the axes and rescaling the frequency axis. 

\subsubsection{Rank-abundance distribution in ecology}

The steady-state clone size distribution given in equation \ref{bk_any_clone_size} can be approximated for large clone size $k$ and large $m$ to give a gamma distribution and logseries distribution respectively, both of which have a long history as descriptions of species abundance in ecology \cite{McGill2007}. 

In the following expressions we replace $N_0$ with $m$, since at steady-state the total number of clones remains fixed at $m$.

We find the following expression for $b_k$ in the limit of large clone size (large $k$) by taking a series expansion as $k \to \infty$ and keeping the first term.

\begin{equation}
b_k \approx \frac{m\left(1-\frac{\beta}{\mu}\right)^{\frac{D}{\beta m}}}{\Gamma\left( \frac{D}{\beta m} \right)} \text{e}^{-\text{ln}(\mu/\beta)k} \left(\frac{1}{k}\right)^{1-\frac{D}{\beta m}}
\label{gamma}
\end{equation}

This is a gamma distribution with shape parameter $\frac{D}{\beta m}$ and rate parameter $\text{ln}(\mu/\beta)$. Note that $\left(1-\frac{\beta}{\mu}\right)^{\frac{D}{\beta m}} \approx \text{ln}(\mu/\beta)^{\frac{D}{\beta m}}$, consistent with the canonical form of the gamma distribution. The additional factor of $m$ in equation \ref{gamma} arises because we treat $b_k$ as the \textit{number} of clones of size $k$; to normalize $b_k$ we would divide by $m$, the total possible number of unique clones.

The gamma distribution has been used to describe species abundance in a number of ecological situations \cite{Dennis1984, Engen1996, Diserud2000, Plotkin2002}. For example, Dennis and Patil \cite{Dennis1984} arrive at a gamma distribution as ``the approximate stationary distribution for the abundance of a population fluctuating around a stable equilibrium,'' and Plotkin and Muller-Landau \cite{Plotkin2002} use a gamma distribution to fit species abundance distributions on a tropical island. 

For practical purposes the gamma distibution given by equation \ref{gamma} is a good approximation to the true distribution for all the parameter values we considered in our simulation.

When the total number of unique spacer types $m$ is large, our model is effectively an infinite alleles model in which each newly acquired spacer is assumed to be completely unique. In the limit of large $m$, we find the following expression for $b_k$.

\begin{equation}
b_k \approx \frac{D}{\beta} \frac{1}{k} \left(\frac{\beta}{\mu}\right)^{k}
\end{equation}

Up to a constant, this is a log-series distribution, made famous by Fisher et al. \cite{Fisher1943} and appearing many times since \cite{Chisholm2010}.

\label{third:app}

\section{Mean-field steady-state solutions}
\label{app3-1}
\subsection{$e=0$ model (no adaptive immunity)}

Equations \ref{eqn:nb0dot_1} to \ref{eqn:cdot_1} describe the full model. If spacer effectiveness $e=0$, the model reduces to three dimensions: bacteria $n_B$, phages $n_V$, and nutrients $C$. Equations \ref{eqn:nvdot_a} to \ref{eqn:cdot_a} describe this simpler model.

\begin{equation}
\frac{dn_V}{dt} = 
	-\alpha n_B n_V
	+\alpha Bp_V n_Bn_V
	- Fn_V \label{eqn:nvdot_a} 
\end{equation}

\begin{equation}
\frac{dn_B}{dt} = 
	gCn_B
	-\alpha p_Vn_Vn_B
	-Fn_B    \label{eqn:nbdot_a} 
\end{equation}

\begin{equation}
\frac{dC}{dt} = 
	FC_0
	- g Cn_B
	-FC 	 \label{eqn:cdot_a}
\end{equation}

Solving equations \ref{eqn:nvdot_a} to \ref{eqn:cdot_a} at steady state gives the following fixed points.

\subsubsection{Trivial fixed point}

There is a trivial fixed point where bacteria and phages are both zero.

\begin{itemize}%[noitemsep]
\item[] $n_B^* = 0$ 
\item[] $n_V^* = 0$ 
\item[] $C^* = C_0$ 
\end{itemize}

The eigenvalues of the Jacobian at this fixed point are $1-f,-f$, and $-f$, where $f=F/(gC_0)$. This means that this fixed point is stable for $f>1$. $f>1$ is a reasonable stability condition: this is the case where the flow rate is too high for bacteria to persist.

\subsubsection{Phages unable to persist}

\begin{itemize}%[noitemsep]
\item[] $n_B^* = C_0(1-f)$ 
\item[] $n_V^* = 0$ 
\item[] $C^* = C_0f$ 
\end{itemize}

$0<f<1$ is required for physical existence of this fixed point. 

The eigenvalues of the Jacobian at this fixed point are $f-1$, $-f$, and $-\frac{(f-1) p (B p_V-1)+f p_V}{p_V}$, where $p=p_V\alpha/g$. The first two are negative under the requirement for existence. The third is negative for $Bp_V < \frac{gf}{(1-f)\alpha} +1$. If this stability condition is satisfied, phages cannot persist in the population --- they will be driven to extinction.

\subsubsection{All populations finite and stable}

If all variables are non-zero, the fixed point is 

\begin{itemize}
\item[] $\frac{n_B^*}{C_0} = \frac{f p_V}{p (-1 + B p_V)}$  \label{eqn:nbstar}
\item[] $\frac{n_V^*}{C_0} = \frac{(1-f) p (B p_V-1)-f p_V}{p (p (B p_V-1)+p_V)}$ \label{eqn:nvstar}
\item[] $\frac{C^*}{C_0} = \frac{p (B p_V-1)}{p (B p_V-1)+p_V}$ \label{eqn:cstar}
\end{itemize}

The condition for existence is

\begin{itemize}%[noitemsep]
\item[] $Bp_V > \frac{gf}{(1-f)\alpha} +1$
\end{itemize}

The eigenvalues are

\begin{itemize}%[noitemsep]
\item[] $-f$
\item[] $-\frac{\sqrt{f} \sqrt{4 (f-1) p^2 (B p_V-1)^2+4 f p p_V (B p_V-1)+f p_V^2}+f p_V}{2 p (B p_V-1)}$
\item[] $\frac{\sqrt{f} \sqrt{4 (f-1) p^2 (B p_V-1)^2+4 f p p_V (B p_V-1)+f p_V^2}-f p_V}{2 p (B p_V-1)}$
\end{itemize}

The first is always negative. The second is negative for

\begin{itemize}%[noitemsep]
\item[] $ \frac{4 p^2 (B p_V-1)^2}{(2 p (B p_V-1)+p_V)^2} \leq f < 1$
\end{itemize}

The third is negative for
\begin{itemize}%[noitemsep]
\item[]$\frac{4 p^2 (B p_V-1)^2}{(2 p (B p_V-1)+p_V)^2} \leq f < \frac{p (B p_V-1)}{p (B p_V-1)+p_V} = \frac{C^*}{C_0}$
\end{itemize}

The upper limit on $f$ is the same as the existence condition (requiring all be solutions $>0$).

\subsection{Nonlinear bacterial growth rate}
\label{app3-2}
Instead of the growth rate for $n_B$ being $gC$, we check what happens when the growth rate is a Hill function of the form $\frac{gkC}{C+k}$, where $k$ is the nutrient concentration at which bacterial growth rate is at half maximum. If $k >>C$, the linear approximation used in our results is valid and $\frac{gkC}{C+k}\approx gC$. 

Solving for the non-trivial steady-state variables in the case when bacteria have no CRISPR spacers, we find that $n_B^*$ is unchanged:

\begin{equation}
n_B^* = \frac{F}{\alpha(Bp_V-1)} \label{eqn:nb_hill}
\end{equation}

$C^*$ and $n_V^*$, however, now depend on $k$:

\begin{equation}
C^* = \frac{1}{2}\left(C_0-k-\frac{gkn_B^*}{F}\right) + \frac{1}{2}\sqrt{\left(C_0-k-\frac{gkn_B^*}{F}\right)^2 +4C_0k} \label{eqn:c_hill}
\end{equation}

\begin{equation}
n_V^* = \frac{gkC^*}{C^*+k}-F \label{eqn:nv_hill}
\end{equation}

This solution for $C$ reduces to the linear growth rate solution (section \ref{eqn:cstar}) when $k$ is large. This can be seen by expanding the square root in $C^*$ and keeping terms up to order $\frac{1}{k^3}$:

$$\frac{C^*}{C_0}\approx 1-\frac{1}{p(B-1/p_V)}+\frac{1}{p^2(B-1/p_V)^2} \approx \frac{p(B-1/p_V)}{p(B-1/p_V)+1}$$

The stability condition for bacteria and phage coexistence now depends on $k$. $k$ must be greater than the following parameter combination in order for phages to persist.

$$k>\frac{F (F + \alpha (C_0 - B C_0 p_V))}{F g + \alpha  (F - C_0 g) (B p_V-1)} = C_0\frac{f(fg+\alpha(1-Bp_V))}{fg+\alpha(1-f)(1-Bp_V)}$$

\begin{figure}
\centering
\includegraphics[width=0.7\textwidth]{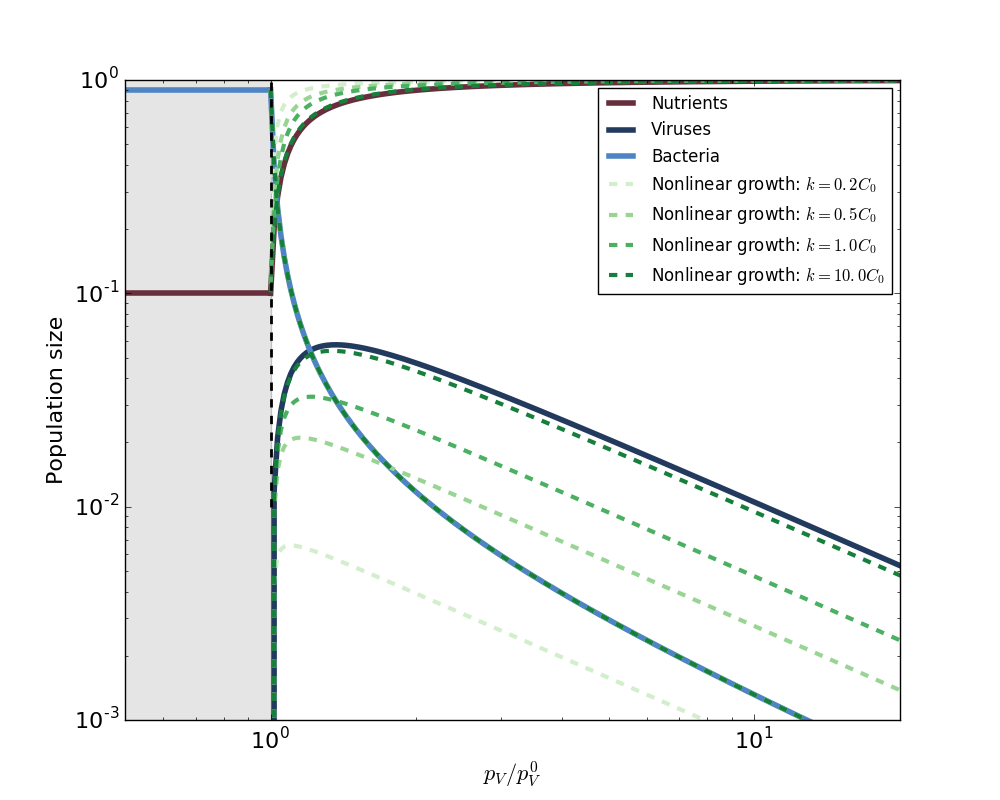} 
\caption{Solid lines: solutions to equations \ref{eqn:nvdot_a} to \ref{eqn:cdot_a} with linear growth for $n_B$. Green dashed lines: solutions to equations \ref{eqn:nb_hill} to \ref{eqn:nv_hill} for different values of $k$.} \label{hill_growth}
\end{figure}

For $f = F/(gC_0) = 0.1, B = 170, p_V = 0.02, g = 2.4\times 10^{-11}, C_0 = 10^9$, and $\alpha = 2\times 10^{-10}$, $k/C_0$ must be greater than $\approx 0.11$. SI Figure \ref{hill_growth} compares the full nonlinear growth solutions (equations \ref{eqn:nb_hill} to \ref{eqn:nv_hill}) to the solutions for linear growth (equations \ref{eqn:nbstar}). Provided $k$ is large enough that phages can persist, the picture is not qualitatively different, and in the low-nutrient limit ($k>>C$), the two solutions are very nearly the same.

\subsection{$e \neq 0$ model (adaptive immunity)}
\label{app3-3}

If $e > 0$, then the system is fully four-dimensional and all four variables are coupled. 

\subsubsection{Trivial fixed points}

The two partially trivial fixed points are the same as in the case when $e = 0$, since if $n_V = 0$, then $\nu = n_B^s/n_B = 0$ at steady state. The stability and existence conditions are also the same; effectively $\nu$ becomes uncoupled and the system is reduced to three dimensions if $n_V = 0$. 

\subsubsection{Non-trivial fixed point}

For convenience we define rescaled population sizes $x = n_B/C_0$, $y = n_V/C_0$, and $z = C/C_0$. 
Solving in the case where all dynamical variables are non-trivial, we get

\begin{equation}
z^* = \frac{p (B p_V (e \nu^* -1)+1)}{p (B p_V (e \nu^* -1)+1)-p_V}\label{zstar}
\end{equation}

\begin{equation}
x^* = \frac{f p_V}{p}\frac{1}{B p_V (1-e \nu^* )-1} \label{xstar}
\end{equation}

\begin{equation}
y^* =\frac{(f-1) p (B p_V (e \nu^* -1)+1)-f p_V}{p (e \nu^* -1) (p (B p_V (e \nu^* -1)+1)-p_V)}\label{ystar}
\end{equation}

And an implicit cubic equation for $\nu$, where $R=r/(gC_0)$.:

\begin{equation}
\begin{gathered}
0=(1-\nu)\left[-p_V\nu e-\eta(1-p_V)\right]\left[(1-f)p(p_VB(1-e\nu)-1)-fp_V\right] \\
+R\nu p_V(1-e\nu)(Bpp_V(1-e\nu)-p+p_V)
\end{gathered} \label{implicit_v}
\end{equation}

This cubic equation is analytically solvable, but the full solutions in terms of all parameters are cumbersome. 

Only one of the three solutions of equation \ref{implicit_v} is physical in the parameter range we use (real-valued and properly bounded):

\begin{equation}
\begin{gathered}
\nu^*=-\frac{\left(1+i \sqrt{3}\right) \sqrt[3]{\sqrt{\left(-27 a^2 d+9 a b c-2 b^3\right)^2+4 \left(3 a c-b^2\right)^3}-27 a^2 d+9 a b c-2 b^3}}{6 \sqrt[3]{2} a} \\
+\frac{\left(1-i \sqrt{3}\right)
   \left(3 a c-b^2\right)}{3\ 2^{2/3} a \sqrt[3]{\sqrt{\left(-27 a^2 d+9 a b c-2 b^3\right)^2+4 \left(3 a c-b^2\right)^3}-27 a^2 d+9 a b c-2 b^3}}-\frac{b}{3 a}
\end{gathered} \label{nustar}
\end{equation}

where the coefficients are

\begin{align}
a&=B e^2 f p p_V^2 (f+R-1) \\ 
b&=-e f p_V (p (f (B (p_V (e+\eta +1)-\eta )-1) \\ \nonumber
&+B (\eta -p_V (e+\eta -2 R+1))-R+1)+p_V (f+R)) \\
c&=f p \left[B p_V^2 (e (f-1) (\eta +1)+(f-1) \eta +R) \right. \\ \nonumber
&-\left. (e-1) (f-1) p_V (B \eta +1)-(2 B+2) (f-1) \eta  p_V+(f-1) \eta -p_V (f+R-1)\right] \\ \nonumber
&+f p_V (e f   p_V-f \eta +p_V (f \eta +R))\\
d &= -f \eta  (p_V-1) ((f-1) p (B p_V-1)+f p_V)
\end{align}

\begin{figure}
\centering
\includegraphics[width=1.0\textwidth]{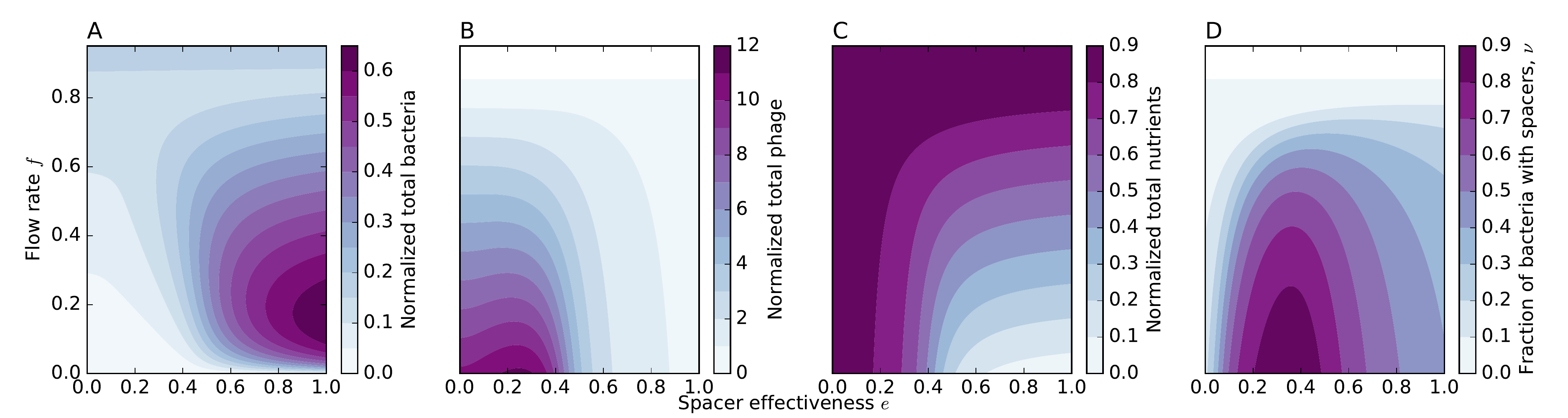} 
\caption{$x$, $y$, $z$, and $\nu$ (A-D respectively) vs. $f$ and $e$ with $R = 0.04$, $\eta = 0.0001$, $B= 170$, $p_V = 0.01$, $\alpha = 2\times 10^{-10}$, and $gC_0 = 0.024$.} \label{contours-f}
\end{figure}

Total bacteria, phage, nutrients, and the fraction of bacteria with spacers are plotted for a range of parameters in SI Figure \ref{contours-f} and SI Figure \ref{large-alpha}.

\begin{figure}
\centering
\includegraphics[width=1.0\textwidth]{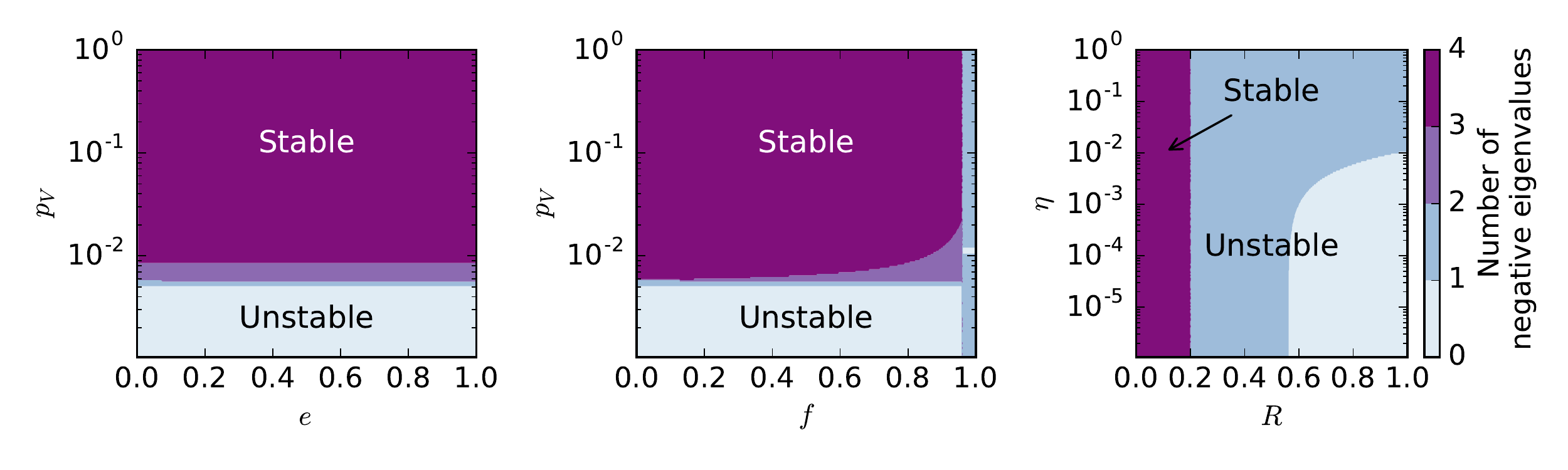} 
\caption{The number of eigenvalues with a negative real part for various parameter combinations ($p_V$ and $e$, $p_V$ and $f$, and $R$ and $\eta$). The unstable regions in the first two plots reflect parameter combinations for which phages cannot persist. In the third plot, Equation \ref{nustar} becomes unstable for large $R$, but one of the other roots takes its place as a stable and physical solution in this regime (confirmed numerically).} \label{stability}
\end{figure}

This fixed point is stable for a wide range of parameters, which we explored numerically. SI Figure \ref{stability} shows the number of negative eigenvalues vs. parameters; where all four eigenvalues have a negative real part, this fixed point is stable.

\begin{figure}
\centering
\includegraphics[width=0.6\textwidth]{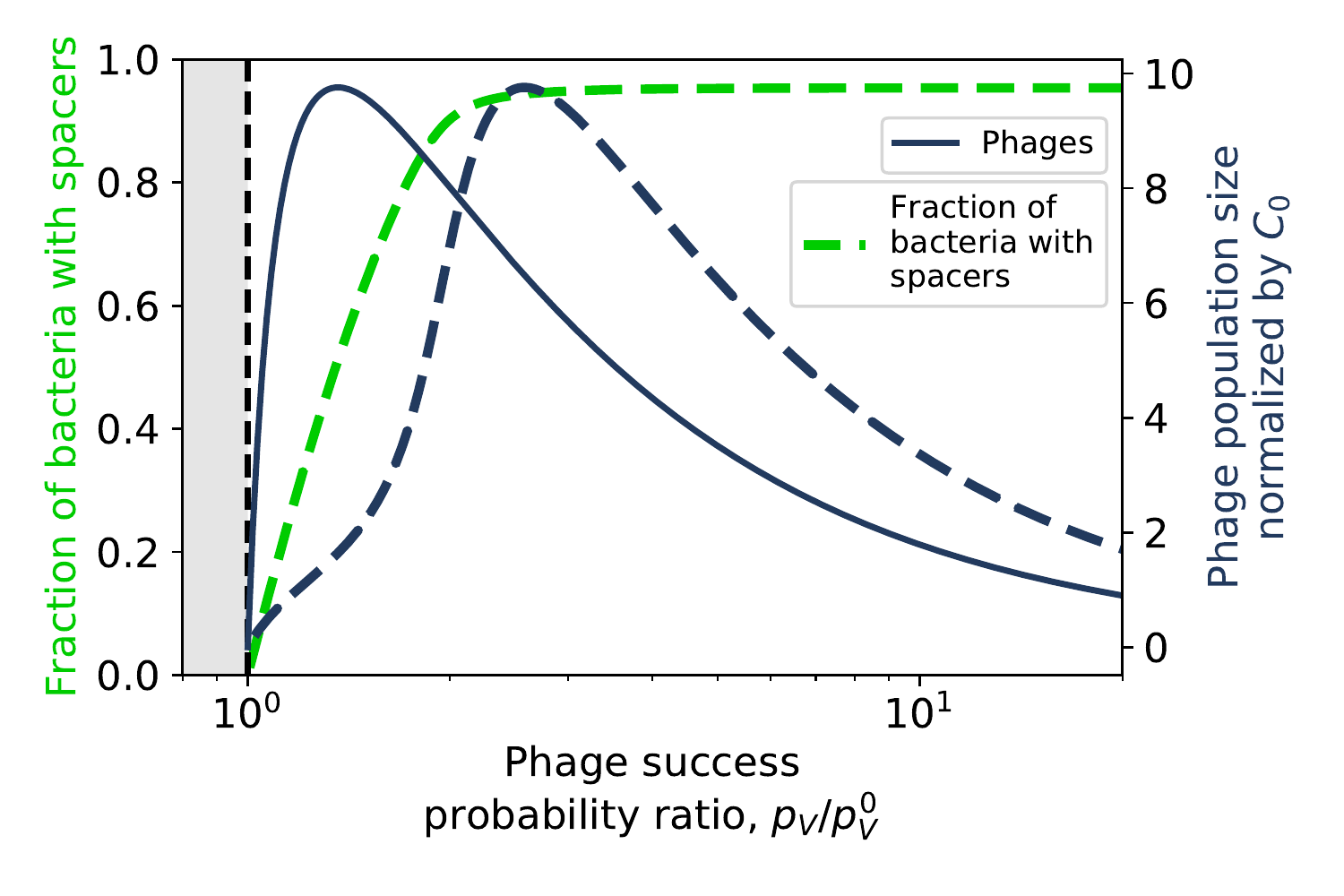}
\caption{Phage and the fraction of bacteria with spacers at steady state as a function of the probability of phage success ($p_V$) for a model without CRISPR ($e=0$, solid line) and for a model where bacteria have CRISPR systems and are able to acquire spacers ($e=0.5$, dashed lines). The phage population size is normalized by the inflow nutrient concentration $C_0$ (y-axis labels on the right).
Below $p_V = p_V^0$, phages cannot persist and the fraction of bacteria with spacers is $0$.}
\label{nu_phage_vs_pv}
\end{figure}

We observed that the minimum success probability $p_V^0 = \frac{1}{B}\left(\frac{gf}{(1-f)\alpha} +1\right)$ required for phages to invade a bacterial culture is independent of $e$, which parametrizes adaptive immunity. To understand this, note that the fraction of bacteria with spacers ($\nu = n_B^s/n_B$) $ = 0$ whenever $n_V$, the number of phages is $0$, since spacers are continually lost with a small rate $r$ but cannot be acquired if there are no phages. As a result, $p_V^0$ is independent of $e$ since there are no bacteria with spacers at the point of phage extinction. 
SI Figure \ref{nu_phage_vs_pv} shows $\nu$ and phages with and without adaptive immunity, illustrating that both $\nu$ and $n_V$ go to zero at $p_V = p_V^0$.

\subsubsection{Large $\alpha$ limit}

\begin{figure}
\centering
\includegraphics[width=1.0\textwidth]{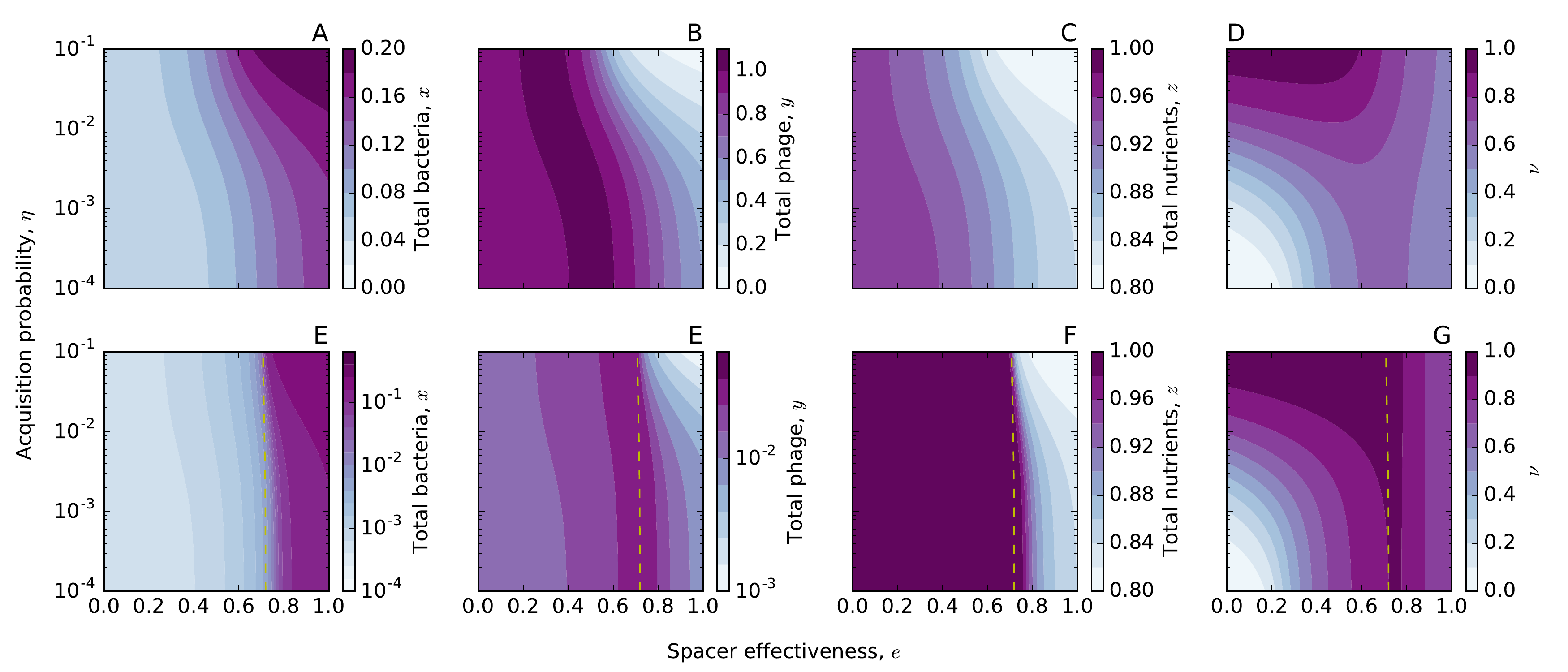} 
\caption{$x$, $y$, $z$, and $\nu$ vs. $\eta$ and $e$ for $\alpha \gtrsim \frac{gf}{(1-f)(Bp_V -1)}$ (top row) and $\alpha >> \frac{gf}{(1-f)(Bp_V -1)}$ (bottom row). The yellow dashed line is the approximate value of $e$ at which both $\nu$ (G) and $y$ (F) are maximized (equation \ref{approx_xi_crit}) which agrees well with the full solution for large $\alpha$.} \label{large-alpha}
\end{figure}

For large $\alpha$ ($\alpha >> \alpha_0$, where $\alpha_0 = \frac{gf}{(1-f)(Bp_V -1)}$), we can find an approximate value of $e$, $e^*$, at which $\nu$ and $n_V$ peak (equation \ref{approx_xi_crit}). This solution is plotted as a yellow dashed line in SI Figure \ref{large-alpha}.

\begin{equation}
e^* = \frac{1}{Bp_V} + \frac{R(1-Bp_V)}{Bp_V(f-1)(1-B(\eta+p_V)+B\eta p_V)}
\label{approx_xi_crit}
\end{equation} 

\label{fourth:app}

\section{Spacer dynamics}
\label{app4-1}
\begin{figure}
\centering
\includegraphics[width=0.6\textwidth]{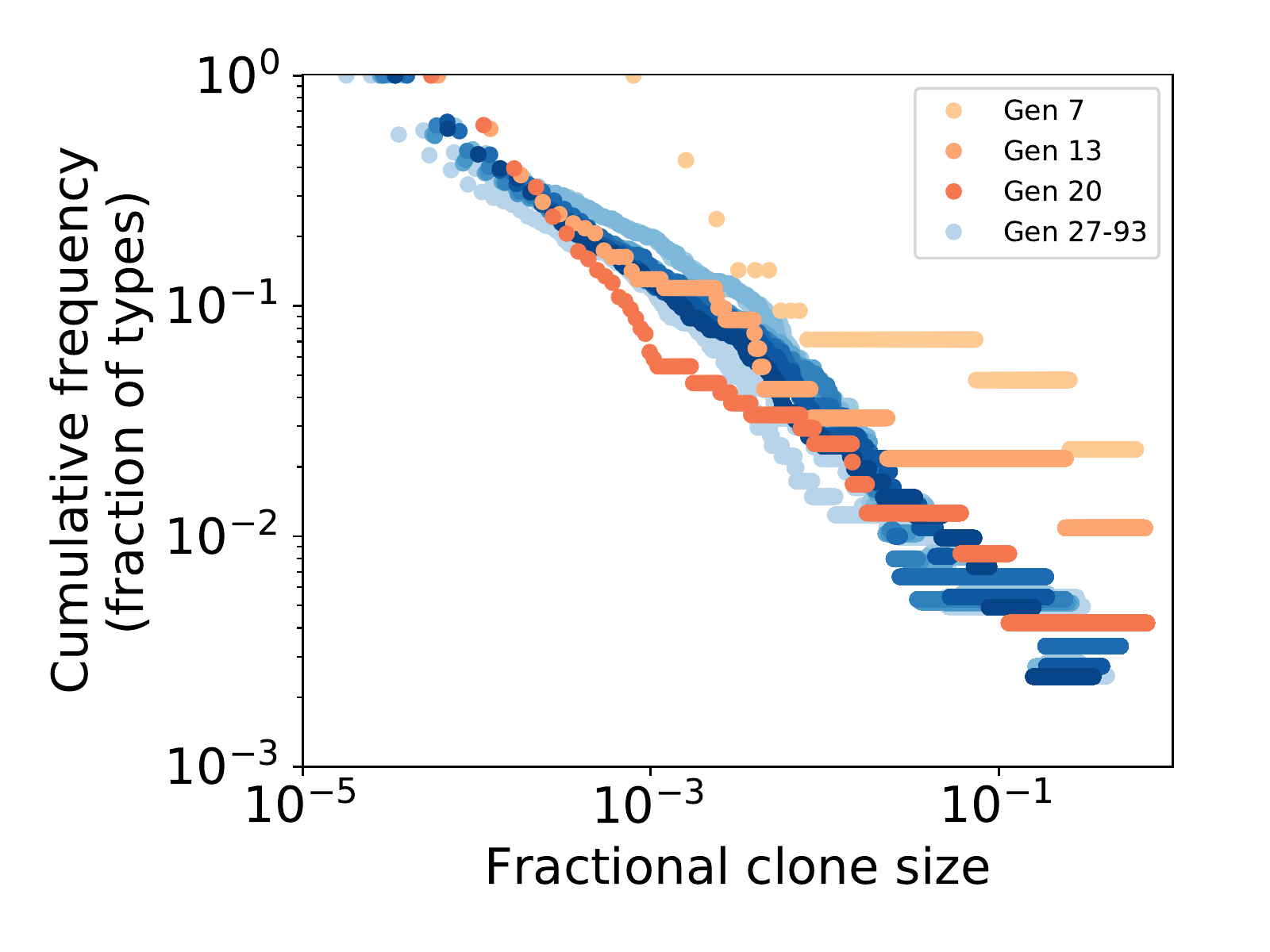} 
\caption{Cumulative frequency of spacer types (clones) as a function of normalized clone size. Darker blue indicates earlier times.} \label{cumulative_dist}
\end{figure}

In our analysis of data from \cite{Paez-Espino2013}, we found that spacer abundance distributions were stable in time after three days and were broad, spanning four orders of magnitude. This distribution $\rho(v)$ is created by summing all spacer types of a particular abundance: $\rho(v) = \sum_i\delta(n_B^i-v)$. The normalized cumulative distribution, $\sum_{v}^{\infty}\rho(v)/\sum_{0}^{\infty}\rho(v)$, is plotted in SI Figure \ref{cumulative_dist}. The corresponding rank-abundance distribution is plotted in Figure \ref{simulation_exp}C in the main text.

\begin{figure}
\centering
\includegraphics[width=0.8\textwidth]{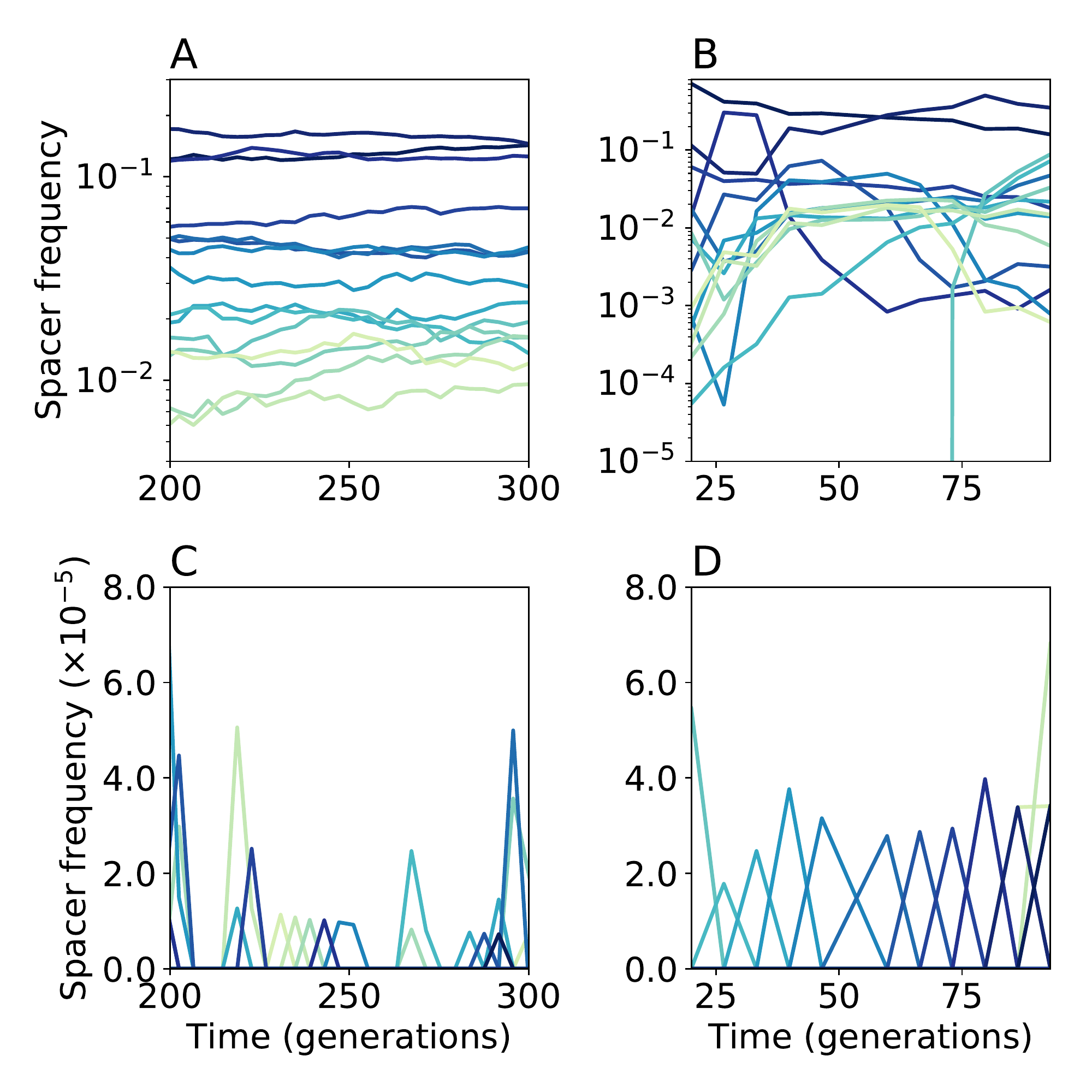} 
\caption{Spacer type trajectories vs. time for experimental data from \cite{Paez-Espino2013} (B and D) and for data from our simulations (A and D). Colours indicate different spacer types. (A and B) show the largest 15 spacer types vs. time and (C and D) show the lowest 15 unique spacer type trajectories vs. time. Both large and small abundance spacers experience turnover in the experimental data, while in simulations the large abundance spacers are approximately stable once the system has reached population-level steady state.} \label{spacer_turnover}
\end{figure}

Individual spacer types experience continual turnover, both in our simulations and in experimental data from \cite{Paez-Espino2013}. In the experimental data, both high-abundance and low-abundance spacers can change in abundance by an order of magnitude or more between time points, while in our simulations we find that the large abundance spacers are approximately stable once the system has reached a population-level steady state (SI Figure \ref{spacer_turnover}).

The observed turnover in large spacer types in the experimental data may reflect additional stochasticity not accounted for in our model, changes in fitness for individual spacer types over time, or the fact that the sequenced spacers are strongly undersampled. There are $\approx 10^8$ to $10^9$ bacteria at the end of each day in the experiment, and there are $\approx 3 \times 10^4$ spacers recovered from sequencing each day. The data is undersampled by a factor of $\approx 10^4$, and apparent turnover may result from this. 

\begin{figure}
\centering
\includegraphics[width=1.0\textwidth]{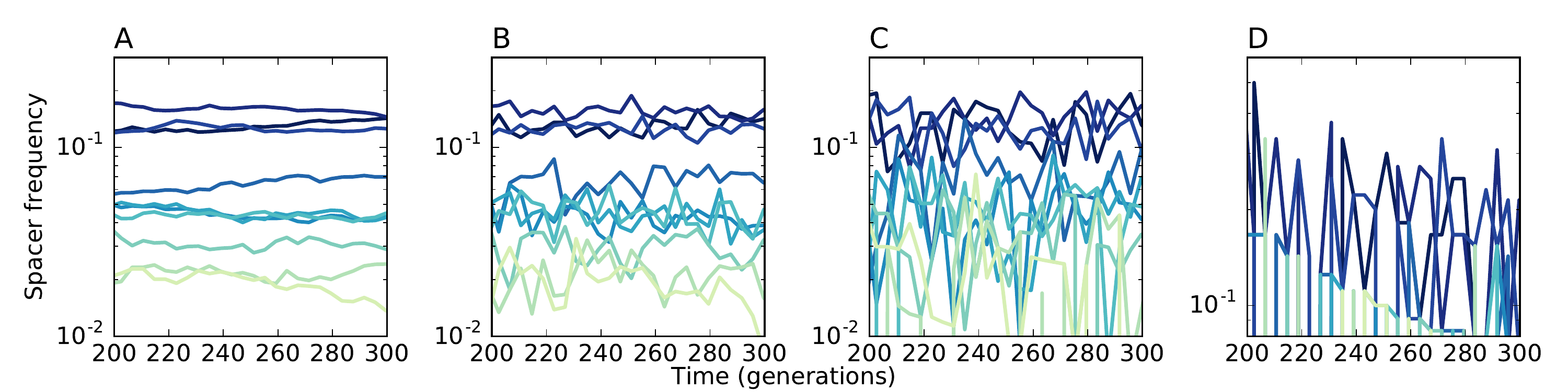} 
\caption{A comparison of the top 10 spacer types from the original simulation (A) with a randomly sampled subset of the simulated data (B, C, D). Spacers are sampled without replacement at every 50th simulation time point. Data is undersampled by a factor of $10^2$ (B), $10^3$ (C) and $10^4$ (D). } \label{undersampling}
\end{figure}

SI Figure \ref{undersampling} compares the original simulation data with data undersampled by a factor of $10^2$, $10^3$, or $10^4$. The mean fractional abundance over time for a particular type appears mostly unaffected by the undersampling, but there is indeed more variability when the degree of undersampling is higher. At an undersampling factor of $10^4$, spacer counts are in the ones and tens, much lower than than counts of $\approx 10^3$ or $10^4$ in the experimental data. Variability in the experimental data is over more orders of magnitude between time points than in the undersampled simulated data. 

This undersampling of simulated data only considered that fewer organisms are sequenced than are present in the population and does not take into account that the experiment was performed with 100:1 serial dilutions and so each time point was seeded with a random subsample from the previous time point. 

\subsection{Time to extinction}
\label{app4-2}

\begin{figure}
\centering
\includegraphics[width=1.0\textwidth]{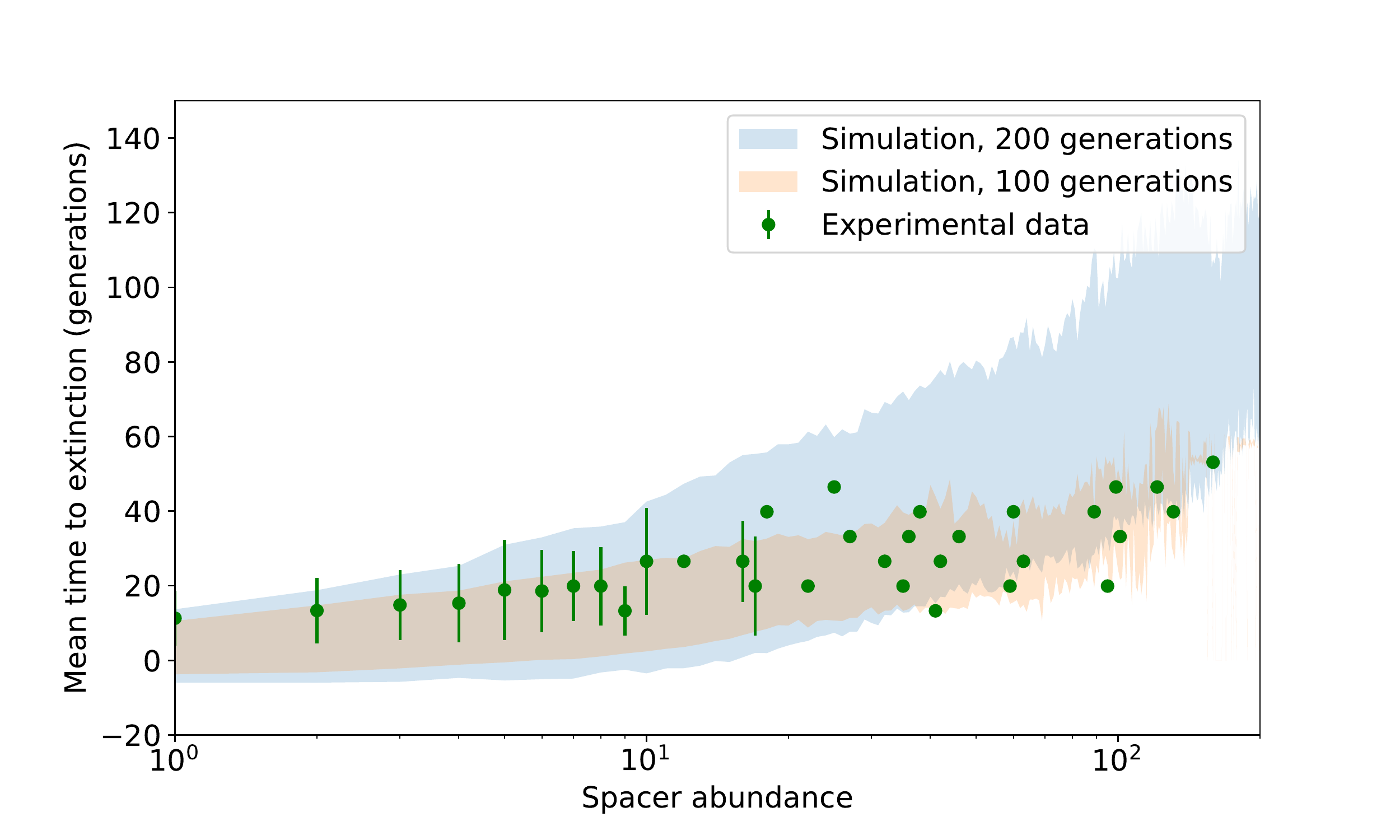} 
\caption{Standard deviation of mean time to extinction for a short and long simulated dataset with $\eta = 10^{-5}$ and $e=0.387$ (shaded areas), and mean time to extinction for experimental data (green points). Errorbars for experimental data are standard deviation of mean time to extinction. Time in generations for the experimental data is time in days $\times 6.64$, assuming exponential growth between daily 100-fold dilutions. } \label{time_to_extinction}
\end{figure}

\begin{figure}
\centering
\includegraphics[width=1.0\textwidth]{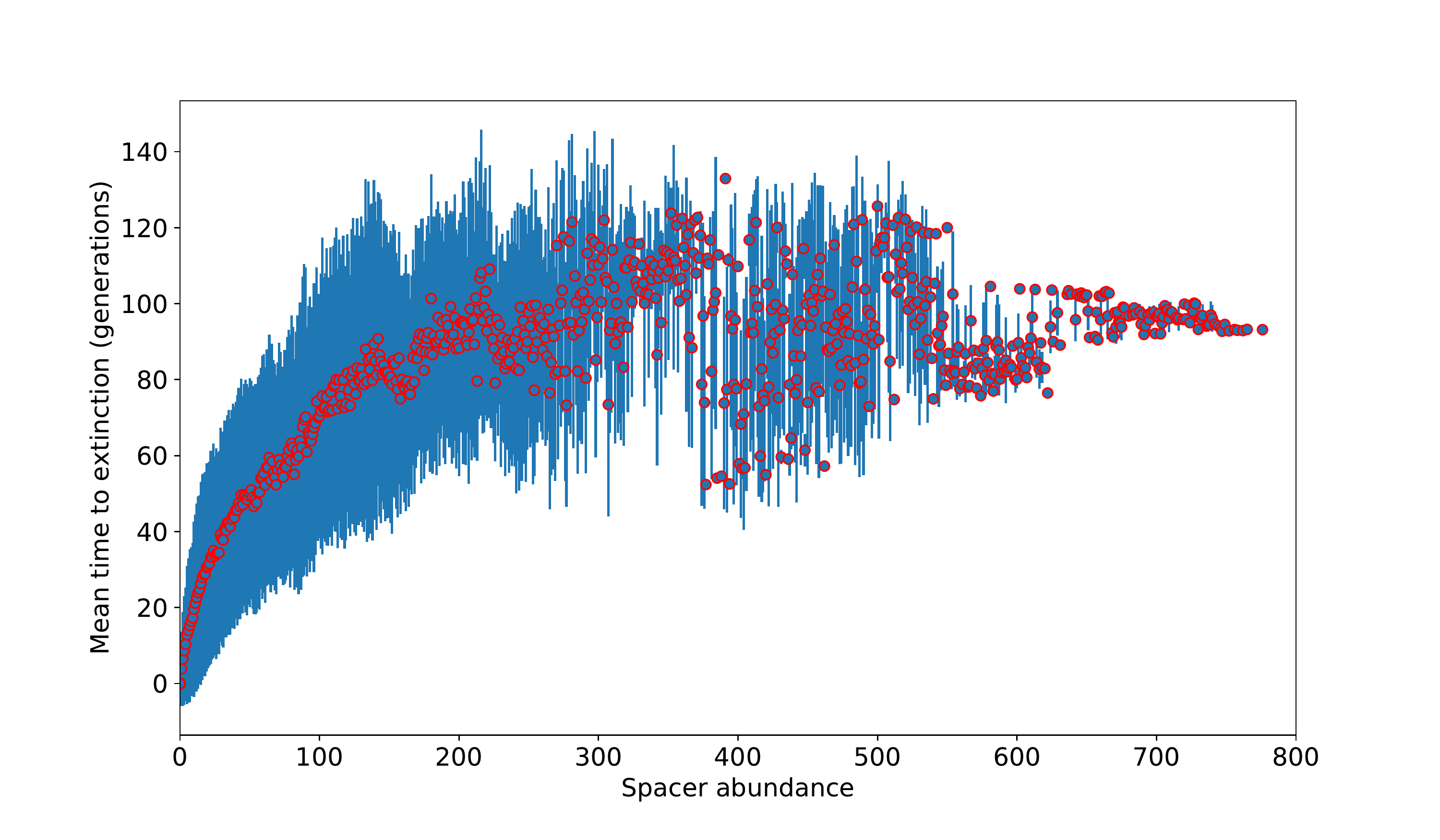} 
\caption{Mean time to extinction for simulated data with $\eta = 10^{-5}$ and $e=0.387$ (red points). Errorbars (blue lines) are standard deviation of mean time to extinction.} \label{time_to_extinction_sim}
\end{figure}

To further investigate ongoing turnover in individual spacer types in our simulations and the experimental data, we calculated the mean time to extinction as a function of spacer abundance. Only data at steady state was used, beginning at Day 4 in the experimental data and generation 200 in the simulated data. For each spacer type that went extinct during the simulation or experiment, the time remaining to extinction was recorded as a function of its abundance at each time point after steady state, and the average and standard deviation over all types were calculated at each abundance. Figure \ref{time_to_extinction} shows the standard deviation envelope for simulated data overlaid with experimental data, indicating that for both simulations and experiment spacers continue to experience turnover at steady-state and that the simulated time to extinction closely matches the experimental observations. Note that the longest observed time to extinction can never exceed the length of the simulation or experiment, meaning that shorter measurement windows will result in shorter average times to extinction. To illustrate this effect we calculate two time to extinction distributions for a short simulation of similar length to the experiment (generation 300 to 400) and a longer simulation (generation 300 to 500). Figure \ref{time_to_extinction_sim} shows that the mean time to extinction is finite even for high abundance spacers in the simulated data. 

\label{fifth:app}

\section{Regulation of CRISPR-Cas}
\label{app5-1}
\subsection{Extent of bistability}
We add regulation of CRISPR-Cas to our model by making spacer effectiveness $e$ a function of bacterial cell density, assuming Cas expression to also be a sigmoidal function of cell density. Many bacterial behaviours controlled by quorum sensing are threshold-dependent: cells must switch between discrete states such as motile and non-motile, biofilm and free-living, virulent and non-virulent. In many quorum sensing systems, production of the autoinducer molecule is under positive feedback and increases nonlinearly with increasing cell density, and so many of the resulting changes in gene expression are switch-like \cite{Miller2001}. For this reason we assume that spacer effectiveness depends strongly on cell density. 

\begin{figure}
\centering
\includegraphics[width=1.0\textwidth]{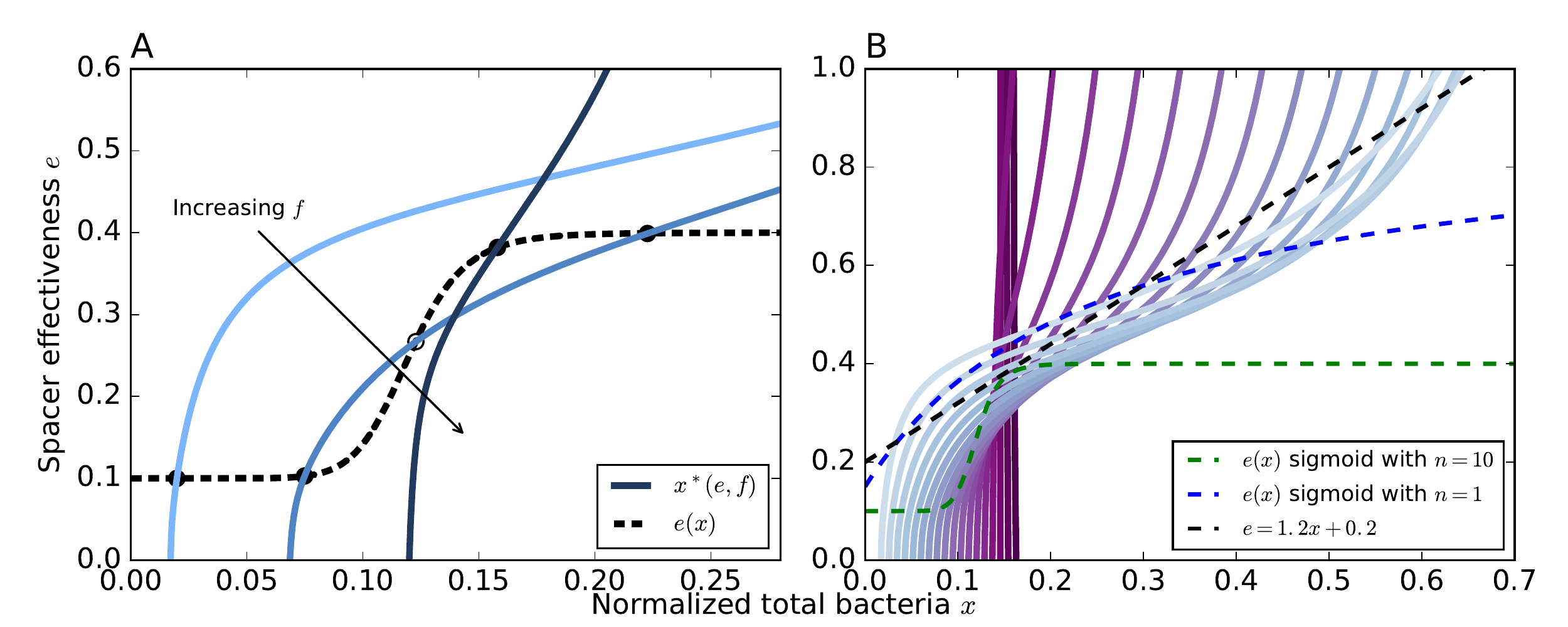} 
\caption{(A) The original dependence of bacterial population size at steady-state on spacer effectiveness $e$ and normalized flow rate $f$ is plotted for three values of $f$ (solid blue lines). 
We model upregulation from quorum sensing by introducing a density-spacer effectiveness (dashed black line), $e(x) = e_{min} + (e_{max} - e_{min})\left(\frac{x^n}{x^n + x_0^n}\right)$,
 so that spacer effectiveness is no longer a constant parameter. 
Any intersection of the dashed line with a solid line is a fixed point; fixed points are indicated with solid circles (stable) and open circles (unstable).
(B) Spacer effectiveness $e$ vs. bacterial population size at steady-state for different values of $f$ (solid lines). Line colour darkens as $f$ increases. Three different choices of $e(x)$ are plotted (dashed lines), all of which intersect some of the solid curves in three places, indicating bistability.} \label{e_vs_x}
\end{figure}

However, we observe bistability for a wide range of parameters and note that spacer effectiveness does not necessarily need to be a sharp function of $x$, where $x = n_B/C_0$. SI Figure \ref{e_vs_x} illustrates the additional dependence of $e$ on $x$ --- wherever $e(x)$ intersects the original solution, there is a fixed point. SI Figure \ref{e_vs_x}B shows that even a linear $e(x)$ can intersect the original solution in three places for certain parameters, in this case for certain values of $f$. Any curve that intersects one of the solid lines in three places will result in bistability. 

\begin{figure}
\centering
	\includegraphics[width=0.8\textwidth]{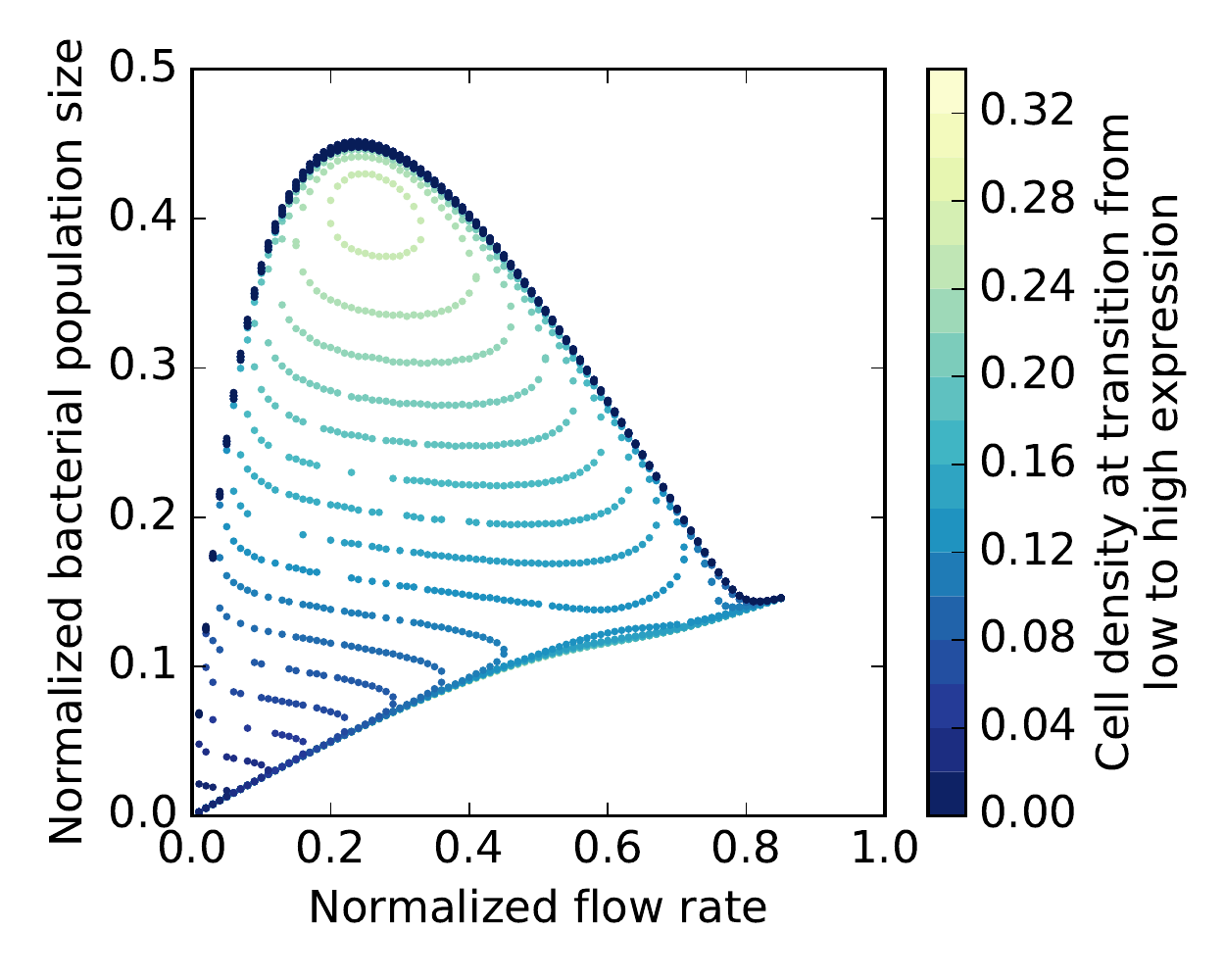} 
	\caption{Fixed points (bacterial population size) as a function of flow rate $f$ for different values of the transition point between low and high expression. As the transition point increases (lighter colours), the bistability changes from an `S' shape to a circle and a line. This bifurcation happens at a transition point of approximately $x=0.15$.} \label{sigmoid_centre_2d}
\end{figure}

\begin{figure}
	\includegraphics[width=0.8\textwidth]{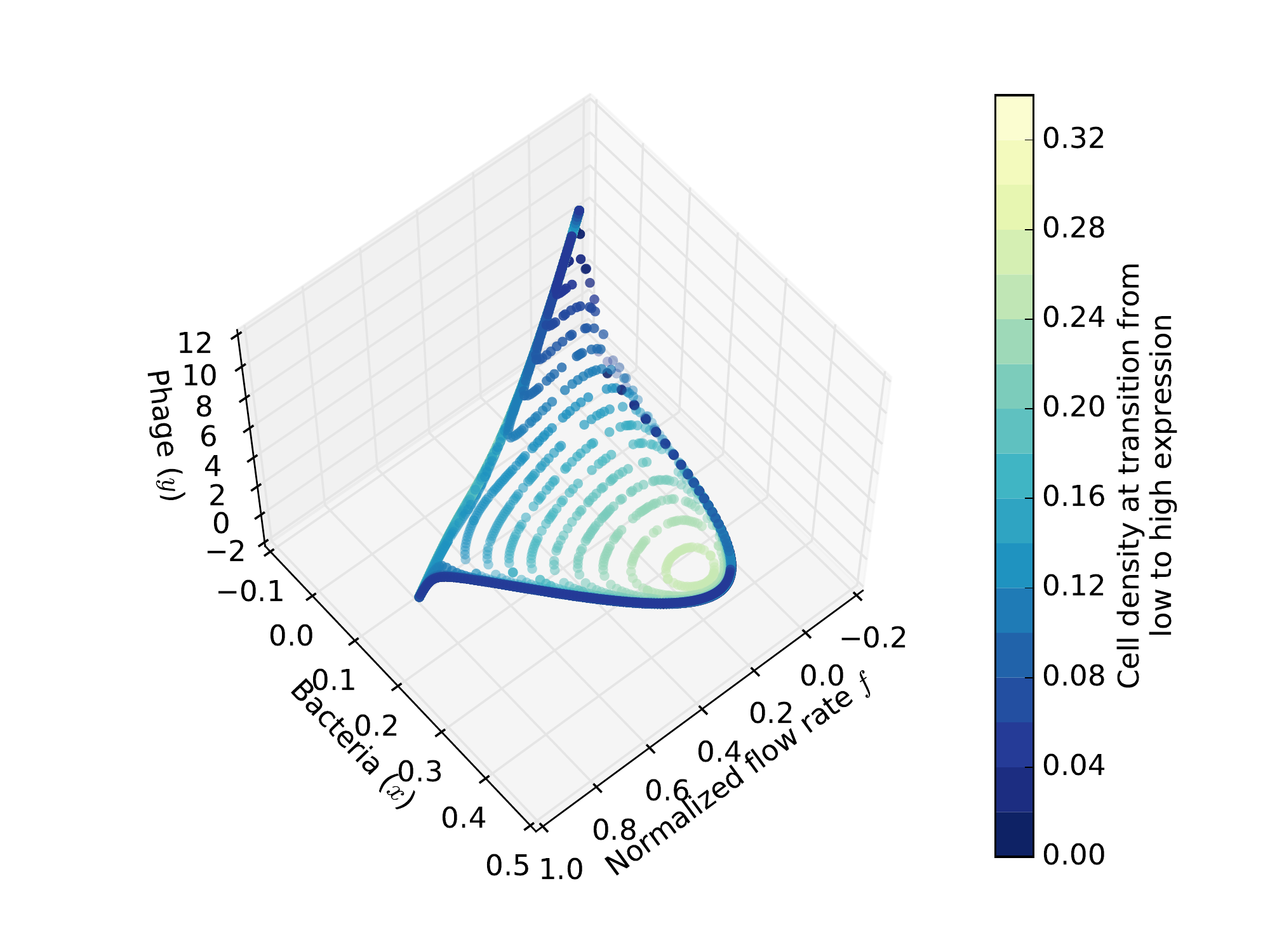} 
	\caption{Fixed points (bacterial population size and phage population size) as a function of flow rate $f$ for different values of the transition point between low and high expression.} \label{sigmoid_centre_3d}
\end{figure}

Changing the precise location of the transition from low to high spacer effectiveness does not change the existence of bistability, but it does cause an interesting bifurcation. SI Figures \ref{sigmoid_centre_2d} and \ref{sigmoid_centre_3d} show in two and three dimensions what happens to the fixed points as the transition point $x_0$ is scanned from $0$ to $0.3$. For a transition point at low cell density, the unstable fixed points are adjacent to the low expression stable fixed points at one end and the high expression stable fixed points at the other end, making hysteresis possible. However, as the transition point increases to higher cell density, the two ends meet and form a closed loop with just the high expression state. In this situation, bistability still exists, but the system can never jump from the low expression state to the high expression state without being placed there since there is one continuous low expression stable state across the entire range of $f$.

\subsection{Bistability across system variables}
\label{app5-2}

\begin{figure}
\centering
\includegraphics[width=1.0\textwidth]{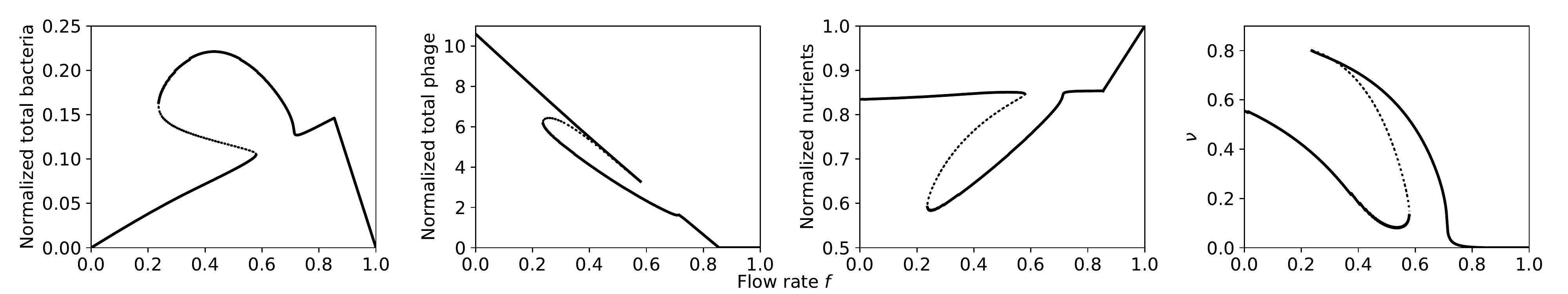} 
\caption{Bacteria $x$, phages $y$, nutrients $z$, and the fraction of bacteria with spacers $\nu$ as a function of $f$ in a parameter regime showing bistability. The solid black lines indicates a stable fixed point and the dashed black line indicates an unstable fixed point. } \label{bistability_all}
\end{figure}

Bistability affects all four dynamical variables in our model. SI Figure \ref{bistability_all} shows each variable at steady state vs. flow rate $f$ in a regime with bistability.

\subsection{Adding regulation to acquisition, loss, and growth rate}
\label{app5-3}

We model CRISPR-Cas regulation by making spacer effectiveness density-dependent, but it is reasonable that up-regulation of CRISPR-Cas would affect other system parameters as well. In particular, spacer acquisition rates would likely increase since acquisition relies on the Cas protein machinery as does interference \cite{Barrangou2014}. Additionally, spacer loss is thought to happen by homologous recombination and to occur in tandem with acquisition \cite{Deveau2008, Weinberger2012a}. 

\begin{figure}
\centering
\includegraphics[width=0.7\textwidth]{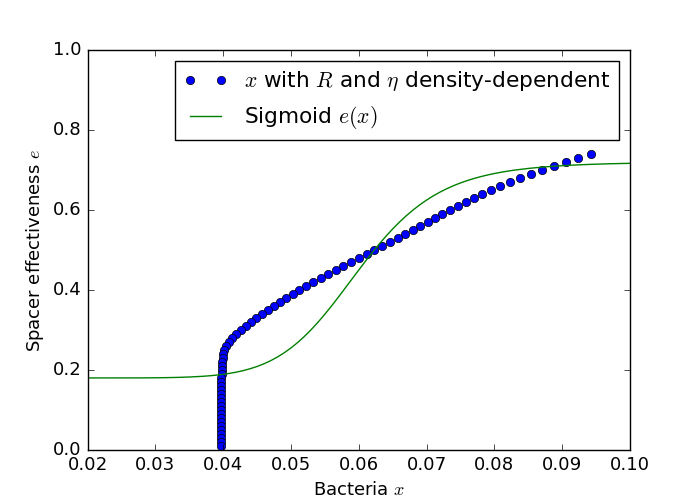} 
\caption{The dependence of bacterial population size $x$ at steady-state on spacer effectiveness $e$ when $r$ and $\eta$ are both sharp functions of density (blue dots). A monotonic function for spacer effectiveness as a function of $x$ (green solid line) can still only intersect in at most three places, qualitatively giving the same bistability result. } \label{sigmoid_r_eta}
\end{figure}

We added a sharp sigmoidal density dependence to both spacer acquisition probability and spacer loss rate. SI Figure \ref{sigmoid_r_eta} shows the resulting steady-state bacterial population size as a function of spacer effectiveness. The result is still monotonically increasing, which means that a monotonic function for spacer effectiveness as a function of $x$ can still only intersect in at most three places, qualitatively giving the same bistability result.

Measurements of the fitness cost of CRISPR in \textit{Streptococcus thermophilus} identified Cas protein expression as having a fitness cost \cite{Gandon2015}, making it reasonable that bacteria would down-regulate Cas expression in times when CRISPR is not needed. \cite{Gandon2015} measured a selective advantage of $0.11$ for S. \textit{thermophilus} with a \textit{cas9} or \textit{csn2} gene knockout in direct competition with wild type but did not observe a difference in maximum growth rate. This definition of selective advantage corresponds to the difference in average exponential growth rate per hour for each strain. We incorporated a Cas-expression-dependent decrease in bacterial growth rate in our model and investigated its effect on bistability. Here we model Cas expression as a theta function (discrete `off' and `on' states) with the switch occurring at $x_C = 0.06$ (arbitrarily chosen):

\begin{equation}
  g(x) =
  \begin{cases}
  g_1 & x \le x_C \\
  g_0 & x > x_C
  \end{cases} \label{g_reg}
\end{equation}

The growth rate $gC_0$ depends on the Cas expression state with $g_1C_0$ being the growth rate per minute without Cas expression and $g_0C_0$ being the growth rate with Cas expression, where $g_0 < g_1$. A selective advantage of $0.11$ gives $g_0 = g_1 - 0.11/(60C_0)$.

\begin{figure}
\centering
	\includegraphics[width=0.8\textwidth]{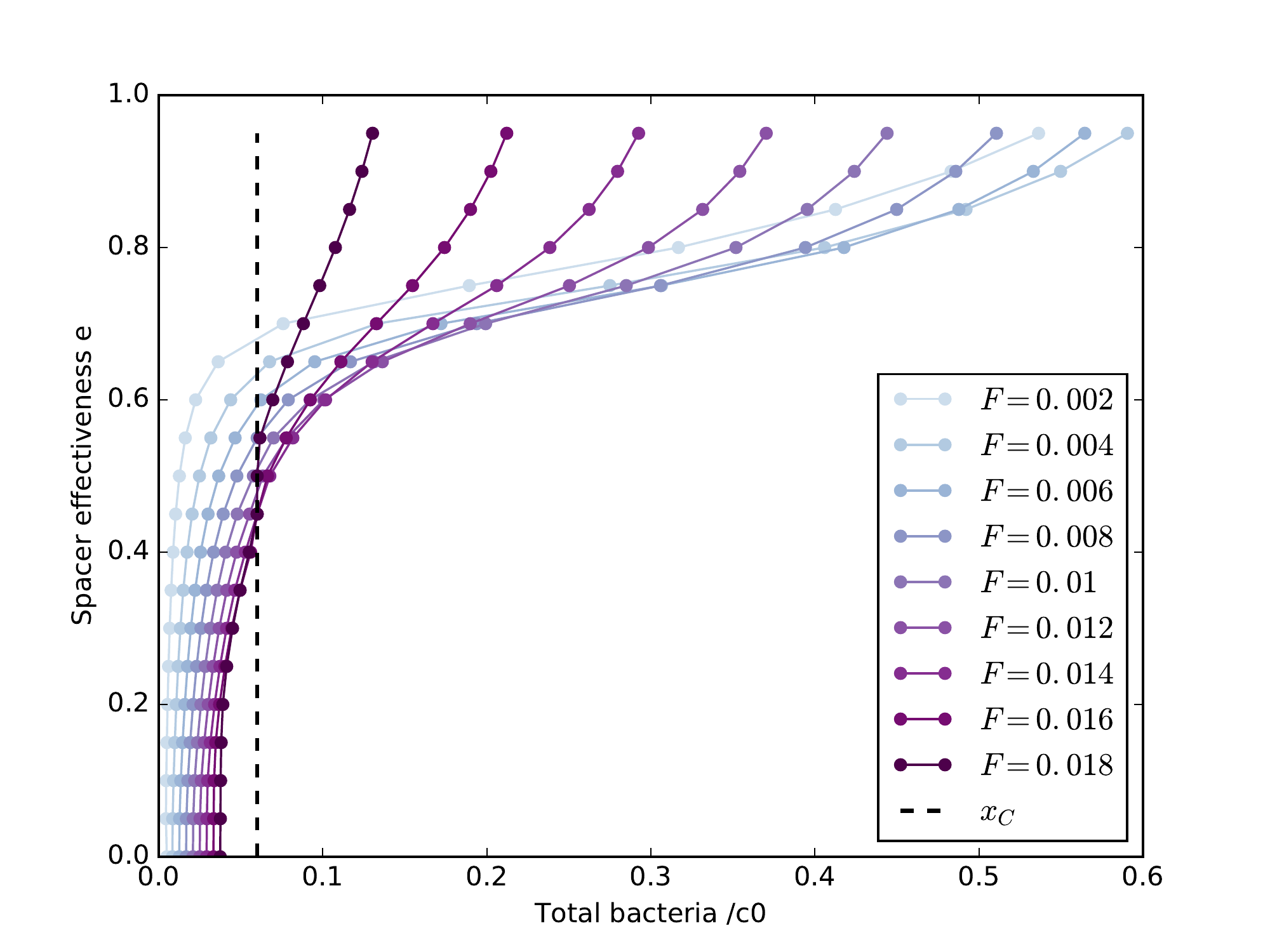} 
	\caption{The dependence of bacterial population size $x$ at steady-state on spacer effectiveness $e$ when $g$ is a sharp function of cell density $x$. The value of $x$ at which regulation is turned on or off is indicated by the black dashed line. Lines are plotted for $F$ instead of $f = F/(gC_0)$ because $f$ depends on $g$. Plotted is bacterial population size at steady state where the growth disadvantage for Cas expression is $g_0 = g_1 - 0.11/(60C_0)$, calculated from the measured selection coefficient in \cite{Gandon2015}.} \label{regulation_g1}
\end{figure}

\begin{figure}
\centering
	\includegraphics[width=0.8\textwidth]{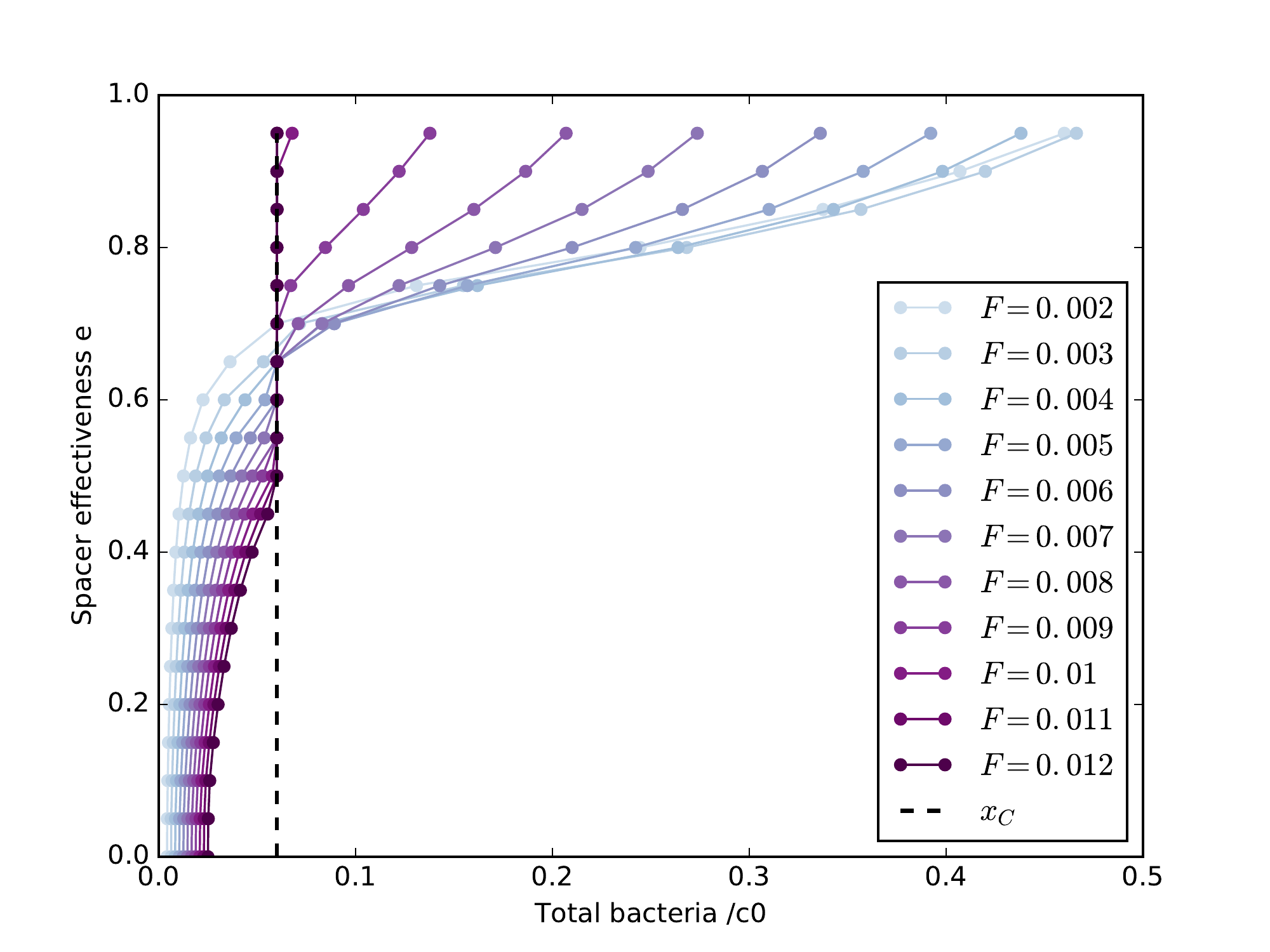} 
	\caption{The dependence of bacterial population size $x$ at steady-state on spacer effectiveness $e$ when $g$ is a sharp function of cell density $x$. The value of $x$ at which regulation is turned on or off is indicated by the black dashed line. Plotted is bacterial population size at steady state where the growth disadvantage for Cas expression is $g_0 = 0.5g_1$.} \label{regulation_g2}
\end{figure}

SI Figures \ref{regulation_g1} and \ref{regulation_g2} show the resulting change in steady-state bacterial population size as a function of spacer effectiveness for two different growth rate dependences on expression. For even a 50 percent reduction in growth rate at high Cas expression, the resulting curves are not qualitatively altered, and as before, a monotonic curve for $e(x)$ can intersect in at most three places to give bistability.

\subsection{Experimentally measuring regulation }
\label{app5-4}

While we chose parameters that are reasonable for S. \textit{thermophilus}, it is unlikely that our quantitative results match experimental conditions for different organisms. Our prediction is that in an appropriate parameter range, an experiment measuring bacterial population density as a function of flow rate may exhibit hysteresis as the flow rate is first increased and then decreased, allowing the bacteria-phage population to reach steady state after each change in flow rate. It is easy to imagine however that the transition determining high or low Cas expression may not automatically align with the cell densities in the chemostat. The first experimental step is to measure the true Cas expression as a function of cell density for \textit{Pseudomonas}, as done in \cite{Hoyland-Kroghsbo2016}. In their experiment, \textit{cas3} expression increased by a factor of about 10 for a 10-fold increase in cell density (from $\approx 8 \times 10^7$ to $\approx 8 \times 10^8$ CFU/mL). 

Next, the concentration of nutrients in the inflow medium $C_0$ can be used to tune the cell density to one at which CRISPR would naturally be highly expressed at a high flow rate. Then the flow rate $F$ can independently tune the position along the bifurcation diagram in SI Figure \ref{regulation}. In this way an experimental population of \textit{Pseudomonas} can be tweaked to qualitatively align with our model. 

\begin{figure}
\centering
\includegraphics[width=0.7\textwidth]{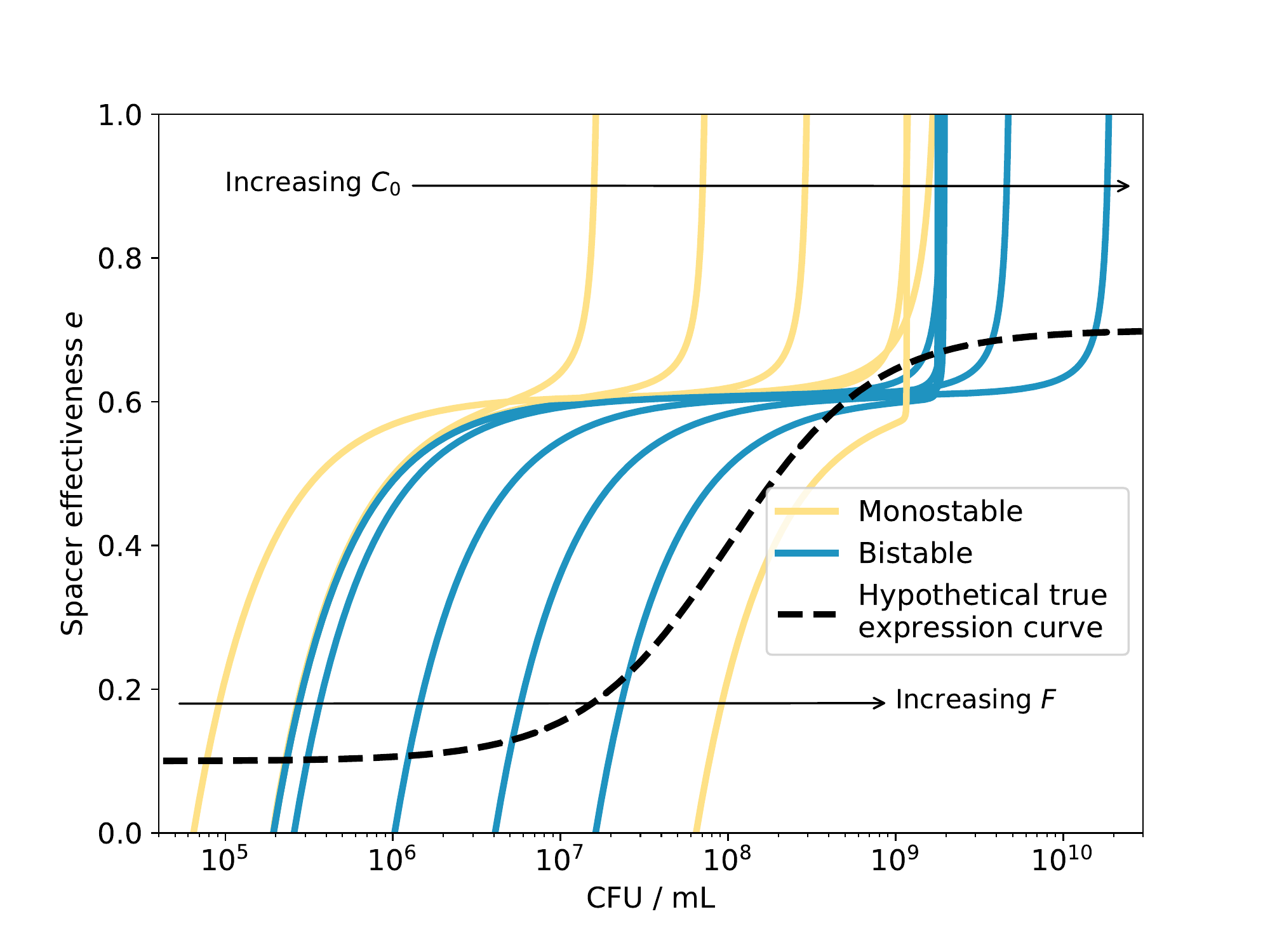} 
\caption{The dependence of bacterial population size $x$ at steady-state on spacer effectiveness $e$ in the model for different values of $F$ and $C_0$ (solid lines). For a given measured dependence of \textit{cas} expression on cell density (black dashed line, for example), $F$ and $C_0$ can tune whether the system is monstable or bistable by changing the number of intersections between the two curves.} \label{e_vs_nb}
\end{figure}

SI Figure \ref{e_vs_nb} shows the steady-state bacterial concentration vs. spacer effectiveness in our model as $C_0$ and $F$ are varied. Provided the true Cas expression is a sharp enough function of density and that the low expression state is below the plateau in effectiveness in SI Figure \ref{e_vs_nb}, it will be possible to choose $C_0$ and $F$ such that the system is bistable. The position of the plateau in effectiveness at which the bacterial density changes sharply is controlled by $Bp_V$: as $Bp_V$ increases, the plateau moves to higher effectiveness. $B$ and $p_V$ are properties specific to the phage and may change with the particular phage species used.  

\subsubsection{Significance of regulation in natural populations}

In natural populations, multiple states may define different ecological niches as seen in structured populations from microbial mats \cite{Ward2006} to the human microbiome \cite{Huttenhower2012}. Biofilms are an example of both dense and structured communities of bacteria and are found in many natural environments such as hot springs \cite{Ward2006} and acid mine drainage \cite{Andersson2008a} and in many clinically relevant environments such as medical implants, lungs of cystic fibrosis patients, and dental plaques \cite{Madigan2006}. Because of their protective polysaccharide coating, biofilms are often difficult to treat with antibiotics \cite{Madigan2006}, and phage therapy has been proposed as a potential treatment for antibiotic-resistant bacterial colonies. H\o yland-Kroghsbo \textit{et al.} posited that upregulation of CRISPR-Cas could pose a challenge to potential phage therapies for biofilms \cite{Hoyland-Kroghsbo2016}. If such a biofilm-bound population is in the bistable regime we find, there may be a way to prime the population in way that pushes it to the low CRISPR-Cas expression state to utilize phage therapy effectively. More broadly, in a resource-limited environment, for example, a bacterial population may do better to maintain a low density and avoid phage predation while repressing the expression of Cas proteins, but consequently may lose their CRISPR-Cas system entirely. These ecological constraints may shed light on why CRISPR-Cas is neither universal nor uncommon in the microbial world. 


\begin{thebibliography}{10}

\bibitem{Ward2006}
David~M. Ward, Mary~M. Bateson, Michael~J. Ferris, M.~Kuhl, Andrea Wieland,
  Alex Koeppel, and Frederick~M. Cohan.
\newblock {Cyanobacterial ecotypes in the microbial mat community of Mushroom
  Spring (Yellowstone National Park, Wyoming) as species-like units linking
  microbial community composition, structure and function}.
\newblock {\em Philos. Trans. R. Soc. B: Biol. Sci.}, 361(1475):1997--2008,
  2006.

\bibitem{Schwabe2013}
Robert~F Schwabe and Christian Jobin.
\newblock {The microbiome and cancer.}
\newblock {\em Nature Rev. Cancer}, 13(11):800--12, Nov 2013.

\bibitem{Collins2014}
Stephen~M Collins.
\newblock {A role for the gut microbiota in IBS.}
\newblock {\em Nature Rev. Gastroenterol. {\&} Hepatol.}, 11(8):497--505, Aug
  2014.

\bibitem{Korem2015}
T.~Korem, D.~Zeevi, J.~Suez, A.~Weinberger, T.~Avnit-Sagi, M.~Pompan-Lotan,
  E.~Matot, G.~Jona, A.~Harmelin, N.~Cohen, A.~Sirota-Madi, C.~A. Thaiss,
  M.~Pevsner-Fischer, R.~Sorek, R.~Xavier, E.~Elinav, and E.~Segal.
\newblock {Growth dynamics of gut microbiota in health and disease inferred
  from single metagenomic samples}.
\newblock {\em Science}, 349(6252):1101--1106, 2015.

\bibitem{Muhlebach2018}
Marianne~S Muhlebach, Bryan~T Zorn, Charles~R Esther, Joseph~E Hatch, Conor~P
  Murray, Lidija Turkovic, Sarath~C Ranganathan, Richard~C Boucher, Stephen~M
  Stick, and Matthew~C Wolfgang.
\newblock {Initial acquisition and succession of the cystic fibrosis lung
  microbiome is associated with disease progression in infants and preschool
  children}.
\newblock {\em PLOS Pathogens}, 14(1):1--20, 2018.

\bibitem{OToole2017}
George~A. O'Toole.
\newblock {Cystic Fibrosis Airway Microbiome: Overturning the Old, Opening the
  Way for the New}.
\newblock {\em J. Bacteriol.}, 200(4), 2017.

\bibitem{Heidelberg2009}
John~F Heidelberg, William~C Nelson, Thomas Schoenfeld, and Devaki Bhaya.
\newblock {Germ warfare in a microbial mat community: CRISPRs provide insights
  into the co-evolution of host and viral genomes.}
\newblock {\em PloS ONE}, 4(1):e4169, Jan 2009.

\bibitem{Suttle2007}
Curtis~A Suttle.
\newblock {Marine viruses -- major players in the global ecosystem.}
\newblock {\em Nature Rev. Microbiol.}, 5(10):801--812, 2007.

\bibitem{Doron2018}
Shany Doron, Sarah Melamed, Gal Ofir, Azita Leavitt, Anna Lopatina, Mai Keren,
  Gil Amitai, and Rotem Sorek.
\newblock {Systematic discovery of antiphage defense systems in the microbial
  pangenome}.
\newblock {\em Science}, 4120(January), 2018.

\bibitem{Andersson2008a}
Anders~F. Andersson and Jillian~F. Banfield.
\newblock {Virus population dynamics and acquired virus resistance in natural
  microbial communities.}
\newblock {\em Science}, 320(5879):1047--50, May 2008.

\bibitem{Tyson2008}
Gene~W. Tyson and Jillian~F. Banfield.
\newblock {Rapidly evolving CRISPRs implicated in acquired resistance of
  microorganisms to viruses}.
\newblock {\em Environ. Microbiol.}, 10(1):200--207, Jan 2008.

\bibitem{Paez-Espino2013}
David Paez-Espino, Wesley Morovic, Christine~L. Sun, Brian~C. Thomas, Ken-ichi
  Ueda, Buffy Stahl, Rodolphe Barrangou, and Jillian~F Banfield.
\newblock {Strong bias in the bacterial CRISPR elements that confer immunity to
  phage.}
\newblock {\em Nature Communications}, 4:1430, Jan 2013.

\bibitem{Paez-Espino2015}
David Paez-Espino, Itai Sharon, Wesley Morovic, Buffy Stahl, Brian~C Thomas,
  Rodolphe Barrangou, and Jillian~F Banfield.
\newblock {CRISPR Immunity Drives Rapid Phage Genome Evolution in Streptococcus
  thermophilus}.
\newblock {\em mBio}, 6(2):1--9, 2015.

\bibitem{He2010}
Jiankui He and Michael~W. Deem.
\newblock {Heterogeneous Diversity of Spacers within CRISPR (Clustered
  Regularly Interspaced Short Palindromic Repeats)}.
\newblock {\em Phys. Rev. Lett.}, 105(12):128102, Sep 2010.

\bibitem{Weinberger2012a}
Ariel~D. Weinberger, Christine~L. Sun, Mateusz~M. Pluci{\'{n}}ski, Vincent~J.
  Denef, Brian~C. Thomas, Philippe Horvath, Rodolphe Barrangou, Michael~S.
  Gilmore, Wayne~M. Getz, and Jillian~F. Banfield.
\newblock {Persisting viral sequences shape microbial CRISPR-based immunity.}
\newblock {\em PLoS Comp. Biol.}, 8(4):e1002475, Jan 2012.

\bibitem{Childs2012}
Lauren~M Childs, Nicole~L Held, Mark~J Young, Rachel~J Whitaker, and Joshua~S
  Weitz.
\newblock {Multiscale model of CRISPR-induced coevolutionary dynamics:
  diversification at the interface of Lamarck and Darwin.}
\newblock {\em Evolution}, 66(7):2015--2029, Jul 2012.

\bibitem{Haerter2012}
Jan~O Haerter and Kim Sneppen.
\newblock {Spatial structure and Lamarckian adaptation explain extreme genetic
  diversity at CRISPR locus}.
\newblock {\em mBio}, 3(4):1--6, 2012.

\bibitem{Han2013a}
Pu~Han, Liang~Ren Niestemski, Jeffrey~E. Barrick, and Michael~W. Deem.
\newblock {Physical Model of the Immune Response of Bacteria Against
  Bacteriophage Through the Adaptive CRISPR-Cas Immune System}.
\newblock {\em Physical Biology}, 10(2):025004, Apr 2013.

\bibitem{Childs2014}
Lauren~M Childs, Whitney~E England, Mark~J Young, Joshua~S Weitz, and Rachel~J
  Whitaker.
\newblock {CRISPR-induced distributed immunity in microbial populations}.
\newblock {\em PLoS ONE}, 9(7):1--12, 2014.

\bibitem{Bradde2017}
Serena Bradde, Marija Vucelja, Tiberiu Teşileanu, and Vijay Balasubramanian.
\newblock {Dynamics of adaptive immunity against phage in bacterial
  populations}.
\newblock {\em PLoS Comp. Biol.}, 13(4):1--16, 2017.

\bibitem{Han2017}
Pu~Han and Michael~W Deem.
\newblock {Non-classical phase diagram for virus bacterial coevolution mediated
  by clustered regularly interspaced short palindromic repeats}.
\newblock {\em J. R. Soc. Interface}, 14(127):20160905, 2017.

\bibitem{Weinstein2009}
Joshua~A Weinstein, Ning Jiang, R.~A. White, Daniel~S Fisher, and Stephen~R
  Quake.
\newblock {High-Throughput Sequencing of the Zebrafish Antibody Repertoire}.
\newblock {\em Science}, 324(5928):807--810, 2009.

\bibitem{Zarnitsyna2013}
Veronika~I Zarnitsyna, Brian~D Evavold, Louis~N Schoettle, Joseph~N Blattman,
  and Rustom Antia.
\newblock {Estimating the diversity, completeness, and cross-reactivity of the
  T cell repertoire}.
\newblock {\em Frontiers Immunol.}, 4(485):1--11, 2013.

\bibitem{Desponds2016}
Jonathan Desponds, Thierry Mora, and Aleksandra~M Walczak.
\newblock {Fluctuating fitness shapes the clone-size distribution of immune
  repertoires.}
\newblock {\em Proc Natl Acad Sci USA}, 113(2):274--9, Jan 2016.

\bibitem{Houte2016}
Stineke van Houte, Alice K.~E. Ekroth, Jenny~M. Broniewski, H{\'{e}}l{\`{e}}ne
  Chabas, Ben Ashby, Sylvain Gandon, Steve~Paterson {Mike Boots4}, Angus~J.
  Buckling, and Edze~R. Westra.
\newblock {The diversity-generating benefits of a prokaryotic adaptive immune
  system}.
\newblock {\em Nature}, 532(7599):385--388, 2016.

\bibitem{Hoyland-Kroghsbo2016}
Nina~M. H{\o}yland-Kroghsbo, Jon Paczkowski, Sampriti Mukherjee, Jenny
  Broniewski, Edze Westra, Joseph Bondy-Denomy, and Bonnie~L. Bassler.
\newblock {Quorum sensing controls the Pseudomonas aeruginosa CRISPR-Cas
  adaptive immune system.}
\newblock {\em Proc Natl Acad Sci USA}, 114(1):201617415, 2016.

\bibitem{Patterson2016}
Adrian~G Patterson, Simon~A Jackson, Corinda Taylor, Rita Przybilski, Raymond
  H~J Staals, Peter~C Fineran, Adrian~G Patterson, Simon~A Jackson, Corinda
  Taylor, Gary~B Evans, George P~C Salmond, Rita Przybilski, Raymond H~J
  Staals, and Peter~C Fineran.
\newblock {Quorum Sensing Controls Adaptive Immunity through the Regulation of
  Multiple CRISPR-Cas Systems}.
\newblock {\em Molecular Cell}, 64(6):1102--1108, 2016.

\bibitem{Heilmann2010}
Silja Heilmann, Kim Sneppen, and Sandeep Krishna.
\newblock {Sustainability of virulence in a phage-bacterial ecosystem}.
\newblock {\em J. Virol.}, 84(6):3016--22, 2010.

\bibitem{Levin2010}
Bruce~R Levin.
\newblock {Nasty viruses, costly plasmids, population dynamics, and the
  conditions for establishing and maintaining CRISPR-mediated adaptive immunity
  in bacteria.}
\newblock {\em PLoS Genet.}, 6(10):e1001171, Oct 2010.

\bibitem{Haerter2011}
Jan~O Haerter, Ala Trusina, and Kim Sneppen.
\newblock {Targeted Bacterial Immunity Buffers Phage Diversity}.
\newblock {\em J. Virol.}, 85(20):10554--10560, 2011.

\bibitem{Weinberger2012c}
Ariel~D. Weinberger, Yuri~I. Wolf, Alexander~E. Lobkovsky, Michael~S. Gilmore,
  and Eugene~V. Koonin.
\newblock {Viral diversity threshold for adaptive immunity in prokaryotes.}
\newblock {\em mBio}, 3(6):1--10, 2012.

\bibitem{Iranzo2013a}
Jaime Iranzo, Alexander~E Lobkovsky, Yuri~I Wolf, and Eugene~V. Koonin.
\newblock {Evolutionary dynamics of the prokaryotic adaptive immunity system
  CRISPR-Cas in an explicit ecological context}.
\newblock {\em J. Bacteriol.}, 195(17):3834--3844, 2013.

\bibitem{Levin2013}
Bruce~R. Levin, Sylvain Moineau, Mary Bushman, and Rodolphe Barrangou.
\newblock {The Population and Evolutionary Dynamics of Phage and Bacteria with
  CRISPR-Mediated Immunity}.
\newblock {\em PLoS Genet.}, 9(3):e1003312, Jan 2013.

\bibitem{Santos2014}
S{\'{i}}lvio~B. Santos, Carla Carvalho, Joana Azeredo, and Eug{\'{e}}nio~C.
  Ferreira.
\newblock {Population dynamics of a Salmonella lytic phage and its host:
  Implications of the host bacterial growth rate in modelling}.
\newblock {\em PLoS ONE}, 9(7), 2014.

\bibitem{Berezovskaya2014}
Faina~S Berezovskaya, Yuri~I Wolf, Eugene~V Koonin, and Georgy~P Karev.
\newblock {Pseudo-chaotic oscillations in CRISPR-virus coevolution predicted by
  bifurcation analysis}.
\newblock {\em Biology Direct}, 9(1):13, 2014.

\bibitem{Westra2015}
Edze~R Westra, Stineke Van~houte, Sam Oyesiku-Blakemore, Ben Makin, Jenny~M
  Broniewski, Alex Best, Joseph Bondy-Denomy, Alan Davidson, Mike Boots, and
  Angus Buckling.
\newblock {Parasite exposure drives selective evolution of constitutive versus
  inducible defense}.
\newblock {\em Current Biology}, 25(8):1043--1049, 2015.

\bibitem{Ali2017}
Qasim Ali and Lindi~M. Wahl.
\newblock {Mathematical modelling of CRISPR-Cas system effects on biofilm
  formation}.
\newblock {\em J. Biol. Dyn.}, 11(S2):264--284, 2017.

\bibitem{Weissman2017}
Jake~L Weissman, Rayshawn Holmes, Rodolphe Barrangou, Sylvain Moineau,
  William~F Fagan, Bruce Levin, and Philip~LF Johnson.
\newblock {Immune Loss as a Driver of Coexistence During Host-Phage
  Coevolution}.
\newblock {\em bioRxiv}, page 105908, 2017.

\bibitem{Djordjevic2012}
Marko Djordjevic, Magdalena Djordjevic, and Konstantin Severinov.
\newblock {CRISPR transcript processing: a mechanism for generating a large
  number of small interfering RNAs}.
\newblock {\em Biology Direct}, 7(1):24, 2012.

\bibitem{Djordjevic2013}
Marko Djordjevic.
\newblock {Modeling bacterial immune systems : Strategies for expression of
  toxic – but useful – molecules}.
\newblock {\em BioSystems}, 112(2):139--144, 2013.

\bibitem{Guzina2017}
Jelena Guzina, Anđela Rodi{\'{c}}, Bojana Blagojevi{\'{c}}, and Marko
  Đorđevi{\'{c}}.
\newblock {Modeling and bioinformatics of bacterial immune systems:
  understanding regulation of CRISPR/Cas and restriction-modification systems}.
\newblock {\em Biologia Serbica}, 39(1):112--122, 2017.

\bibitem{Barrangou2014}
Rodolphe Barrangou and Luciano~A. Marraffini.
\newblock {CRISPR-Cas systems: Prokaryotes upgrade to adaptive immunity}.
\newblock {\em Molecular Cell}, 54(2):234--244, Apr 2014.

\bibitem{Emerson2013}
Joanne~B. Emerson, Karen Andrade, Brian~C. Thomas, Anders Norman, Eric~E.
  Allen, Karla~B. Heidelberg, and Jillian~F. Banfield.
\newblock {Virus-host and CRISPR dynamics in archaea-dominated hypersaline Lake
  tyrrell, Victoria, Australia}.
\newblock {\em Archaea}, 2013:370871, 2013.

\bibitem{Pride2011}
David~T Pride, Christine~L Sun, Julia Salzman, Nitya Rao, Peter Loomer, Gary~C
  Armitage, Jillian~F Banfield, and David~A Relman.
\newblock {Analysis of streptococcal CRISPRs from human saliva reveals
  substantial sequence diversity within and between subjects over time.}
\newblock {\em Genome Res.}, 21(1):126--36, Jan 2011.

\bibitem{Dennis1984}
B.~Dennis and G.~P. Patil.
\newblock {The gamma distribution and weighted multimodal gamma distributions
  as models of population abundance}.
\newblock {\em Mathematical Biosciences}, 68(2):187--212, 1984.

\bibitem{Engen1996}
S.~Engen and R.~Lande.
\newblock {Population dynamic models generating the lognormal species abundance
  distribution}.
\newblock {\em J. Theor. Biol}, 132(2):169--183, 1996.

\bibitem{Diserud2000}
O.~H. Diserud and S.~Engen.
\newblock {A general and dynamic species abundance model, embracing the
  lognormal and the gamma models}.
\newblock {\em The American Naturalist}, 155(4):497--511, 2000.

\bibitem{Plotkin2002}
Joshua~B. Plotkin and Helene~C. Muller-Landau.
\newblock {Sampling the species composition of a landscapre}.
\newblock {\em Ecology}, 83(12):3344--3356, 2002.

\bibitem{Levy2015}
Sasha~F. Levy, Jamie~R. Blundell, Sandeep Venkataram, Dmitri~A. Petrov,
  Daniel~S. Fisher, and Gavin Sherlock.
\newblock {Quantitative evolutionary dynamics using high-resolution lineage
  tracking}.
\newblock {\em Nature}, 519(7542):181--186, Feb 2015.

\bibitem{Deveau2008}
H{\'{e}}l{\`{e}}ne Deveau, Rodolphe Barrangou, Josiane~E. Garneau, Jessica
  Labont{\'{e}}, Christophe Fremaux, Patrick Boyaval, Dennis~A. Romero,
  Philippe Horvath, and Sylvain Moineau.
\newblock {Phage response to CRISPR-encoded resistance in Streptococcus
  thermophilus}.
\newblock {\em J. Bacteriol.}, 190(4):1390--1400, Feb 2008.

\bibitem{Sun2013}
Christine~L. Sun, Rodolphe Barrangou, Brian~C. Thomas, Philippe Horvath,
  Christophe Fremaux, and Jillian~F. Banfield.
\newblock {Phage mutations in response to CRISPR diversification in a bacterial
  population}.
\newblock {\em Environ. Microbiol.}, 15(2):463--470, Feb 2013.

\bibitem{Han2013}
Pu~Han, Liang~Ren Niestemski, Jeffrey~E Barrick, and Michael~W Deem.
\newblock {Physical model of the immune response of bacteria against
  bacteriophage through the adaptive CRISPR-Cas immune system.}
\newblock {\em Physical biology}, 10(2):025004, Apr 2013.

\bibitem{England2013}
Whitney~E. England and Rachel~J. Whitaker.
\newblock {Evolutionary causes and consequences of diversified CRISPR immune
  profiles in natural populations}.
\newblock {\em Biochem. Soc. Trans.}, 41(6):1431--1436, 2013.

\bibitem{Gore2009}
Jeff Gore, Hyun Youk, and Alexander {Van Oudenaarden}.
\newblock {Snowdrift game dynamics and facultative cheating in yeast}.
\newblock {\em Nature}, 459(7244):253--256, 2009.

\bibitem{Eldar2010}
Avigdor Eldar and Michael~B Elowitz.
\newblock {Functional roles for noise in genetic circuits.}
\newblock {\em Nature}, 467(7312):167--173, 2010.

\bibitem{Norman2015}
Thomas~M Norman, Nathan~D Lord, Johan Paulsson, and Richard Losick.
\newblock {Stochastic Switching of Cell Fate in Microbes}.
\newblock {\em Annual Review of Microbiology}, 69(1):381--403, 2015.

\bibitem{Tarnita2015}
Corina~E Tarnita, Alex Washburne, Ricardo Martinez-Garcia, Allyson~E Sgro, and
  Simon~a Levin.
\newblock {Fitness tradeoffs between spores and nonaggregating cells can
  explain the coexistence of diverse genotypes in cellular slime molds.}
\newblock {\em Proc Natl Acad Sci USA}, 112(9):2776--81, 2015.

\bibitem{Symmons2016}
Orsolya Symmons and Arjun Raj.
\newblock {What's Luck Got to Do with It: Single Cells, Multiple Fates, and
  Biological Nondeterminism}.
\newblock {\em Molecular Cell}, 62(5):788--802, 2016.

\bibitem{Shaffer2017}
Sydney~M Shaffer, Margaret~C Dunagin, Stefan~R Torborg, Eduardo~A Torre,
  Benjamin Emert, Clemens Krepler, Marilda Beqiri, Katrin Sproesser, Patricia~A
  Brafford, Min Xiao, Elliott Eggan, Ioannis~N Anastopoulos, Cesar~A.
  Vargas-Garcia, Abhyudai Singh, Katherine~L Nathanson, Meenhard Herlyn, and
  Arjun Raj.
\newblock {Rare cell variability and drug-induced reprogramming as a mode of
  cancer drug resistance}.
\newblock {\em Nature}, 546(7658):431--435, 2017.

\bibitem{Bunin2016}
Guy Bunin.
\newblock {Interaction patterns and diversity in assembled ecological
  communities}.
\newblock {\em arXiv}, 2016.

\bibitem{Tikhonov2016}
Mikhail Tikhonov.
\newblock {Community-level cohesion without cooperation}.
\newblock {\em eLife}, 5:e15747, 2016.

\bibitem{Tikhonov2017}
Mikhail Tikhonov and Remi Monasson.
\newblock {Collective Phase in Resource Competition in a Highly Diverse
  Ecosystem}.
\newblock {\em Phys. Rev. Lett.}, 118(4):1--5, 2017.

\bibitem{Biroli2017}
Giulio Biroli, Guy Bunin, and Chiara Cammarota.
\newblock {Marginally Stable Equilibria in Critical Ecosystems}.
\newblock {\em arXiv}, 2017.

\bibitem{Parikka2017}
Kaarle~J Parikka, Marc {Le Romancer}, Nina Wauters, and St{\'{e}}phan Jacquet.
\newblock {Deciphering the virus-to-prokaryote ratio (VPR): Insights into
  virus-host relationships in a variety of ecosystems}.
\newblock {\em Biological Reviews}, 92:1081--1100, 2017.

\bibitem{Burnham2002}
K.P. Burnham and D.R. Anderson.
\newblock {\em {Model Selection and Multimodel Inference}}.
\newblock Springer-Verlag New York, 2002.

\bibitem{Brussow2002}
Harald Br{\"{u}}ssow and Roger~W. Hendrix.
\newblock {Phage Genomics: Small is beautiful}.
\newblock {\em Cell}, 108(1):13--16, Jan 2002.

\bibitem{Held2013}
Nicole~L Held, Lauren~M Childs, Michelle Davison, Joshua~S Weitz, Rachel~J
  Whitaker, and Devaki Bhaya.
\newblock {CRISPR-Cas systems to probe ecological diversity and host-viral
  interactions}.
\newblock In {\em CRISPR-Cas Systems}, pages 221--250. 2013.

\bibitem{Payet2013}
J{\'{e}}r{\^{o}}me~P. Payet and Curtis~A Suttle.
\newblock {To kill or not to kill: The balance between lytic and lysogenic
  viral infection is driven by trophic status}.
\newblock {\em Limnology and Oceanography}, 58(2):465--474, 2013.

\bibitem{Kasman2002}
Laura~M Kasman, Alex Kasman, Caroline Westwater, Joseph Dolan, Michael~G
  Schmidt, and James~S Norris.
\newblock {Overcoming the phage replication threshold: a mathematical model
  with implications for phage therapy.}
\newblock {\em J. Virol.}, 76(11):5557--64, 2002.

\bibitem{Knowles2016}
B~Knowles, C~B Silveira, B~A Bailey, K~Barott, V~A Cantu, A.~G.
  Cobian-Gu{\"{e}}mes, F~H Coutinho, E~A Dinsdale, B~Felts, K~A Furby, E~E
  George, K~T Green, G~B Gregoracci, A~F Haas, J~M Haggerty, E~R Hester,
  N~Hisakawa, L~W Kelly, Y~W Lim, M~Little, A~Luque, T.~McDole-Somera,
  K.~McNair, L.~S. {De Oliveira}, S~D Quistad, N~L Robinett, E~Sala, P~Salamon,
  S~E Sanchez, S~Sandin, G.~G.Z. Silva, J~Smith, C~Sullivan, C.~Thompson,
  M.~J.A. Vermeij, M~Youle, C~Young, B~Zgliczynski, R~Brainard, R~A Edwards,
  J~Nulton, F~Thompson, and F~Rohwer.
\newblock {Lytic to temperate switching of viral communities}.
\newblock {\em Nature}, 531(7595):466--470, 2016.

\bibitem{Cao2006}
Yang Cao, Daniel~T. Gillespie, and Linda~R. Petzold.
\newblock {Efficient step size selection for the tau-leaping simulation
  method.}
\newblock {\em J. Chem. Phys}, 124(4):044109, 2006.

\bibitem{Gilles2011}
Andr{\'{e}} Gilles, Emese Megl{\'{e}}cz, Nicolas Pech, St{\'{e}}phanie
  Ferreira, Thibaut Malausa, and Jean-Fran{\c{c}}ois Martin.
\newblock {Accuracy and quality assessment of 454 GS-FLX Titanium
  pyrosequencing}.
\newblock {\em BMC Genomics}, 12(1):245, 2011.

\bibitem{Lucchini1999}
Sacha Lucchini.
\newblock {\em {Genetic Diversity of Streptococcus thermophilusPhages and
  Development of}}.
\newblock PhD thesis, Swiss Federal Institute of Technology Zurich, 1999.

\bibitem{Delbruck1940}
M.~Delbr{\"{u}}ck.
\newblock {Adsorption of bacteriophage under various physiological conditions
  of the host}.
\newblock {\em J. Gen. Physiol.}, 23(5):631--42, 1940.

\bibitem{Vaningelgem2004}
F.~Vaningelgem, M.~Zamfir, T.~Adriany, and Luc {De Vuyst}.
\newblock {Fermentation conditions affecting the bacterial growth and
  exopolysaccharide production by Streptococcus thermophilus ST 111 in
  milk-based medium}.
\newblock {\em J. Appl. Microbiol.}, 97(6):1257--1273, 2004.

\bibitem{VanKampen1981}
N.~G. van Kampen.
\newblock {\em {Stochastic Processes in Physics and Chemistry}}.
\newblock Elsevier, third edition, 1981.

\bibitem{McGill2007}
Brian~J. McGill, Rampal~S. Etienne, John~S. Gray, David Alonso, Marti~J.
  Anderson, Habtamu~Kassa Benecha, Maria Dornelas, Brian~J. Enquist, Jessica~L.
  Green, Fangliang He, Allen~H. Hurlbert, Anne~E. Magurran, Pablo~A. Marquet,
  Brian~A. Maurer, Annette Ostling, Candan~U. Soykan, Karl~I. Ugland, and
  Ethan~P. White.
\newblock {Species abundance distributions: Moving beyond single prediction
  theories to integration within an ecological framework}.
\newblock {\em Ecology Letters}, 10(10):995--1015, 2007.

\bibitem{Fisher1943}
R.~A. Fisher, A.~Steven Corbet, and C.~B. Williams.
\newblock {The relation between the number of species and the number of
  individuals in a random sample of an animal population}.
\newblock {\em J. Animal Ecology}, 12(1):42--58, 1943.

\bibitem{Chisholm2010}
Ryan~A. Chisholm and Stephen~W. Pacala.
\newblock {Niche and neutral models predict asymptotically equivalent species
  abundance distributions in high-diversity ecological communities}.
\newblock {\em Proc Natl Acad Sci USA}, 107(36):15821--15825, 2010.

\bibitem{Miller2001}
Melissa~B. Miller and Bonnie~L. Bassler.
\newblock {Quorum Sensing in Bacteria}.
\newblock {\em Annual Review of Microbiology}, 55(1):165--199, 2001.

\bibitem{Gandon2015}
Sylvain Gandon, Pedro~F. Vale, Guillaume Lafforgue, Francois Gatchitch, Rozenn
  Gardan, and Sylvain Moineau.
\newblock {Costs of CRISPR-Cas-mediated resistance in Streptococcus
  thermophilus}.
\newblock {\em Proc Biol Sci}, 282(1812):20151270, 2015.

\bibitem{Huttenhower2012}
Curtis Huttenhower and et~al.
\newblock {Structure, function and diversity of the healthy human microbiome}.
\newblock {\em Nature}, 486(7402):207--214, Jun 2012.

\bibitem{Madigan2006}
Michael~T. Madigan and John~M. Martinko.
\newblock {\em {Brock biology of microorganisms}}.
\newblock Pearson Prentice Hall, 11 edition, 2006.

\end{thebibliography}
\end{document}